%% file: SciPostPhysCore.tex
\documentclass{article}
\usepackage{arxiv}

\usepackage[bitstream-charter]{mathdesign}

\usepackage{subcaption}
 \usepackage{amsmath}
 \usepackage{url} 
 \usepackage{natbib}
 
\newcommand{\vc}[1]{{\bf #1 }}
\newcommand{\bt}{\pmb{\theta}}

\usepackage{algorithm}
\usepackage{algorithmic}
\usepackage{tikz}
\usepackage{tcolorbox}

\newenvironment{algo}[1]{
\begin{tcolorbox}
\textbf{Algorithm: #1}
\vspace{4pt}
\begin{algorithmic}
}{
\end{algorithmic}
\end{tcolorbox}
}

\begin{document}

\title{A thermodynamic approach to Approximate Bayesian Computation with multiple Summary Statistics}

\renewcommand{\shorttitle}{Thermodynamic ABC with multiple Summary Statistics}

\maketitle

\begin{center}
\textbf{
Carlo Albert\textsuperscript{1$\star$},
Simone Ulzega\textsuperscript{2},
Simon Dirmeier\textsuperscript{3,4},
Andreas Scheidegger\textsuperscript{1},
Alberto Bassi\textsuperscript{1,4} and 
Antonietta Mira\textsuperscript{5}
}\end{center}

\begin{center}
{\bf 1} Swiss Federal Institute of Aquatic Science and Technology
\\
{\bf 2} Zurich University of Applied Sciences
\\
{\bf 3} Swiss Data Science Center
\\
{\bf 4} ETH Zurich
\\
{\bf 5} Università della Svizzera italiana and University of Insubria

$\star$ {\small carlo.albert@eawag.ch}
\end{center}

\begin{abstract}
Bayesian inference with stochastic models is often difficult because their likelihood functions involve high-dimensional integrals.
Approximate Bayesian Computation (ABC) avoids evaluating the likelihood function and instead infers model parameters by comparing model simulations with observations using a few carefully chosen summary statistics and a tolerance that can be decreased over time.
Here, we present a new variant of simulated-annealing ABC algorithms, drawing intuition from non-equilibrium thermodynamics. We associate each summary statistic with a state variable (energy) quantifying its distance from the observed value, as well as a temperature that controls the extent to which the statistic contributes to the posterior.
We derive an optimal annealing schedule on a Riemannian manifold of state variables based on a minimal-entropy-production principle. 
We validate our approach on standard benchmark tasks from the simulation-based inference literature as well as on challenging real-world inference problems, and show that it is highly competitive with the state of the art.
\end{abstract}

\vspace{\baselineskip}

\section{Introduction}

Bayesian inference algorithms that rely on many likelihood evaluations are impractical when each evaluation requires integrating over a very high-dimensional space, as is common for stochastic models.
{\em Simulation-based inference} (SBI) methods avoid direct likelihood evaluation and approximate the posterior by comparing many simulated model outputs produced under diverse parameter values.
Since the number of model parameters is usually far smaller than the dimensionality of the output, a relatively small set of {\em summary statistics} often suffices to approximate the posterior well. This provides the key efficiency advantage of simulation-based inference algorithms.

There are two classes of simulation-based algorithms. The first uses neural density estimators, Machine Learning (ML) models that attempt to learn either the posterior density or the likelihood directly. At the same time, they typically also attempt to learn a minimal set of near-sufficient summary statistics.
The second class comprises algorithms that {\em sample} from an approximate posterior by comparing simulated summary statistics with those observed.
Since the comparison typically relies on a strictly positive tolerance, which introduces an additional layer of approximation, these sampling methods are known as {\em Approximate Bayesian Computation} (ABC).
While ML methods can be very fast, they often suffer from biases that are hard to control. Furthermore, they typically involve a large number of hyperparameters that require non-trivial tuning. ABC algorithms, on the other hand, tend to be simulation-intensive, but their bias can be controlled through the tolerance between simulated and observed statistics.
Furthermore, ABC algorithms based on {\em simulated annealing} (SABC, \cite{Albert_2013_ABC, Albert_2015_ABC}) are largely self-tuning, as they lower the tolerance adaptively during the course of the algorithm based on a {\em minimal entropy production} principle.
One shortcoming of existing ABC algorithms is that the user must choose a metric on the space of summary statistics. This choice can greatly influence the efficiency of the algorithm.
While standardizing the summary statistics with respect to the prior can balance their contribution to the acceptance probability to some extent, their prior distributions can still differ markedly. A large skewness of certain statistics, for instance, could mean that they quickly lose their influence once the few “bad samples” have been removed during the course of the algorithm.
The original SABC algorithm “rectifies” the user-defined metric such that it follows a uniform distribution under the prior. Here, we propose to rectify the univariate distances {\em individually} for each summary statistic. We expect this to balance the contribution of different summary statistics to the acceptance probability of proposed moves throughout the algorithm.
Furthermore, we propose a variant of SABC in which each summary statistic is equipped with its own dynamic tolerance (or temperature). The rationale is that more informative summary statistics are expected to converge faster: because they couple more strongly to the parameters, they benefit more from the convergence toward the posterior. Consequently, their tolerances should decay more rapidly and thereby accelerate overall convergence toward the true posterior.

Ideally, summary statistics should encode most of the information relevant for constraining the parameters while filtering out as much noise as possible.
Such statistics can be found by means of ML models (e.g., \cite{chen2020neural}, \cite{albert2022AE}, \cite{chen2023learning}).In practice, however, the information relevant for constraining the parameters will be distributed quite unevenly across the statistics, and may even change over the course of the annealing. To some extent, our algorithms can counterbalance this unevenness and thus enable an optimal extraction of information for each model simulation. Furthermore, our method allows the convergence of each summary statistic to be monitored individually. In cases where the observations are out of sample, this can help identify the features that the model struggles to fit. Allowing those features to be less compliant with the observations by using individual temperatures can also improve robustness.

We benchmark our new SABC variants against an ABC variant that is not based on annealing, as well as state-of-the-art ML approaches, and show that they are generally competitive and can be superior in hard cases.
Finally, we apply them to three challenging real-world scenarios: a high-dimensional model from epidemiology, a stochastic model with intractable likelihood from neuroscience, and a case study with real data from solar physics.

\section{Theory}\label{Sect:Theory}

Consider a stochastic model with an intractable probability density $f(\vc x|\bt)$ that depends on model parameters $\bt$.
We are given a low-dimensional vector of summary statistics, $\vc s(\vc x)$, and denote the induced joint prior over statistics and parameters by $f(\vc s,\bt)=f(\vc s|\bt)f(\bt)$.
We are interested in the {\em posterior}, that is the joint prior conditioned on the summary statistics of observed data, $f(\bt|\vc s(\vc x_{obs}))$.
If we assume that the statistics are {\em sufficient}, this posterior will be identical to the full posterior conditioned on all the data, $f(\bt|\vc x_{obs})$.
However, we do not impose this requirement here.

SABC algorithms initialize a population of particles drawn from the joint prior $f(\vc s,\bt)$ and iteratively update them by sampling from the transition probability
\begin{multline}\label{eq:oldupdate}
q(\vc s',\bt'|
\vc s,\bt):=
k(\bt'|\bt)f(\vc s'|\bt')\min(1,\exp[-\beta^e(\rho(\vc s')-\rho(\vc s))]f(\bt')/f(\bt))
+\delta(\bt-\bt')\delta(\vc s-\vc s')r(\vc s,\bt)
\,,
\end{multline}
with appropriate rejection probability $r(\vc s,\bt)$.
That is, we propose a new particle position by first drawing a random jump in parameter space from the symmetric distribution $k(\bt'|\bt)$, then {\em simulating} a new model output $\vc x' \sim f(\vc x'|\bt')$ and computing its summary statistics $\vc s'=\vc s(\vc x')$. The proposal is accepted with the usual Metropolis probability, based on the ratio of the prior densities of the parameters and on the change in distance between the simulated summary statistics and the target $\vc s_{obs}$, as measured by a user-specified metric $\rho(\cdot)$.
The inverse temperature $\beta^e$ is gradually increased, forcing the particle population to become increasingly compliant with the data. \cite{Albert_2013_ABC} show that convergence to the posterior $f(\bt|\vc s_{obs})$ is guaranteed provided the annealing is sufficiently slow, following a power law determined by the number of summary statistics.
In practice, we prefer to use an annealing schedule that adapts the temperature based on the current average distance of the particles from the target. The adaptive SABC algorithm reparametrizes the distance function $\rho(\cdot)$ using an {\em energy} function in such a way that a minimal entropy production principle can be invoked for an optimal annealing (\cite{Albert_2013_ABC}).

When multiple summary statistics are involved, defining a single metric $\rho(\vc s)$ that balances their influence on the Metropolis probability can be difficult.
Our proposed solution is to assign a separate energy to each statistic.
First, we replace the statistics $s^i$, $i=1,\dots,n$, with their distances $\rho^i=\rho^i(s^i)$ to the observed values $s^i_{obs}$.
Next, we reparametrize these distances using their cdfs under the prior marginals $f(\rho^i)$, and define the new energy coordinates $u^i$ as
\begin{equation}\label{eq:rho}
    u^i(\rho^i)
    :=
    \int_{\tilde\rho^i<\rho^i}
    f(\tilde\rho^i)d\tilde\rho^i
    \,.
\end{equation}
This choice enables us to derive the temperatures directly from the observed energies (see eq.~\ref{betacorr} below).
In practice, this reparameterization can be approximated during initialization, when a large prior sample of parameters and corresponding summary statistics is generated.
We introduce an individual temperature, $\beta^e_i$ ($i=1,\dots,n$), for each summary statistic and replace update rule (\ref{eq:oldupdate}) by
\begin{multline}\label{eq:q}
q(\vc u',\bt'|
\vc u,\bt):=
k(\bt'|\bt)f(\vc u'|\bt')\min\left(
1,\exp\left[
-\sum_i \beta^e_i({u'}^i-u^i)
\right]
\frac{f(\bt')}{f(\bt)}
\right)
+\delta(\bt-\bt')\delta(\vc u-\vc u')r(\vc u,\bt)
\,
\end{multline}
(sequentially or in parallel) to a large number of particles.
If annealing proceeds slowly enough relative to the mixing in parameter space, we may invoke the {\em endoreversibility assumption}, whereby the population approximates the distribution at any point in time
\begin{equation}\label{eq:pi}
\pi_{\boldsymbol{\beta}}(\vc u,\bt)
=
Z^{-1}(\boldsymbol{\beta})f(\vc u,\bt)\exp\left[
-\sum_i\beta_iu^i
\right]\,,
\end{equation}
where the inverse {\em internal} temperatures, $\beta_i$, are slightly smaller than the respective external ones $\beta^e_i$. 

The adaptive annealing scheme we use is grounded in the principle of {\em minimal entropy production}.
When the particle population is large and only a small fraction of particles is updated in each step, we may assume the observable {\em energy densities}
\begin{equation}\label{eq:U}
    \vc U(\boldsymbol{\beta}):=
    \int \vc u
    \pi_{\boldsymbol{\beta}}(\vc u,\bt)
    d\vc u d\bt
\end{equation}
to be continuously varying in time. 
Under the endoreversibility assumption, the entropy production rate can be expressed as the product of thermodynamic fluxes and forces, as
\begin{equation}
    \dot S_{prod}=\sum_i\dot U^iF_i
    \,,
\end{equation}
where $F_i:=\beta^e_i-\beta_i$ (see \cite{Albert_2013_ABC} and references therein for the thermodynamic background).
Using (\ref{eq:q}) through (\ref{eq:U}) and the {\em master equation} describing the dynamics of (\ref{eq:pi}), we derive the fluxes as \cite{Albert_2013_ABC}
\begin{multline}\label{eq:Udot1}
    \dot {\vc U}(\boldsymbol{\beta},\boldsymbol{\beta}^e)
    =
    Z^{-1}(\boldsymbol{\beta})
    \int
    (\vc u-\vc u')
    k(\bt|\bt')
    f(\vc u|\bt)
    f(\vc u',\bt')
    \\
    \times
     \min\left(
         1,
         \exp\left[
             -\sum_i\beta^e_i(u^i-{u'}^i)
         \right]
         \frac{f(\bt)}{f(\bt')}
         \right)
         \exp\left[
             -\sum_i\beta_i{u'}^i
         \right]
     d\vc u d\bt
     d\vc u' d\bt'
     \,.
\end{multline}
We assume that the annealing is slow enough to warrant the linearity assumption
\begin{equation}
    \dot{\vc U}
    \approx
    L(\vc U)\vc F
    \,.
\end{equation}
Minimal entropy production, for fixed initial and final energies (with the final energies $\vc U=0$), implies that the system should follow a {\em geodesic curve} with respect to the metric given by the negative inverse {\em Onsager matrix}, $g_{ij}(\vc U):=-L_{ij}^{-1}(\vc U)$.
In practice, this metric is typically intractable.
To obtain a first-order approximation, we introduce three assumptions that are only approximately valid.
Specifically, we assume (i) a largely uninformative prior ($f(\bt)\approx 1$), (ii) a broad jump distribution ($k(\bt|\bt')\approx 1$), and (iii) summary statistics chosen such that, at least for small $\vc u$, $f(\vc u)\approx\operatorname{const}$.
Assumption (i) is generally reasonable. Indeed, if the prior is highly informative, gradually lowering the tolerance (as done in sequential Monte Carlo (SMC)-ABC \cite{beaumont_2009_aABC} or SABC) may even be counterproductive, and a simple rejection ABC may be preferable.
Assumption (ii) requires sufficiently fast mixing, which is necessary for the endoreversibility condition to hold.
Assumption (iii) entails that the summary statistics are not strongly correlated and that the prior is broad relative to the posterior—both reasonable expectations.
Under these assumptions, the parameters can trivially be integrated out, and (\ref{eq:Udot1}) is approximated by the integral
\begin{equation}
    \dot{\vc U}
    \approx
    c_0
    \int
    (\vc u-\vc u')
    \min\left(1,\exp\left[-\sum_i\beta^e_i(u^i-{u'}^i)\right]\right)
    \times\exp\left[-\sum_i\beta_i{u'}^i\right]
    \prod_i\beta_i du^id{u'}^i
    \,,
\end{equation}
where $c_0$ is a constant that does not strongly depend on the number of summary statistics. 
Introducing coordinates $u^i_\pm=u^i\pm {u'}^i$, and assuming that $\beta_i\gg1$, we find that 
\begin{equation}\label{eq:udot2}
    \dot{\vc U}
    \approx
    c_0
    \int
    \vc u_-
    \min\left(
    1,
    \exp\left[-\sum_i\beta^e_iu^i_-\right]
   \right)
    \prod_i
    \min\left(
    1,
    \exp\left[\sum_j\beta_j u^j_-\right]
    \right)
    du_-^i
    \,.
\end{equation}
Expanding the r.h.s. around $\boldsymbol{\beta}_e=\boldsymbol{\beta}$, yields (see Appendix \ref{app:proof} for details)
\begin{multline} \label{Theorem}
    L^{ij}(\vc U)
    \approx
    -c_0
    \int
    u_-^iu_-^j
    \chi\left(
    \sum_k\beta_ku_-^k
    \right)
    \prod_l
    \min\left(
    1,
    \exp\left[-\sum_k\beta_k u_-^k\right]
    \right)
    du_-^l
    \\
    =
    -c_n
    \left(\prod_kU^k\right)
    U^iU^j
    (-1+\delta^{ij}(n+1))\,,
\end{multline}
where the coefficients $c_n=c_0 (2n+2)!/((n+1)!(n+2)!)$ are proportional to the {\em Catalan numbers}.
Inverting this matrix yields the metric
\begin{equation}\label{eq:metric}
    g_{ij}(\vc U)
    =
    \frac{1}{c_n(n+1)}
    \frac{1}{U^iU^j}
    \prod_k (U^k)^{-1}
    \left(
        \delta_{ij}
        +
        1
    \right)
    \,.
\end{equation}
The change of coordinates $P^i=\ln U^i$ renders this metric {\em conformally flat}:
\begin{equation}
    g_{ij}(\vc P)
    =
    c_n^{-1}
     e^{-\Phi(\vc P)}\eta_{ij}\,,
\end{equation}
with $\Phi(\vc P)=\sum_iP^i$ and $\eta_{ij}=(\delta_{ij}+1)/(n+1)$.
To determine the geodesics, we need to compute the {\em Christoffel symbols}\footnote{Here and throughout, we adopt the Einstein convention, automatically summing over repeated co- and contravariant indices.}
\begin{equation}
    \Gamma^i_{jk}
    =
    \frac{1}{2}
    g^{il}(g_{lj,k}+g_{lk,j}-g_{jk,l})
    =
    \frac{1}{2}\bigl(-\delta^i_j-\delta^i_k+\eta_{jk}\bigr)
    \,.
\end{equation}
By symmetry, the simplest geodesics in $\vc P$-space are straight lines of unit slope.
Imposing $\dot P^i=\dot P^j$ for all $i,j$, the geodesic equation reduces to
\begin{equation}
    \ddot P^i 
    =
    -\Gamma^i_{jk}\dot P^j\dot P^k
    =
    \frac{n}{2}(\dot P^i)^2 
    \,.
\end{equation}
Thus, $P^i(t)=P^i_0-(2/n)\ln (vnt/2+1)$, where $v$ denotes the velocity, and which reads in $\vc U$ space as
\begin{equation}
    U^i(t)=U^i_0\left(\frac{vn}{2}t+1\right)^{-2/n}\,.
\end{equation}
Hence,
\begin{equation}\label{eq:Udot}
    \dot U^i(t)
    =
    -v(U^i_0)^{-n/2}(U^i(t))^{1+n/2}
    \,,
\end{equation}
from which we see that geodesics are straight lines ending at $\vc U=0$.
Finally, we need to calculate the thermodynamic force, using (\ref{eq:metric}) and (\ref{eq:Udot}).
Setting the initial conditions to $U^i_0=1$, we get
\begin{equation}
    F_i(\vc U)
    =
    -g_{ij}(\vc U)
    \dot U^j
    =
    v
    \frac
    {
        1+\sum_j(U^j/U^i)^{n/2}
    }
    {c_n(n+1)(U^i)^{1+n/2}\prod_j(U^j/U^i)}
    \,,
\end{equation}
from which we finally get the adaptive annealing schedule
\begin{equation}\label{adaptiveForcecorr}
    \beta^e_i(\vc U)
    =
    \beta_i(U^i)
    + v
    \frac
    {
        1+\sum_j(U^j/U^i)^{n/2}
    }
    {c_n(n+1)(U^i)^{1+n/2}\prod_j(U^j/U^i)}
    \,.
\end{equation}
Note that the approximation $\beta_i\approx 1/U^i$ is valid only for $U^i\ll 1$.
However, at the beginning of the algorithm, $U^i=1/2$. 
Including the boundary condition that $u^i\leq 1$, we get the relation
\begin{equation}\label{betacorr}
    U^i=\frac
    {1-e^{-\beta_i}(1+\beta_i)}
    {\beta_i(1-e^{-\beta_i})}\,,
\end{equation}
which yields $U^i=1/2$, for $\beta_i=0$, as it ought to be. 
We therefore suggest to numerically solve eq. (\ref{betacorr}) for determining the current temperatures $\beta_i(U^i)$, and then use eq. (\ref{adaptiveForcecorr}) to update the annealing temperature.

If the trajectory did stay on the diagonal, eq. (\ref{adaptiveForcecorr}) would simplify to \\
$\beta^e_i(\vc U) =
    \beta_i(U^i)
    + (v/c_n)(U^i)^{1+n/2}$.
Because Assumptions (i)–(iii) are not generally satisfied, the trajectories tend to deviate from the diagonal.
Furthermore, allowing certain statistics to converge more quickly can be beneficial, as they typically contain more information about the parameters (see next section).
However, this flexibility may also cause instabilities in certain situations.
To address this, we provide a variant in which all $\beta^e_i$ are constrained to be equal, keeping the trajectory closer to the diagonal.
In that setting, we recommend using the original annealing schedule of \cite{Albert_2013_ABC}. The only change is that the user no longer aggregates distances across summary statistics; instead, the algorithm assigns an energy to each distance individually and sums them.
We refer to this version as {\em SABC single}, and the original formulation as {\em SABC multi}.
The basic (serial) form of SABC multi is presented below. Further refinements we have used in our experiments are given in Appendix \ref{app:benchmark-tasks-experiments}.
For the version with {\em single} temperature, we set all $\beta^e_i=\beta^e$, which is updated adaptively as described in \cite{Albert_2013_ABC} using the energy $U=(1/N)\sum_{i,\alpha}u_\alpha^i$ of the current population.

\begin{algo}{SABC (multi)}
\STATE \textbf{Given:} (i) prior $f(\boldsymbol{\theta})$, (ii) simulator $f(\mathbf{x}\mid\boldsymbol{\theta})$, (iii)
       summary statistics $\mathbf{s}(\mathbf{x})$, (iv) distance functions, 
       $\rho^i(s^i)$, for all components $i$, (v) number of population updates $M$.
\STATE \textbf{Initialize:}
       draw a population of particles $\{(\boldsymbol{\theta}_\alpha,\mathbf{s}_\alpha)\}_{\alpha=1}^N$ 
       by sampling $\boldsymbol{\theta}_\alpha \sim f(\boldsymbol{\theta})$ and 
       $\mathbf{x}_\alpha \sim f(\mathbf{x}\mid\boldsymbol{\theta}_\alpha)$, then 
       computing $\mathbf{s}_\alpha = \mathbf{s}(\mathbf{x}_\alpha)$.
\STATE For each component $i$, define the energy function (according to eq. (\ref{eq:rho}))
\begin{equation}\label{eq:uapprox}
    u^i(\rho) := \#\bigl\{ \alpha \,\big|\, \rho^i(s^i_\alpha) < \rho \bigr\}.
\end{equation}
\STATE Transform initial $\vc s_\alpha$ vectors to $\vc u_\alpha$ vectors. 
\STATE Set the initial inverse temperatures $\beta^e_i \gets 0$.
\vspace{0.5em}
\FOR{$m = 1$ to $MN$}
  \STATE Draw a random particle $(\boldsymbol{\theta}_\alpha,\mathbf{u}_\alpha)$ from the population.
  \STATE Propose a new parameter $\boldsymbol{\theta}' \sim k(\boldsymbol{\theta}'\mid\boldsymbol{\theta}_\alpha)$.
  \STATE Simulate $\mathbf{x}' \sim f(\mathbf{x}\mid\boldsymbol{\theta}')$ and compute
         $\mathbf{s}' = \mathbf{s}(\mathbf{x}')$.
  \STATE Compute distances to the observation
         ${\rho'}^{\,i} = \rho^i({s'}^{\,i})$ and transform them to energies
         ${u'}^{\,i} = u^i({\rho'}^{\,i})$, for all components $i=1,\dots,n$.
  \STATE Accept the proposal with probability
  \[
    \min\left\{
      1,\;
      \exp\!\left[-\sum_i \beta^e_i \bigl({u'}^{\,i} - u_\alpha^i\bigr)\right]
      \frac{f(\boldsymbol{\theta}')}{f(\boldsymbol{\theta}_\alpha)}
    \right\}.
  \]
  \STATE If the proposal is accepted, replace 
         $(\boldsymbol{\theta}_\alpha,\vc u_\alpha)$ by 
         $(\boldsymbol{\theta}',{\vc u'})$ in the population.
  \STATE Update $\beta^e_i$ adaptively using eq. (\ref{adaptiveForcecorr}) where $U^i=(1/N)\sum_{\alpha} u^i_\alpha$ and the $\beta$'s are calculated inverting eq. (\ref{betacorr}).
\ENDFOR
\vspace{0.5em}
\STATE \textbf{return} final population $\{(\boldsymbol{\theta}_\alpha,\mathbf{u}_\alpha)\}_{\alpha=1}^N$
\end{algo}

\input{sections/experiments}

\section*{SABC software packages}

The code is available as a Julia package, \url{https://github.com/Eawag-SIAM/SimulatedAnnealingABC.jl},
and as a Python package, \url{https://github.com/ulzegasi/SimulatedAnnealingABC.git}.

\section*{Acknowledgement}
This work has been supported by the Swiss National Science Foundation (Grant N$^0$ 208249).

\clearpage
\begin{appendix}
\numberwithin{equation}{section}

\input{sections/appendix}

\end{appendix}

\clearpage

\bibliographystyle{apalike}

\bibliography{refs}

\end{document}

%% file: sections/experiments.tex
\section{Experiments}
\subsection{Benchmark tasks}
We evaluate the introduced SABC variants on four benchmark tasks.
Three are standard in the SBI literature: the two moons example, a hyperboloid model, and a mixture model
(\cite{lueckmann2021benchmarking}).
The fourth is a Gaussian mixture model with distractors, mimicking the presence of uninformative summary statistics.
When sampling its posterior with SABC, both the informative and the uninformative components of proposed particles have a decreasing chance of getting accepted as they are zeroing in on the observed values.
However, the informative components have an advantage as they profit from the convergence of the parameters towards the posterior.
This allows them to decay faster than the uninformative statistics when equipped with their own tolerances, and exemplifies the advantage of using multiple temperatures.
More details on the models can be found in Appendix~\ref{app:benchmark-tasks-models}.

To systematically assess estimator variance, we repeat all four benchmark experiments using five different random seeds.
For each seed, we generate ground-truth posterior samples as described in Appendix~\ref{app:benchmark-tasks-experiments} and compare them with the approximate posteriors obtained from each method.
Method performance is then evaluated using classifier two-sample tests (C2ST; \cite{lopezpaz2017revisiting}), maximum mean discrepancy (MMD; \cite{sutherland2017generative}), and the recently suggested H-min distance \citep{zhao2022comparing,dirmeier2023simulation}.

\begin{figure}
    \centering
    \includegraphics[width=\textwidth]{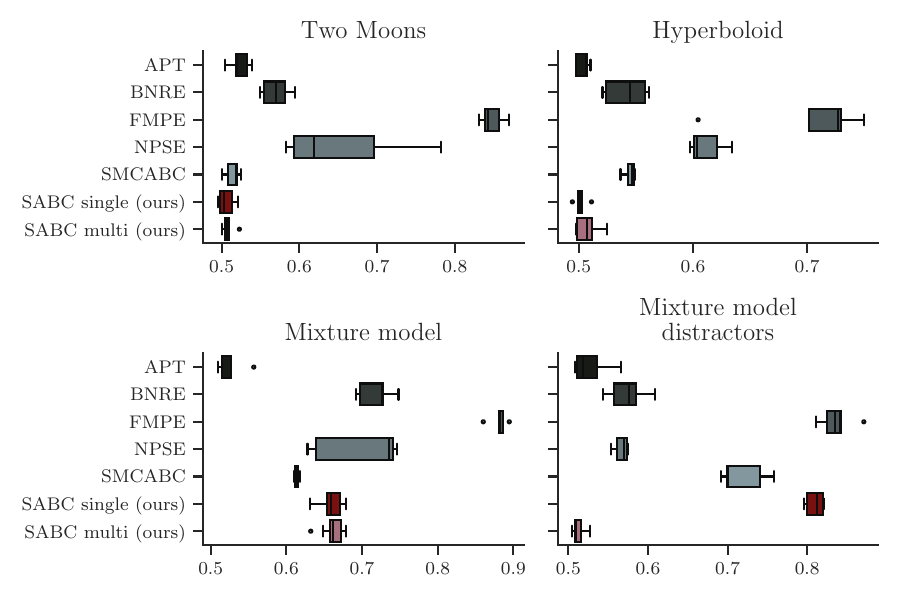}
    \caption{SABC performance on benchmark tasks using a C2ST metric (lower is better, ideal is 0.5).}
    \label{fig:sabc-benchmark-tasks-c2st}
\end{figure}

\begin{figure}
\centering
\begin{subfigure}[b]{0.3\textwidth}
\includegraphics[width=1\textwidth]{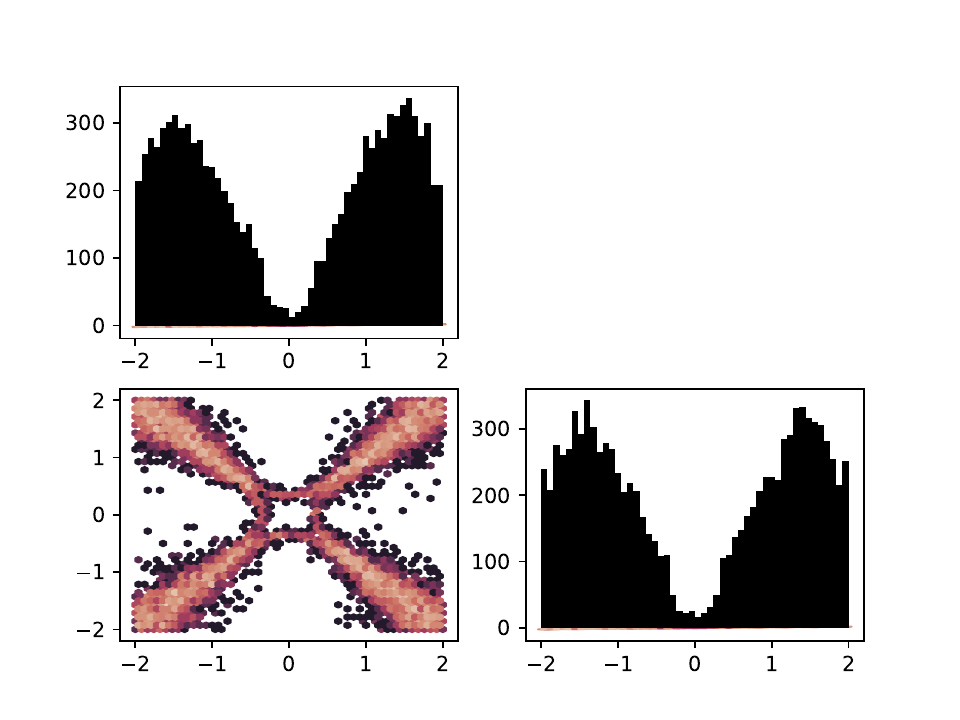}
\caption{Reference.}
\end{subfigure}
\begin{subfigure}[b]{0.3\textwidth}
\includegraphics[width=1\textwidth]{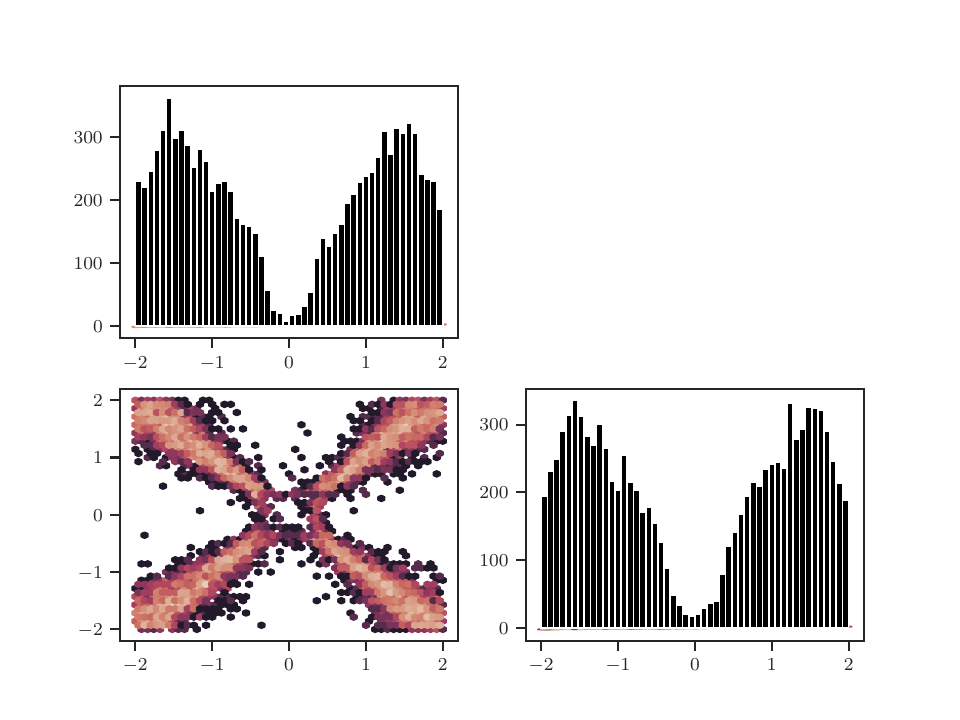}
\caption{APT.}
\end{subfigure}
\begin{subfigure}[b]{0.3\textwidth}
\includegraphics[width=1\textwidth]{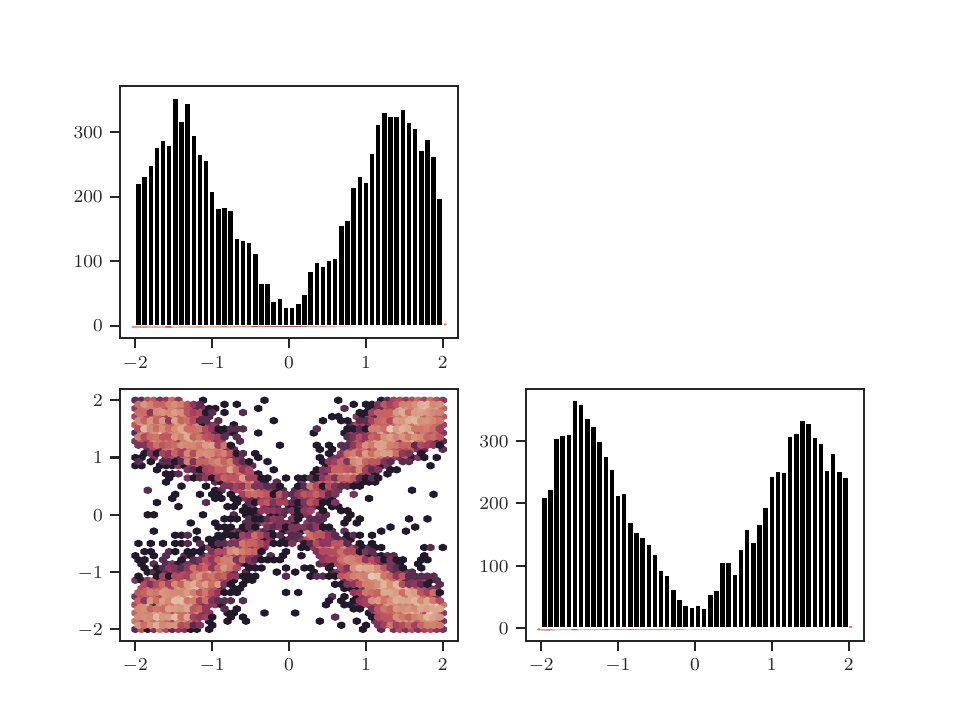}
\caption{BNRE.}
\end{subfigure}
~
\begin{subfigure}[b]{0.3\textwidth}
\includegraphics[width=1\textwidth]{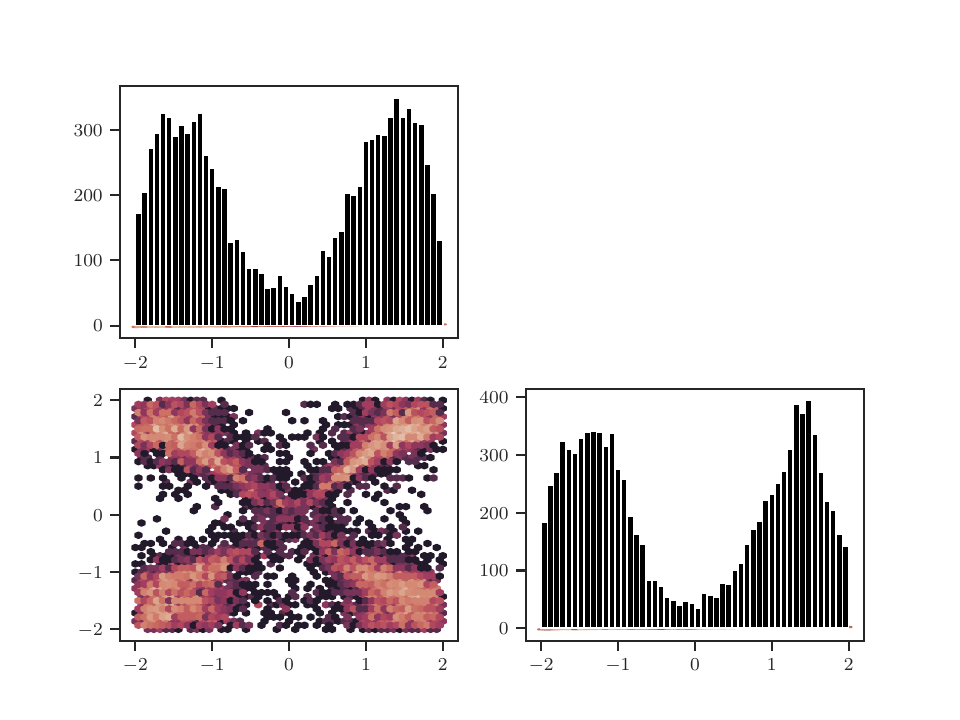}
\caption{FMPE.}
\end{subfigure}
\begin{subfigure}[b]{0.3\textwidth}
\includegraphics[width=1\textwidth]{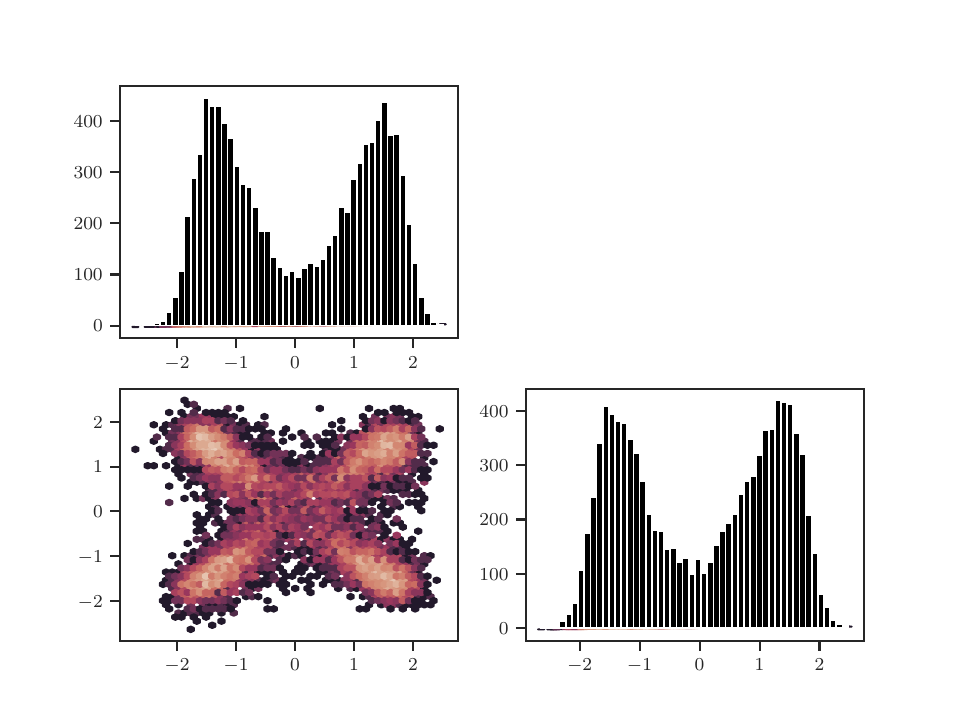}
\caption{NPSE.}
\end{subfigure}
\begin{subfigure}[b]{0.3\textwidth}
\includegraphics[width=1\textwidth]{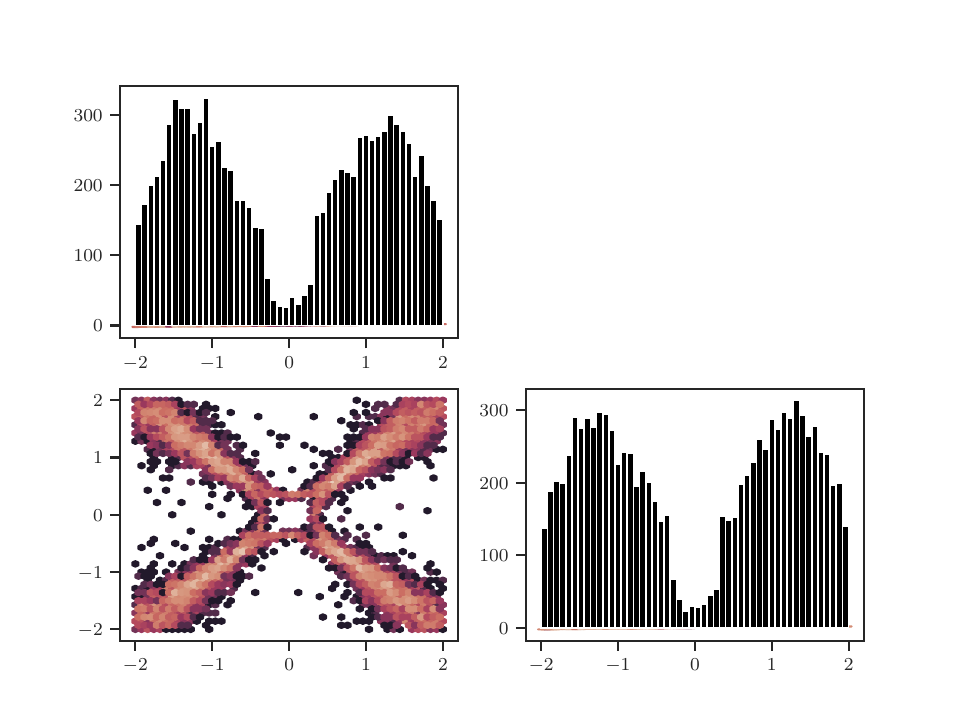}
\caption{SMCABC.}
\end{subfigure}
~
\begin{subfigure}[b]{0.3\textwidth}
\includegraphics[width=1\textwidth]{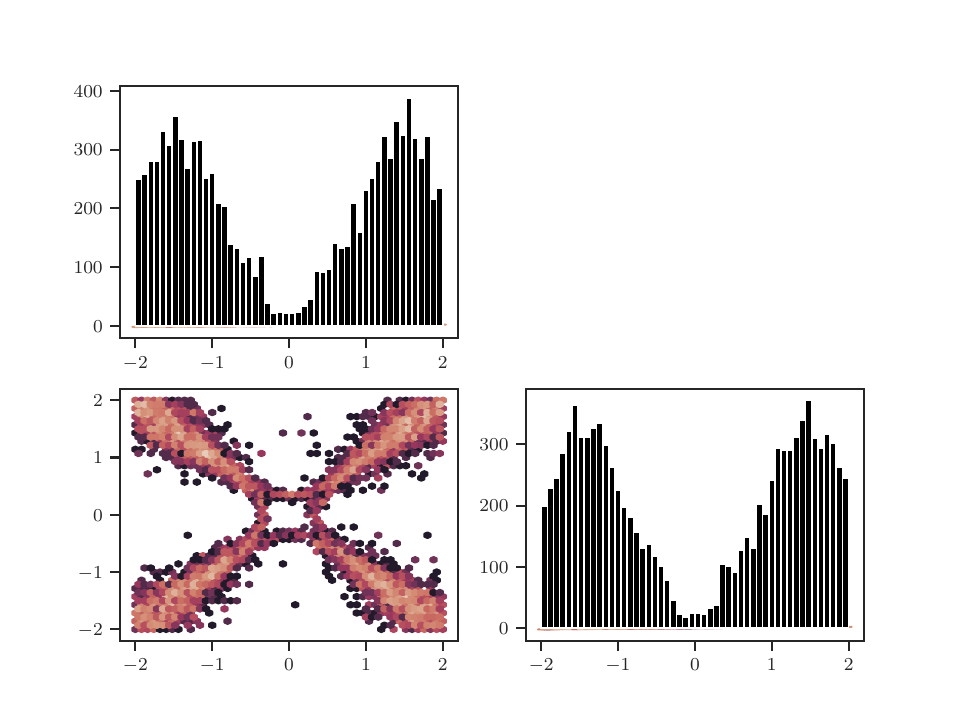}
\caption{SABC single (ours).}
\end{subfigure}
\begin{subfigure}[b]{0.3\textwidth}
\includegraphics[width=1\textwidth]{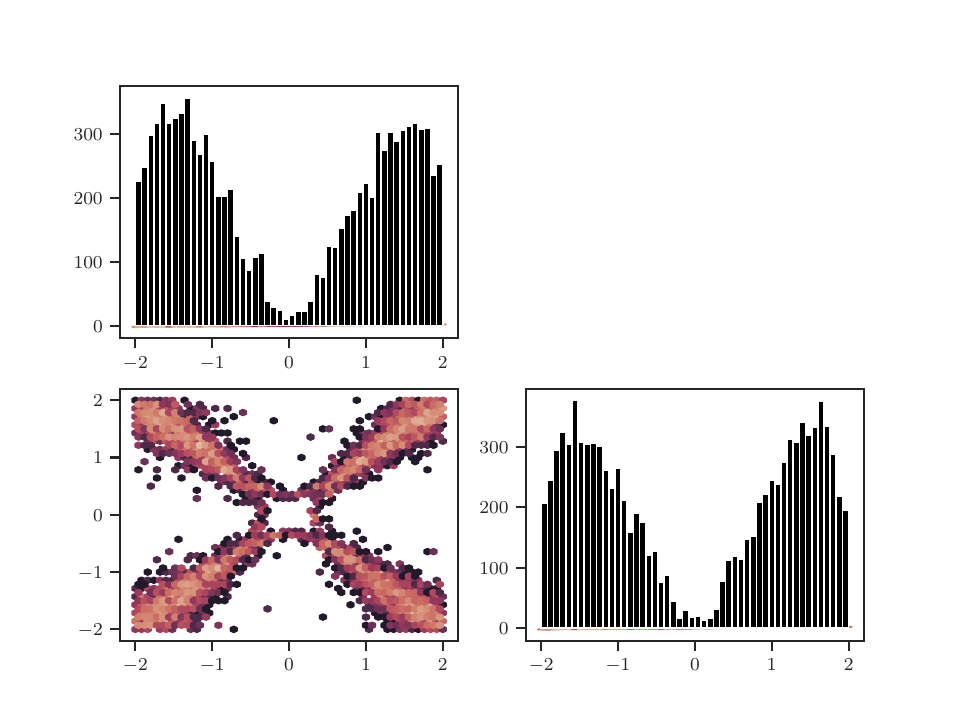}
\caption{SABC multi (ours).}
\end{subfigure}
\caption{Posterior distributions for the hyperboloid example using an arbitrary seed.}
\label{fig:hyperboloid-posteriors}
\end{figure}

We compare SABC (single and multi) against multiple baseline methods: automatic posterior transformation (APT, \cite{greenberg2019automatic}), balanced neural ratio estimation (BNRE, \cite{delaunoy2022towards}), flow matching posterior estimation (FMPE, \cite{wildberger2023flow}), neural posterior score estimation (NPSE, \cite{sharrock2024sequential}), and sequential Monte Carlo approximate Bayesian computation (SMC-ABC, \cite{beaumont_2009_aABC}). We refer to Appendix~\ref{app:benchmark-tasks-experiments} for experimental and implementation details.
Across the four benchmark evaluations, SABC consistently matches or surpasses state-of-the-art performance (Fig.~\ref{fig:sabc-benchmark-tasks-c2st} and Fig.~\ref{app:sabc-bechmark-tasks}), with the exception of the mixture model, where APT demonstrates a clear advantage.
As anticipated, employing multiple temperatures yields substantial benefits when distractors are present (see the bottom right panel of Fig.~\ref{fig:sabc-benchmark-tasks-c2st} and the decay curves in Fig.~\ref{fig:rho-trajectories}).
Qualitatively, the inferred posteriors produced by SABC are on par with those obtained via APT (compare the posteriors in Fig.~\ref{fig:hyperboloid-posteriors} and Appendix~\ref{app:sabc-benchmark-results-all-posteriors}). 

\subsection{SIR}
\begin{figure}
    \centering
    \includegraphics[width=0.5\textwidth]{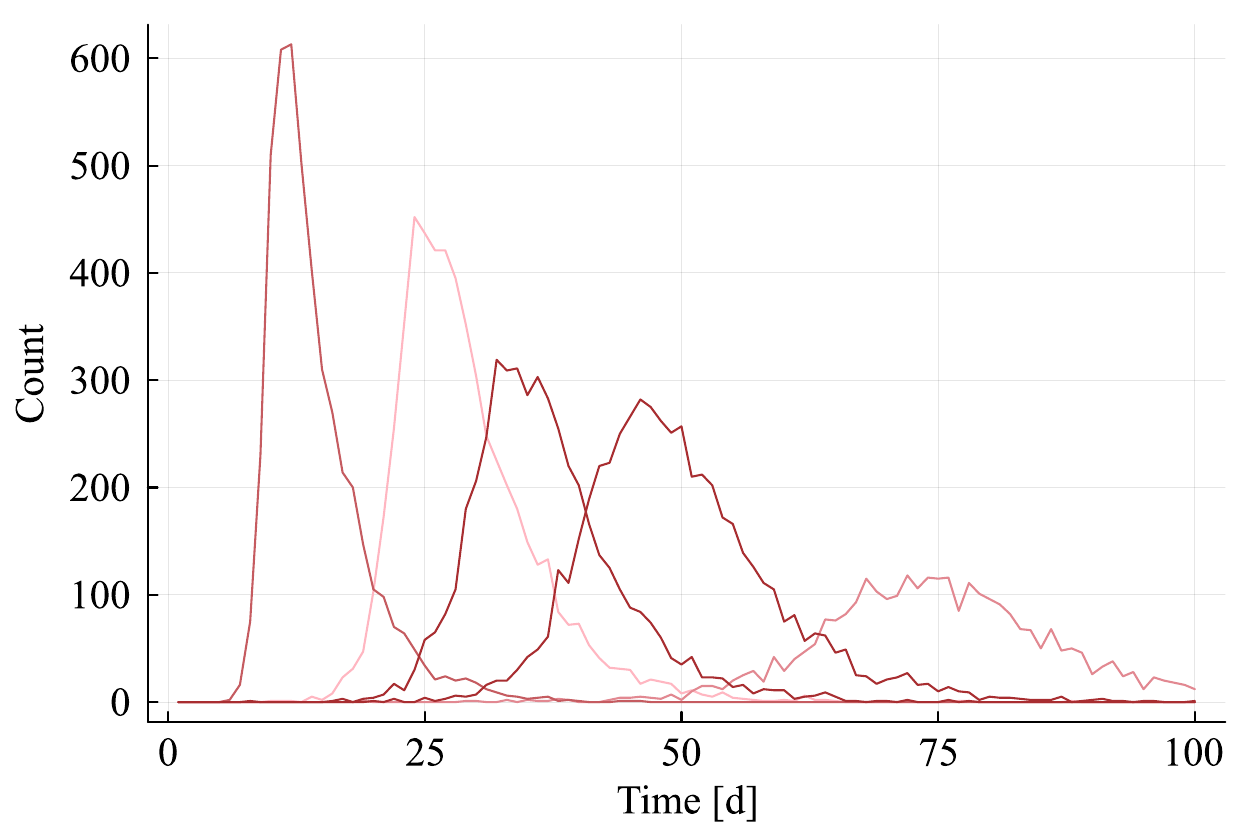}
    \caption{Multiple realizations $Y$ from the SIR model (shown in different colors).}
    \label{fig:sabc-data-sir}
\end{figure}

To illustrate that SABC scales to high-dimensional problems, we test it on the SIR model, a standard model for epidemic dynamics.
Its output is a timeseries of length $100$, which must be compressed using appropriate summary statistics.
To compare SABC against ML approaches that jointly learn summary statistics and the posterior, we pair it with a recent deep-learning method for automatic summary extraction \citep{chen2023learning}. 
Details of the model are given in Appendix~\ref{AppSIR}, its dynamics can be appreciated in Fig.~\ref{fig:sabc-data-sir}.

\begin{figure}
    \centering
    \includegraphics[width=\textwidth]{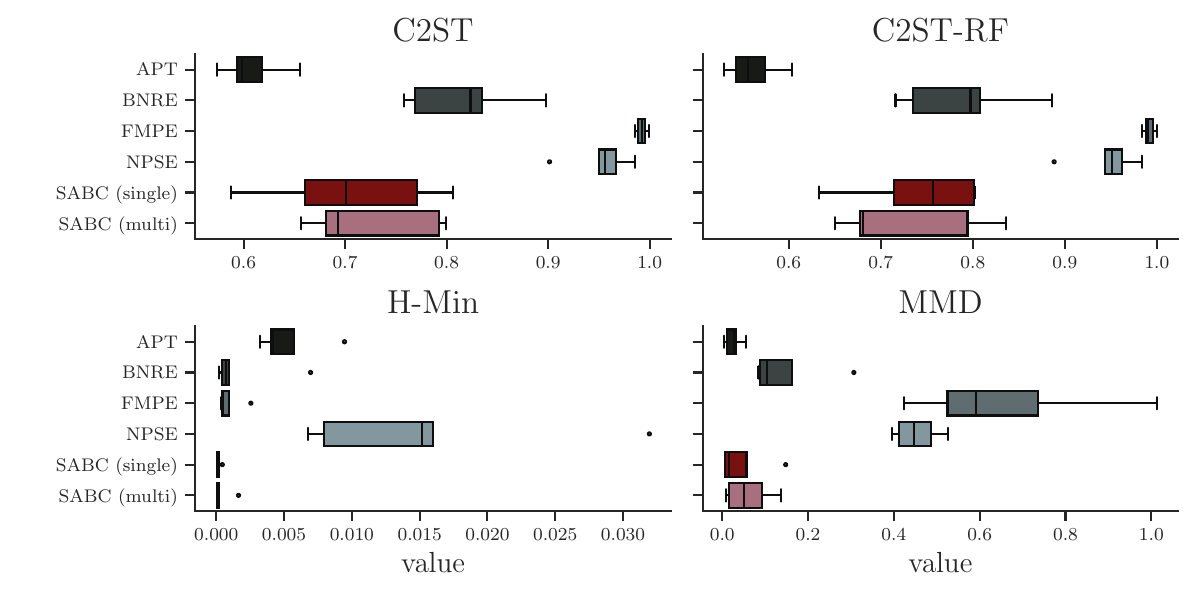}
    \caption{SABC performance on the SIR model.}
    \label{fig:sabc-benchmark-sir}
\end{figure}

\begin{figure}
\centering
\begin{subfigure}[b]{0.3\textwidth}
\includegraphics[width=1\textwidth]{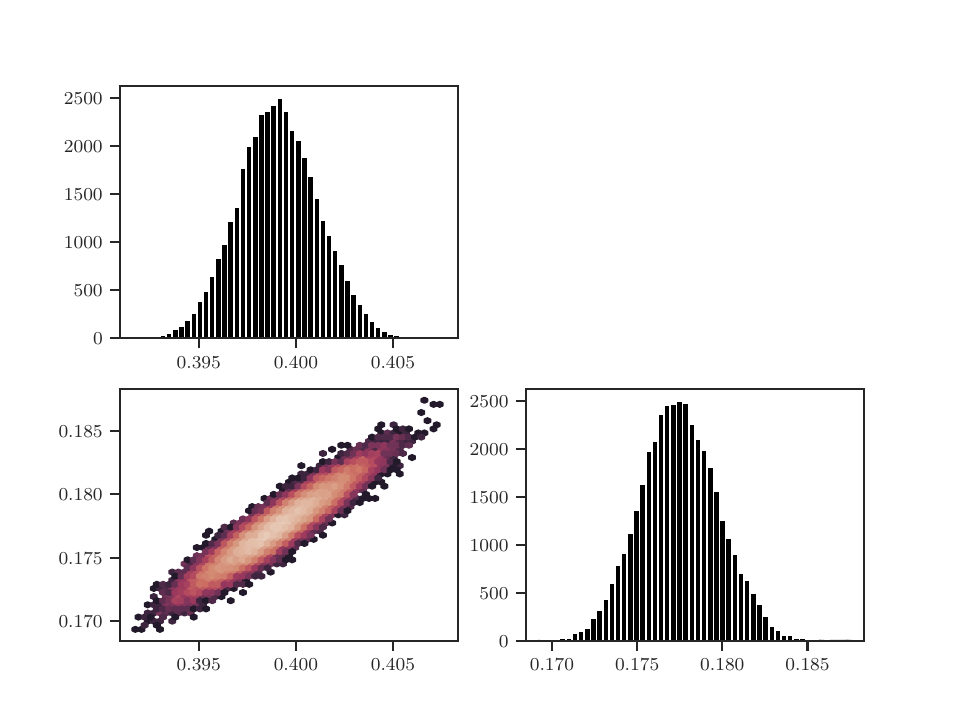}
\caption{Reference.}
\end{subfigure}
\begin{subfigure}[b]{0.3\textwidth}
\includegraphics[width=1\textwidth]{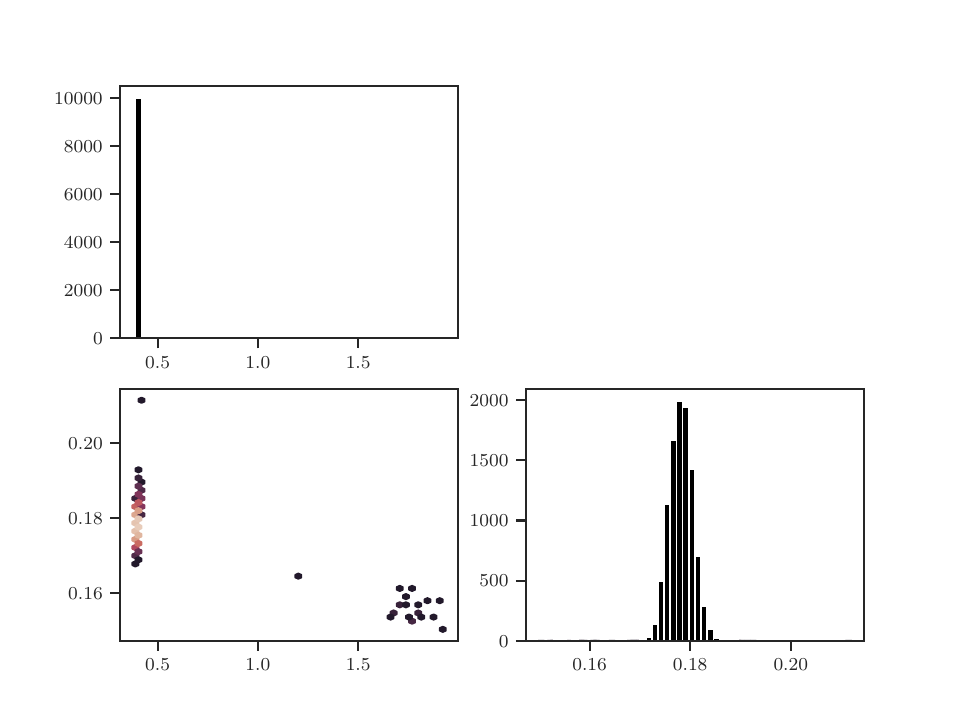}
\caption{APT.}
\end{subfigure}
\begin{subfigure}[b]{0.3\textwidth}
\includegraphics[width=1\textwidth]{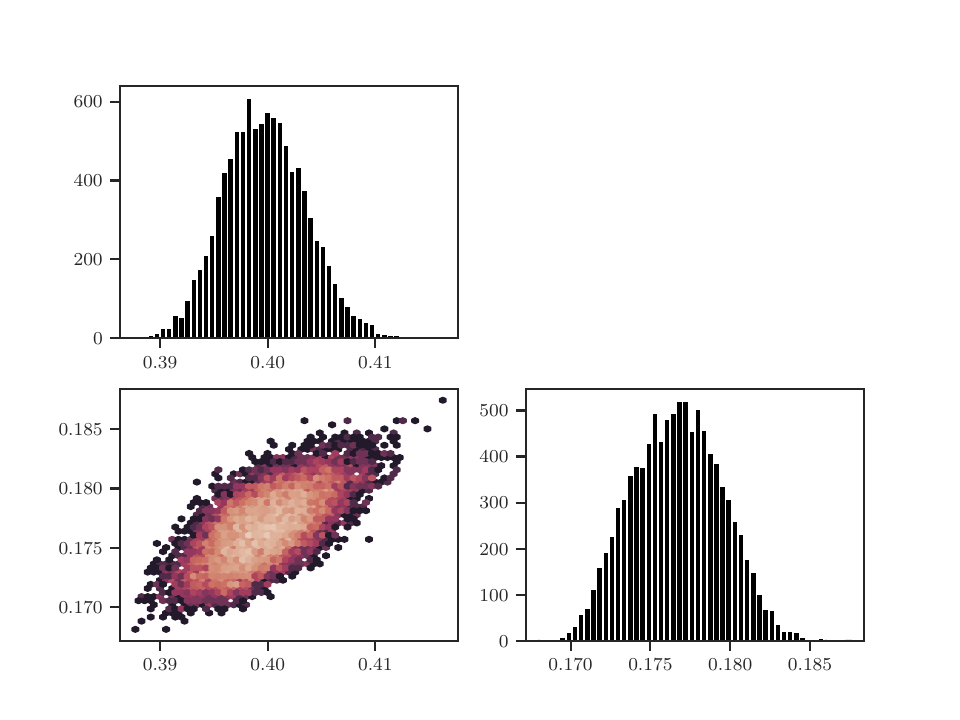}
\caption{BNRE.}
\end{subfigure}
~
\begin{subfigure}[b]{0.3\textwidth}
\includegraphics[width=1\textwidth]{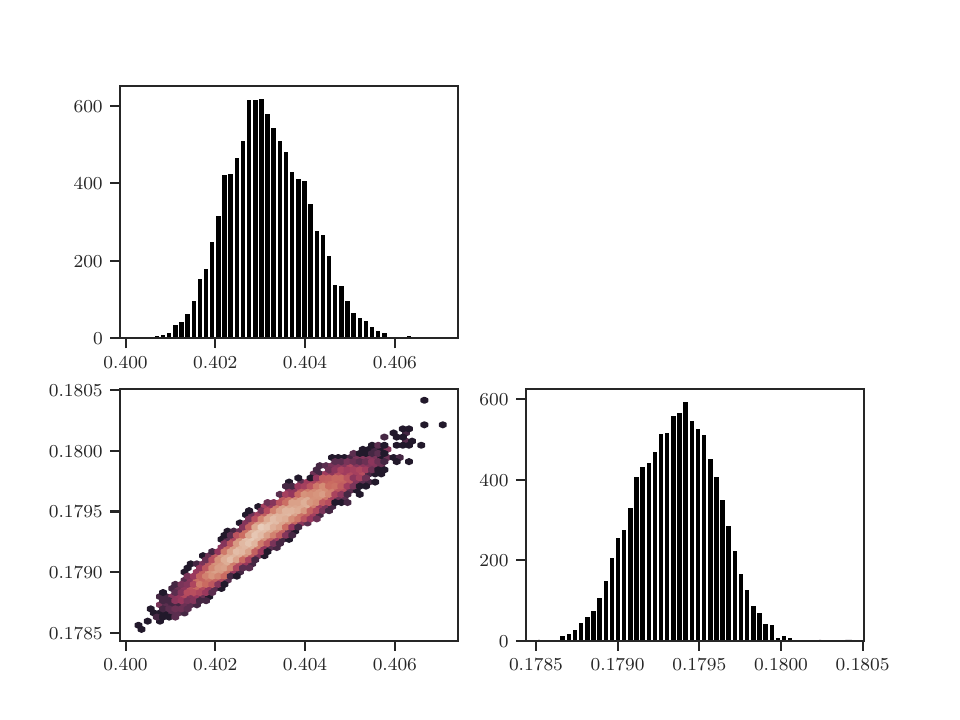}
\caption{FMPE.}
\end{subfigure}
\begin{subfigure}[b]{0.3\textwidth}
\includegraphics[width=1\textwidth]{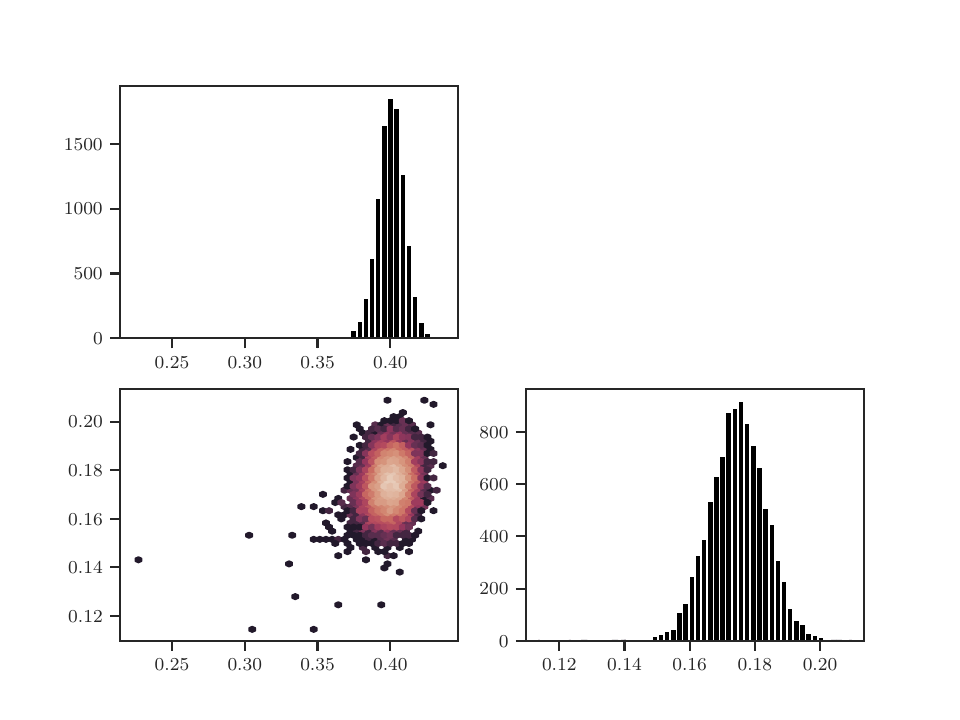}
\caption{NPSE.}
\end{subfigure}
~
\begin{subfigure}[b]{0.3\textwidth}
\includegraphics[width=1\textwidth]{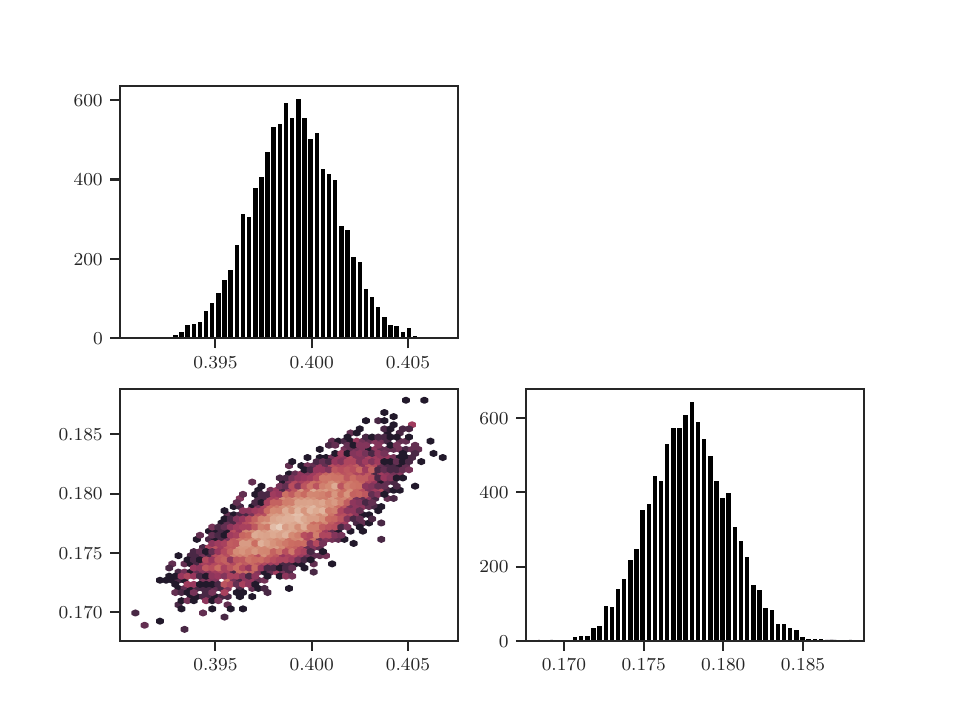}
\caption{SABC single (ours).}
\end{subfigure}
\begin{subfigure}[b]{0.3\textwidth}
\includegraphics[width=1\textwidth]{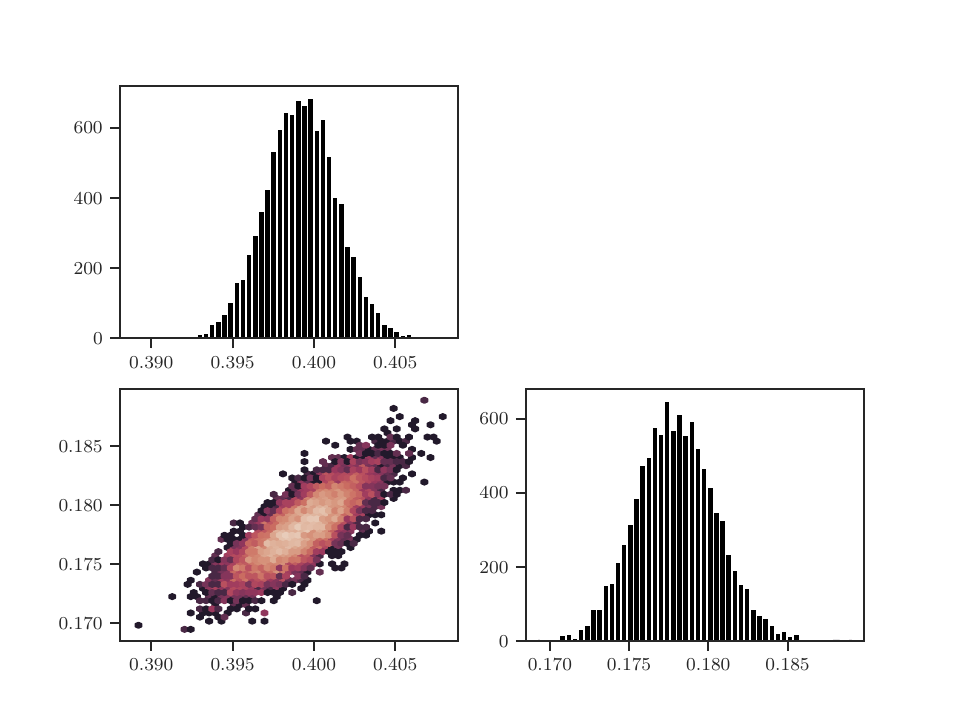}
\caption{SABC multi (ours).}
\end{subfigure}
\caption{Posterior distributions for the SIR model. APT, which appears to be better than SABC in terms of some of the metrics (Fig.~\ref{fig:sabc-benchmark-sir}), produces a cloud of outliers.}
\label{fig:sir-posteriors}
\end{figure}

We evaluate all methods on a synthetic dataset and compare the inferred posteriors with the true posterior, which is available for this problem and has been obtained via MCMC sampling.
Both SABC variants outperform or match all baselines on the H-min and MMD metrics (Fig.~\ref{fig:sabc-benchmark-sir}), although APT performs better under the C2ST metric.
Notably, the posterior plots (Fig.~\ref{fig:sir-posteriors}) show that APT yields a number of samples far from the reference posterior, whereas the SABC posterior distributions lie visually much closer to it.

\subsection{Jansen Rit}
\begin{figure}\label{Fig.JansenRit}
    \centering
    \begin{subfigure}[b]{.48\textwidth}
    \includegraphics[width=\textwidth]{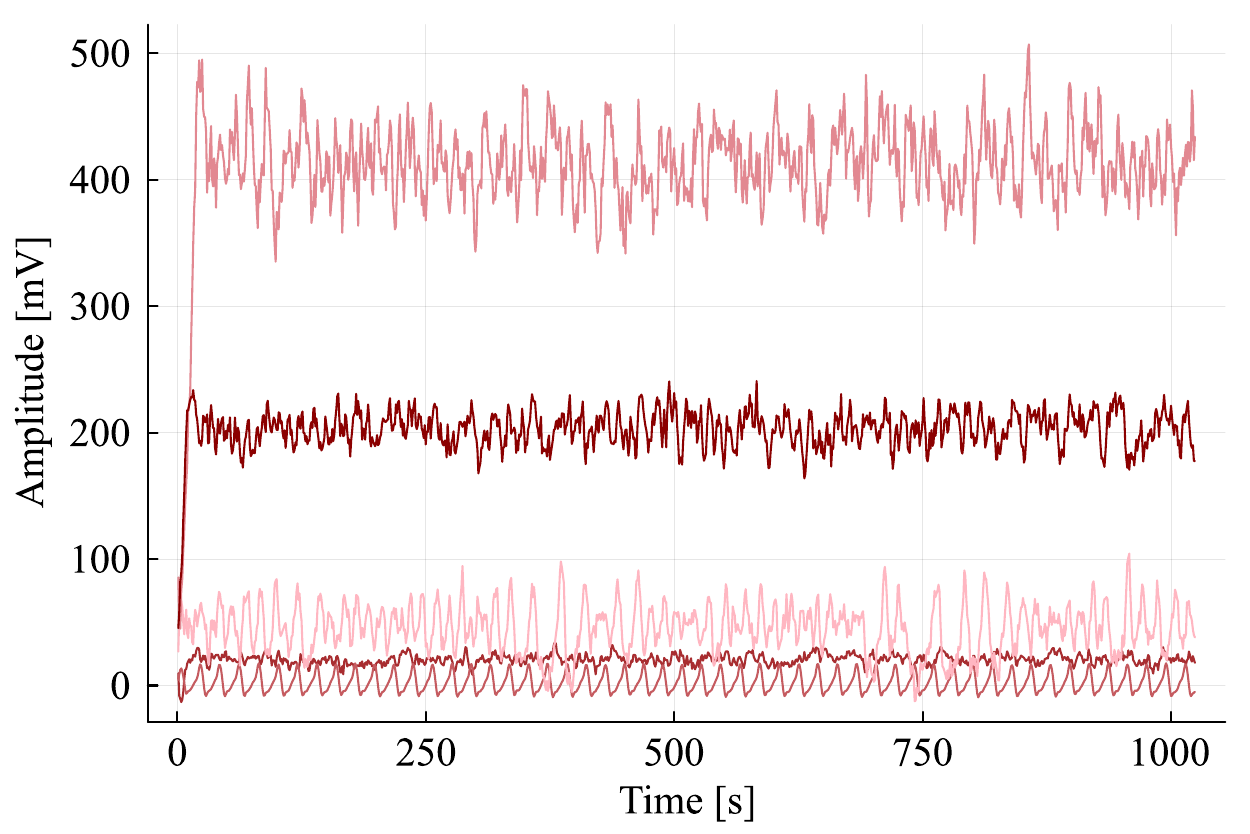}    
    \end{subfigure}    
    \hfill
    \begin{subfigure}[b]{.48\textwidth}
    \includegraphics[width=\textwidth]{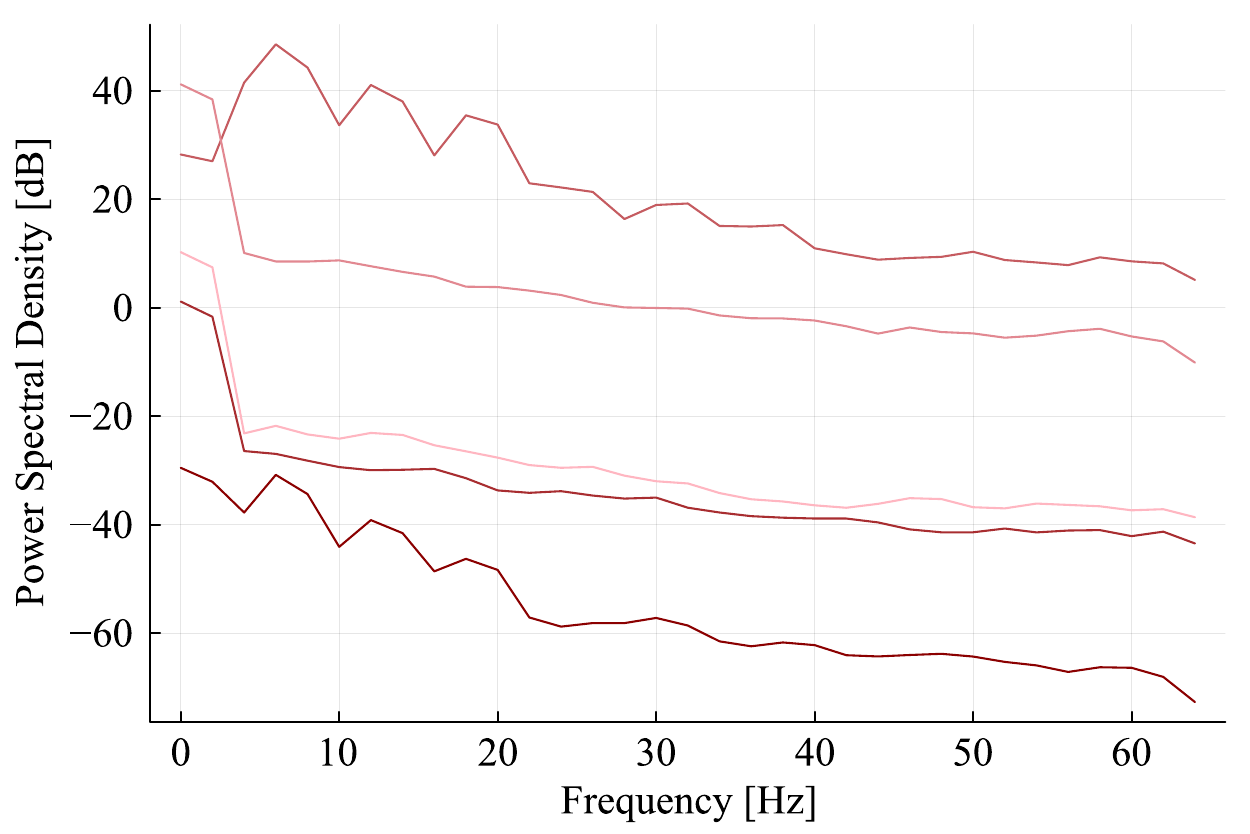}        
    \end{subfigure}    
    \caption{Realizations from the Jansen–Rit model.
The raw signals differ in location and scale (left).
We compute the PSD using $33$ frequency bins to reduce the dimensionality of the data (right).}
    \label{fig:sabc-data-jr}
\end{figure}

To evaluate SABC on a real-world model with an intractable likelihood, we consider a stochastic version of the Jansen–Rit neural mass model (JRNMM; \cite{ableidinger2017stochastic}), widely used in neuroscience as a simulator of EEG recordings.
The model is a six-dimensional first-order stochastic differential equation (SDE) with four unknown parameters to be inferred.
Representative model dynamics are shown in Fig.~\ref{fig:sabc-data-jr}.
The simulated output is a timeseries of length $1024$, from which we extract 33 Fourier components as summary statistics.


\begin{figure}
    \begin{center}    
    \includegraphics[width=.5\textwidth]{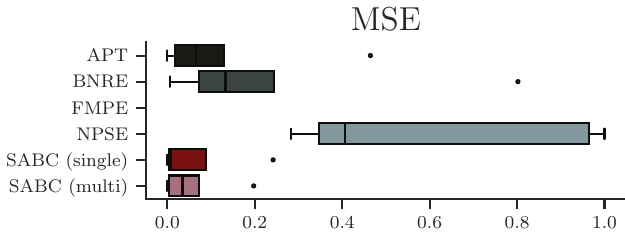}
    \caption{SABC performance on the Jansen-Rit model (sampling using FMPE did not successfully converge in 24h and was left out).}
    \label{fig:sabc-benchmark-jr}
    \end{center}
\end{figure}

\begin{figure}
\centering
\begin{subfigure}[b]{\textwidth}
\includegraphics[width=1\textwidth]{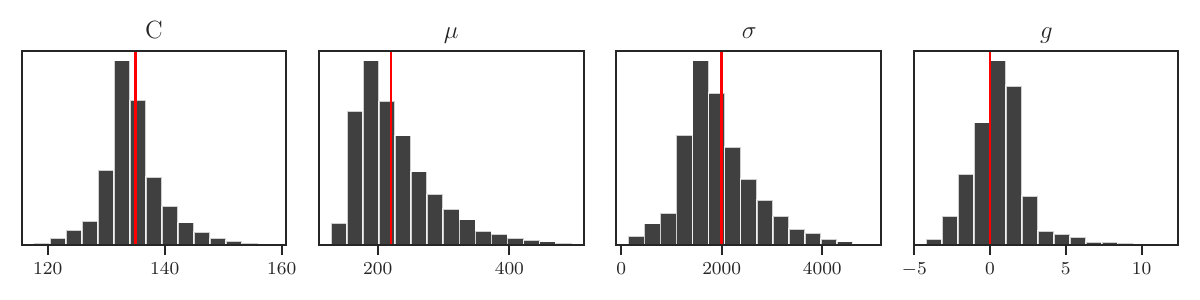}
\caption{APT.}
~
\end{subfigure}
\begin{subfigure}[b]{\textwidth}
\includegraphics[width=1\textwidth]{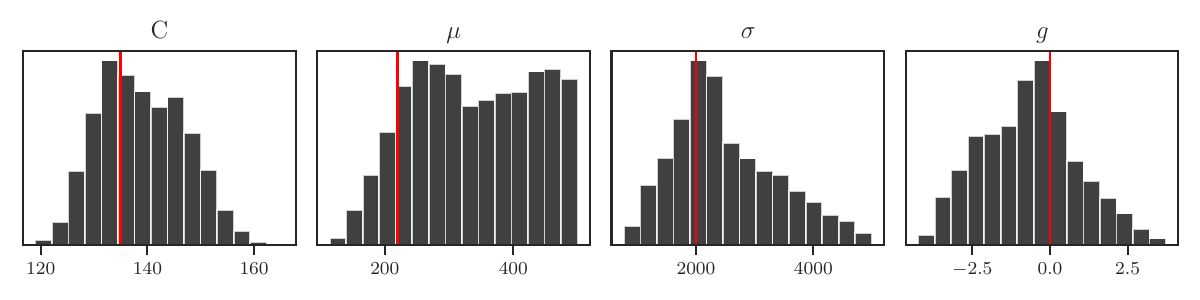}
\caption{BNRE.}
\end{subfigure}
~
\begin{subfigure}[b]{\textwidth}
\includegraphics[width=1\textwidth]{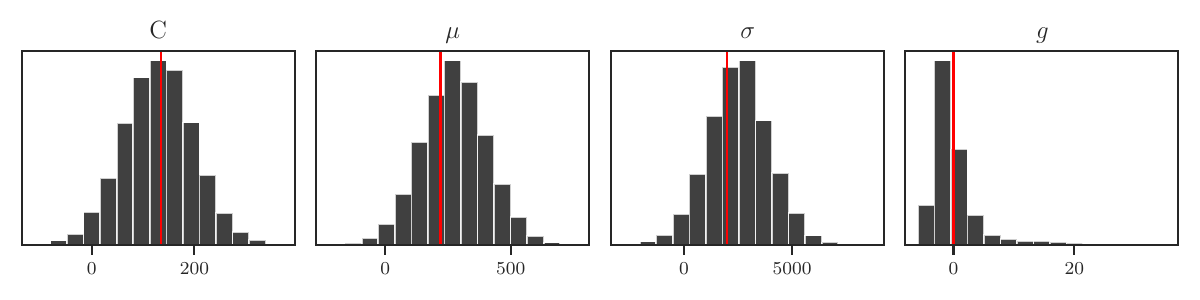}
\caption{NPSE.}
\end{subfigure}
~
\begin{subfigure}[b]{\textwidth}
\includegraphics[width=1\textwidth]{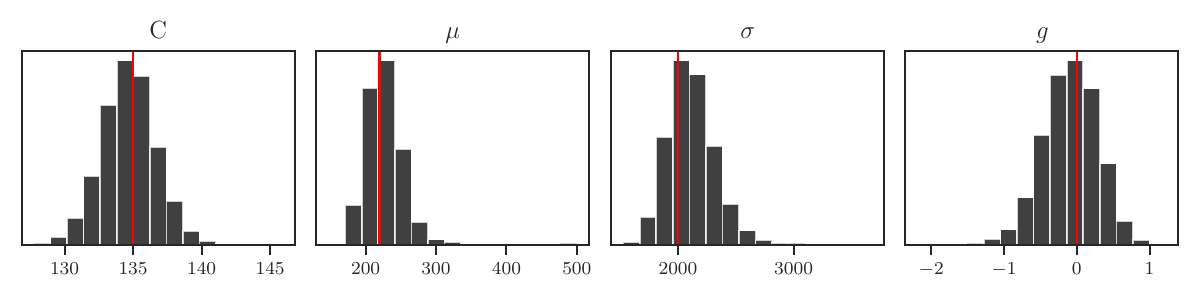}
\caption{SABC single (ours).}
\end{subfigure}
~
\begin{subfigure}[b]{0.8\textwidth}
\includegraphics[width=1\textwidth]{figures/jansen_rit/jansen_rit-sabc-single_eps_scalar-seed_1-posterior.pdf}
\caption{SABC multi (ours).}
\end{subfigure}
\caption{Posterior distributions for the Jansen-Rit model.}
\label{fig:jr-posteriors}
\end{figure}

Because the likelihood for this model is intractable, we assess the SABC posterior against the true prior sample $\bt_{obs}$ used to generate the observation $\vc y_{obs}$ (Figure~\ref{fig:sabc-benchmark-jr}).
In this setting, SABC clearly outperforms all baselines, including APT.
We further visualize posterior distributions for a representative random seed and find that both SABC variants produce markedly more peaked posteriors than the competing methods (Fig.~\ref{fig:jr-posteriors}).

\subsection{Solar Dynamo}

\begin{figure}
    \centering
    \includegraphics[width=0.9\textwidth]{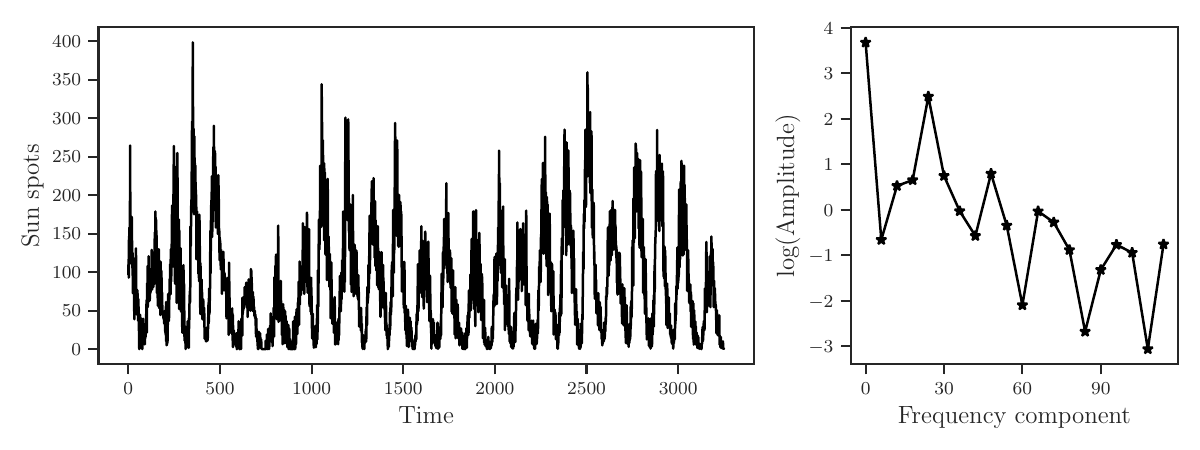}
    \caption{The SN record.
The sunspot dataset comprises $3251$ monthly observations collected between $1749$ and $2019$, exhibiting a characteristic $\sim 11$-year cycle (left).
We use $20$ Fast Fourier transform components at different frequencies as summary statistics.}
    \label{fig:sabc-solardynamo-data}
\end{figure}

We further evaluate SABC on a real-data solar physics case study.
The underlying model is a stochastic delay differential equation describing the evolution of the solar magnetic field strength $B(t)$ (see Appendix~\ref{app:solardynamo-models}).
For observations, we use the official sunspot number (SN) record \citep{SILSO_Sunspot_Number}, a commonly used proxy for the magnetic field, and compute $20$ frequency-based summary statistics (Fig.~\ref{fig:sabc-solardynamo-data}).

We compare SABC ({\em single} outperforming multi in this task) with the sequential posterior estimation approaches APT/SNPE-C and SNLE \citep{greenberg2019automatic,papamakarios_2019_sequentialNeuralLikelihood}, as these showed the strongest performance in our experiments relative to more recent approaches (data not shown; see Appendix~\ref{app:solardynamo-experimental-details} for further details).
Despite substantial practical and computational effort—including evaluations of multiple density estimators and embedding architectures—and despite prior evidence that neural SBI methods scale well to high-dimensional data and parameter spaces, we were unable to obtain meaningful posteriors with APT or SNLE when applied to the raw data.
We therefore apply the same summary statistics to APT and SNLE as to SABC.
Within this setup, we evaluated linear, MLP, and RNN-based embedding networks, and found that a simple linear projection yielded the best posterior predictive distributions (see eq.~(\ref{eq:PPD}) below).
\begin{figure}
\centering
\includegraphics[width=1\textwidth]{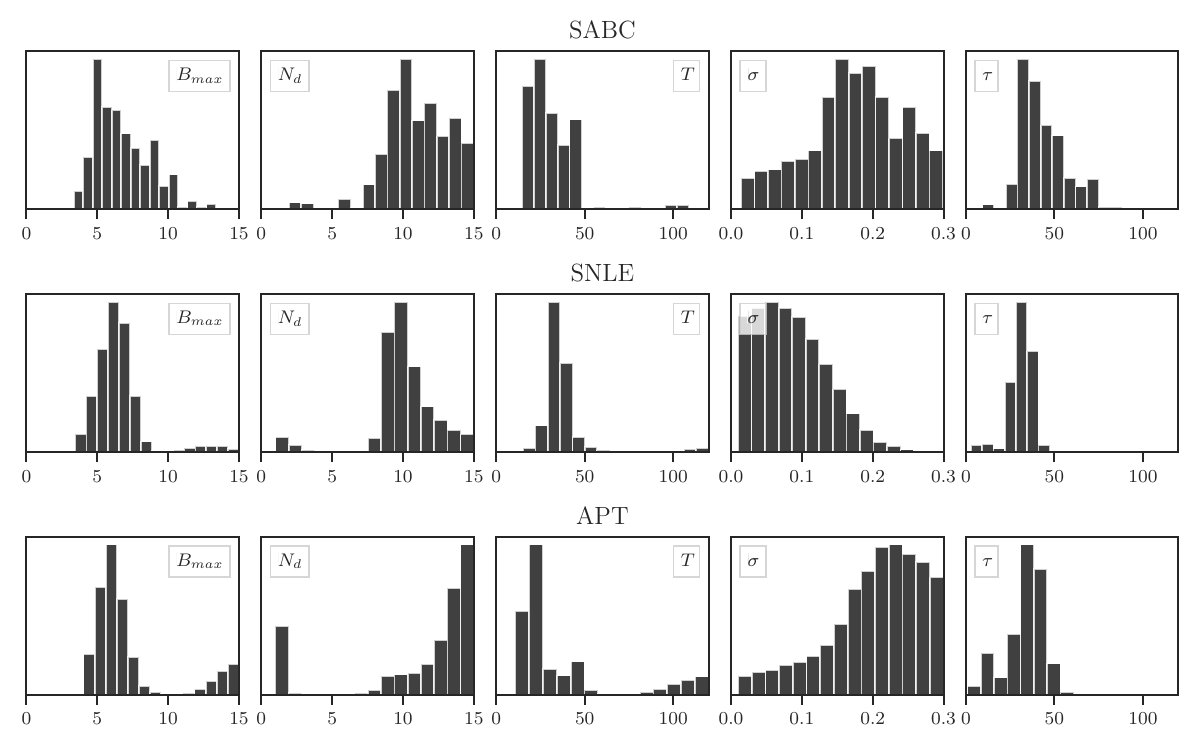}
\caption{Solar dynamo posterior distributions for the SN record of SABC, SNLE and APT.}
\label{fig:sabc-solardynamo-posteriormarginals}
\end{figure}
 The posteriors obtained with all three methods are quite similar (Fig.~\ref{fig:sabc-solardynamo-posteriormarginals} and \ref{fig:sabc-solardynamo-sabc-bivariate}–\ref{fig:sabc-solardynamo-snle-bivariate} in Appendix~\ref{app:solardynamo-additional-results}), demonstrating that SABC performs competitively even on real-world data.
These results are also well aligned with expectations from domain knowledge \citep{ulzega2025shedding}.

\begin{figure}
\centering
\includegraphics[width=1\textwidth]{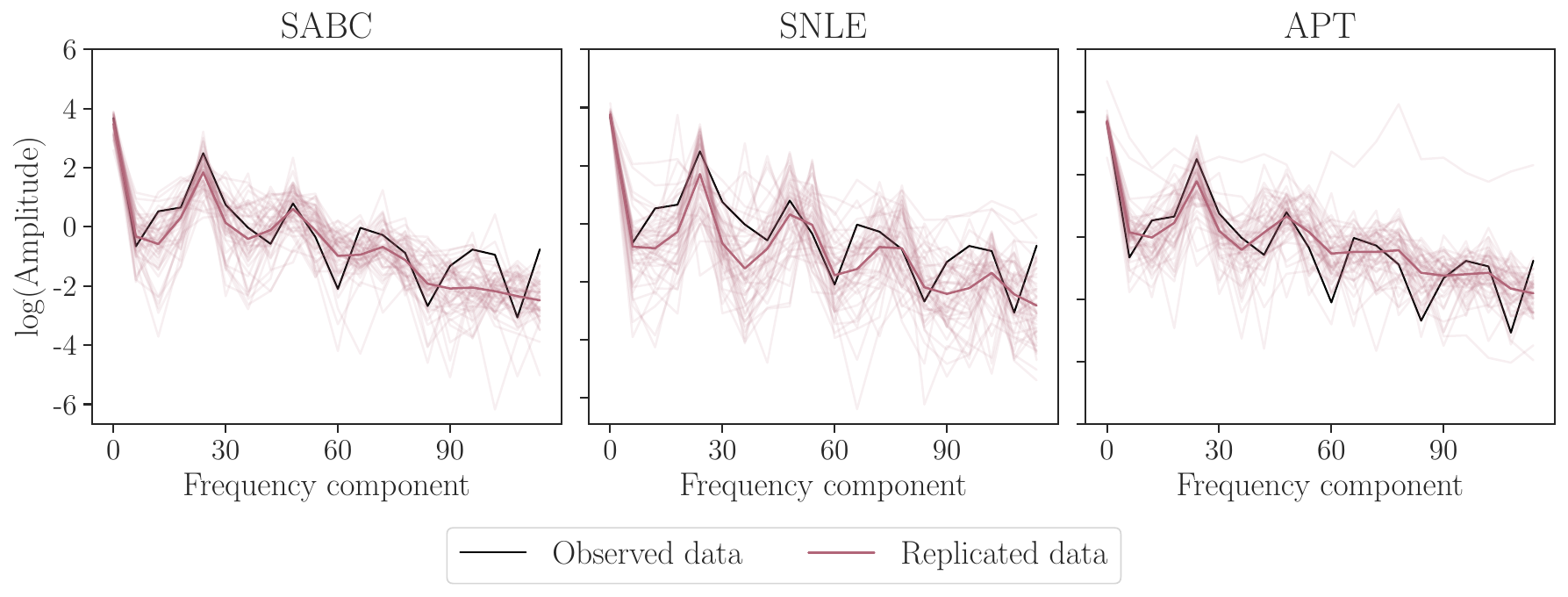}
\caption{SABC posterior predictive distributions. Agreement with the data is best for SABC and APT.
}
\label{fig:sabc-solardynamo-ppd}
\end{figure}

Since the solar dynamo model lacks a tractable likelihood, the metrics used in the previous case studies are not applicable.
We therefore assess the inferred posterior by comparing the posterior predictive distribution (PPD),
\begin{equation}\label{eq:PPD}
f({\vc s}(\vc x) | {\vc s}(\vc x_{obs})) =  \int f({\vc s}(\vc x) | \bt)  f(\bt|\vc s(\vc x_{obs})) d\bt \,,
\end{equation}
to the calibration data.
Fig.~\ref{fig:sabc-solardynamo-ppd} indicates that SABC and APT agree somewhat better with the observed data than SNLE.

As a more challenging out-of-sample evaluation, we calibrated the same dynamo model to a markedly different (and much longer) dataset: a reconstruction of sunspot numbers inferred from $^{14}$C measurements in tree rings
\citep{usoskin2021solarActivityResonstructed}. 
Fig.~\ref{fig:C14data} illustrates that this dataset is far more out of sample than the directly observed sunspots, exhibiting low-frequency structure and artefacts—including negative sunspot numbers—that our model cannot represent. 
Despite these limitations, SABC performs reasonably well in inferring the model parameters, yielding estimates that are remarkably consistent with those obtained from directly observed sunspots (see Fig.~\ref{fig-solar-AllMarginals}).
APT and SNLE, in contrast, yield posteriors that deviate from previous results, and their PPDs reveal difficulties in capturing the main frequency peak (Fig.~\ref{fig-solar-AllPPD}), corroborating the known difficulty these methods face in out-of-sample settings.

A comparison of the energy decay curves for the two datasets (Fig.~\ref{fig-solar-uCurves}) is informative:
In the sunspot data, the energies decay in unison, whereas in the $^{14}$C case two energies remain elevated.
One reflects a low-frequency component that the model fails to reproduce well; the other corresponds to the dominant $\sim 11$-year cycle.
Because this cycle is strongly modulated in both frequency and amplitude, the model struggles to capture it accurately.

\section{Conclusions}

In this paper, we argue that ABC is a viable—and in some cases superior—alternative to neural density estimators (ML methods).
Although ML methods are typically highly efficient, they do not guarantee convergence. In contrast, ABC methods can, at least in principle, approximate the true posterior arbitrarily well given sufficient computational effort.
Moreover, the distances of the most recently accepted proposals to the observations provide an indication of how well the posterior is approximated—something ML methods do not offer.
In scientific settings, where credible uncertainty quantification is essential, this is a compelling advantage of ABC.
Unlike ML approaches, ABC methods, especially SABC variants, are also considerably easier to tune.

However, when multiple summary statistics are involved, the efficiency of SABC becomes highly sensitive to the choice of metric.
We therefore propose using separate energies (rectified distances) and, optionally, separate temperatures (tolerances), which offers several advantages:
\begin{itemize}
    \item 
    This simplifies tuning, since the user no longer needs to specify an appropriate metric.
Separate energies automatically balance the influence of the different summary statistics across the run, and separate temperatures can allow the more informative statistics to converge more rapidly.
    \item 
   It also provides a diagnostic tool: the rate at which the individual energies decay reveals which features (summary statistics) are well matched and which are not.
This insight can in turn guide model refinement or identify data features that may be safely ignored.
    \item 
    It increases robustness: permitting difficult-to-match features to adhere less strictly to the observations (i.e.\ to converge more slowly) can be advantageous when the data are out of sample.
\end{itemize}
We have shown that our new SABC variants are competitive and advantageous through benchmarks and challenging inference tasks across three distinct domains, demonstrating their suitability for problems with high-dimensional outputs, intractable likelihoods, and out-of-sample data.

The main limitation of our method—shared by all ABC approaches—is the need for a very large number of model simulations. When simulators require minutes or more per run, a fast surrogate model may be necessary.
Our implementation of SABC single is nearly identical to the original SABC algorithm, aside from using particle interactions rather than a random walk in parameter space. We nevertheless expect that summing energies will outperform user-defined metrics on summary statistics in most applications.
SABC multi performed very well in our benchmarks, particularly when uninformative statistics were included. Our practical experience with this variant on real-world data, however, is still limited.

%% file: sections/appendix.tex
\section{Proof of eq. (\ref{Theorem})}
\label{app:proof}
After changing variables $y_i := \beta_i u_{\textunderscore}^i$, we are left to prove the following integral:
\begin{equation}\label{integral}
    L^{ij}_n := \int y_i \; y_j\; \theta \biggl(\sum_{k=1}^n y_k \biggr) \prod_{k=1}^n \min(1, e^{-y_k}) \; dy_k  = c_n (-1+ \delta^{ij}(n+1))\;,
\end{equation}
where $\theta(\cdot)$ is the Heaviside function and $c_n=(2n+2)!/((n+1)!(n+2)!)$ are proportional to the Catalan numbers (here we re-define \(c_n\) from the main text by dividing by \(c_0\)). Convergence is guaranteed by the Heaviside function constraint.

Due to symmetry, all the diagonal terms $L^{ii}_n$, for $i=1, \dots, n$, are equal, and so are all the $n(n-1)/2$ off-diagonal terms $L_n^{ij}$ with $i \neq j$. Therefore, we only need to consider the following two terms as function of $n$:

\begin{align}\label{recursion}
    Q_n &:= \int y_1^2 \; \theta \biggl(\sum_{k=1}^n y_k \biggr) \prod_{k=1}^n \min(1, e^{-y_k}) \; dy_k \\
    M_n &:= \int y_1\; y_2 \; \theta \biggl(\sum_{k=1}^n y_k \biggr) \prod_{k=1}^n \min(1, e^{-y_k}) \; dy_k \;.
\end{align}
In order to prove eq. (\ref{integral}) we prove $Q_n = nc_n$ and $M_n = -c_n$ by induction.
It is easy to verify that:
\begin{equation}
    Q_1 = \int_{0}^{\infty} y^2\; e^{-y} \; dy = 2 = c_1\,,
\end{equation}
and
\begin{multline}
    M_2 
    = \int_{y_1 + y_2 > 0} y_1\; y_2 \; \min(1, e^{-y_1})\; \min(1,e^{-y_2}) \; dy_1 \; dy_2 
    = \\
=\int y_1 \min(1,e^{-y_1}) \; dy_1 \int_{y_2 > - y_1} y_2 \; \min(1, e^{-y_2})\; dy_2 = \\
= \int y_1 \min(1,e^{-y_1}) \; dy_1  \biggl\{ \theta(y_1) \biggl( \int_0^{\infty} y_2\; e^{-y_2}\; dy_2 + 
\int_{-y_1}^0 y_2 \; dy_2 \biggr) \\
+ \theta(-y_1) \int_{-y_1}^{\infty} y_2 \; e^{-y_2} \; dy_2\biggr\} =\\
 = \int y_1 \min(1,e^{-y_1}) \; dy_1  \biggl\{  \theta(y_1) \bigl( 1 - y_1^2 /2\bigr) + \theta(-y_1) \bigl( 1- y_1 \bigr) e^{y_1}\biggr\} = \\
= \int_0^{\infty} y_1 \biggl(1- \frac{y_1^2}{2} \biggr) \; e^{-y_1} \; dy_1 +  \int_{-\infty}^0 y_1 \bigl(1 - y_1 \bigr) \; e^{y_1} \; dy_1 = -5 = -c_2\;.
\end{multline}
Now, let us solve $Q_n$ by induction. We have for $n > 1$:
\begin{multline}
    Q_{n+1} := \int y_1^2 \; \theta \biggl(\sum_{k=1}^{n+1} y_k \biggr) \prod_{k=1}^{n+1}\min(1, e^{-y_k}) \; dy_k  = \\
    =\int y_1^2 \; \theta \biggl(\sum_{k=1}^n y_k + y_{n+1}\biggr) \prod_{k=1}^n \min(1, e^{-y_k}) \; dy_k \; \min(1, e^{-y_{n+1}}) \; dy_{n+1}\;.
\end{multline}
Given the theta constraint, we can split the integral into three integration regions
:
\begin{multline}
    Q_{n+1} =\int y_1^2 \; \theta \biggl(\sum_{k=1}^n y_k \biggr) \prod_{k=1}^n \min(1, e^{-y_k}) \; dy_k \int_0^{\infty} e^{-y_{n+1}} \; dy_{n+1} \; + \\
    +\; \int y_1^2 \; \theta \biggl(\sum_{k=1}^n y_k \biggr) \prod_{k=1}^n \min(1, e^{-y_k}) \; dy_k \int_{- \sum_{k=1}^n y_k} ^0 dy_{n+1} \; + \\
    + \; \int  y_1^2 \; \theta \biggl( - \sum_{k=1}^n y_k \biggr) \prod_{k=1}^n \min(1, e^{-y_k}) \; dy_k \int_{-\sum_{k=1}^n y_k}^{\infty} e^{-y_{n+1}}\; dy_{n+1} = \\
    = Q_n +  \int y_1^2 \;  \biggl(\sum_{k=1}^n y_k \biggr) \theta \biggl(\sum_{k=1}^n y_k \biggr) \prod_{k=1}^n \min(1, e^{-y_k}) \; dy_k  + \\
    + \int y_1^2 \; \theta \biggl(-\sum_{k=1}^n y_k \biggr) \prod_{k=1}^n \min(1, e^{-y_k}) \; \prod_{k=1}^n e^{y_k}\; dy_k = \\
    = 2 Q_n + \sum_{k=1}^n \int y_1^2  y_k \; \theta \biggl(\sum_{k=1}^n y_k \biggr) \prod_{k=1}^n \min(1, e^{-y_k}) \; dy_k \,.
\end{multline}
Similarly:
\begin{equation}
    M_{n+1} = 2 M_n + \sum_{k=1}^n \int y_1 y_2  y_k \theta \biggl(\sum_{k=1}^n y_k \biggr) \prod_{k=1}^n \min(1, e^{-y_k}) dy_k \,.
\end{equation}
We introduce the linear and zeroth order terms
\begin{align}\label{recursion}
    O_n &:= \int\; \theta \biggl(\sum_{k=1}^n y_k \biggr) \prod_{k=1}^n \min(1, e^{-y_k}) \; dy_k \;,\\
    L_n &:= \int y_1 \; \theta \biggl(\sum_{k=1}^n y_k \biggr) \prod_{k=1}^n \min(1, e^{-y_k}) \; dy_k \;.
\end{align}
and the $n$ dimensional measure
\begin{equation}
    dY_n := \prod_{k=1}^n \min(1, e^{-y_k})\; dy_k \,.
\end{equation}
Then
\begin{multline}\label{zeroth_linear_rec}
    O_{n+1} = \int dY_{n+1} \; \theta \biggl(\sum^{n+1}_{k=1} y_k \biggr) =   \int dY_n \;  \biggl[ \theta \biggl(\sum^{n}_{k=1} y_k \biggr)  \biggl(\int_{0}^{\infty} e^{-y_{n+1}} dy_{n+1} + \int_{- \sum^{n}_{k=1} y_k}^0 dy_{n+1} \biggr) + \\
    +\theta\biggl(- \sum^{n}_{k=1} y_k \biggr) \int_{- \sum^{n}_{k=1} y_k}^{\infty} e^{-y_{n+1}} dy_{n+1} \biggr]= \\
    =\int dY_n \;  \biggl[ \theta \biggl(\sum^{n}_{k=1} y_k \biggr) \biggl(1+ \sum^{n}_{k=1} y_k \biggr) + \theta\biggl(- \sum^{n}_{k=1} y_k \biggr) e^{\sum^{n}_{k=1} y_k} \biggr] = 2O_n + nL_n\;.
\end{multline}
For the linear term, we have two recursions, depending on whether we define the linear term by integrating \(y_1\) or \(y_{n+1}\) (these two integrals are equal due to permutation symmetry). Namely:
\begin{multline}\label{first_linear_rec}
    L_{n+1} =  \int y_1 dY_{n+1} \; \theta \biggl(\sum^{n+1}_{k=1} y_k \biggr) = \int  y_1 dY_n \;  \biggl[ \theta \biggl(\sum^{n}_{k=1} y_k \biggr) \biggl(1+ \sum^{n}_{k=1} y_k \biggr) \\
    + \theta\biggl(- \sum^{n}_{k=1} y_k \biggr) e^{\sum^{n}_{k=1} y_k} \biggr]\;  
    =  Q_n + (n-1)M_n = \frac{I_n}{n}\;
\end{multline}
where we have defined 
\begin{equation}
    I_n :=  \int \biggl(\sum^{n}_{k=1} y_k \biggr)^2 dY_{n} \; \theta \biggl(\sum^{n}_{k=1} y_k \biggr) = nQ_n + n(n-1)M_n\;. 
\end{equation}
The last equation can be easily proved by expanding the square under the integral.
By defining the linear term with the integration of \(y_{n+1}\), we get:
\begin{multline}\label{second_linear_rec}
    L_{n+1} =  \int y_{n+1} dY_{n+1} \; \theta \biggl(\sum^{n+1}_{k=1} y_k \biggr) =\\
    =\int  dY_n \;  \biggl[ \theta \biggl(\sum^{n}_{k=1} y_k \biggr) \biggl(1- 
 \frac{1}{2} \biggl(\sum^{n}_{k=1} y_k\biggr)^2 \biggr) + \theta \biggl(- \sum^{n}_{k=1} y_k \biggr)\biggl(1- \sum^{n}_{k=1} y_k\biggr) e^{\sum^{n}_{k=1} y_k} \biggr]=\\
 = 2 O_n - \frac{1}{2} I_n + nL_n \;.
\end{multline}
Combining eqs. (\ref{zeroth_linear_rec}), (\ref{first_linear_rec}) and  (\ref{second_linear_rec}), we get a system of coupled recursive relations:
\begin{equation}
    \begin{cases}
     O_{n+1} = 2 O_n + n L_n\;,  \\
     L_{n+1} = \frac{I_n}{n} \;, \\
     L_{n+1} = 2 O_n - \frac{I_n}{2} + n L_n \;, \\ 
    \end{cases}
\end{equation}
which easily gives a relation between the linear and zeroth order term as $(n+1)L_n = 2O_n$. Moreover, we can easily solve for the linear term:
\begin{equation}
    L_{n+1} = \frac{2(n+1)}{n+2} L_n\;,
\end{equation}
which is the same recursion satisfied by \(c_n\). Since $L_1=1$, we obtain
\begin{equation} \label{solved_system}
    \begin{cases}
     L_{n} = c_{n-1}\;, \\
     I_{n} = n c_n \;,\\
     O_{n} = \frac{n+1}{2} c_{n-1}\;. \\
    \end{cases}
\end{equation}
To complete the proof, we have to show that $I_n = Q_n$. This, together with eq. (\ref{solved_system}), would imply that $Q_n = n c_n $ and $M_n  = - c_n$, which are equivalent to eq. (\ref{integral}). Let us consider the integration over $y_{n+1}$:
\begin{multline}
    I_{n+1} =  \int \biggl(\sum^{n+1}_{k=1} y_k \biggr)^2 dY_{n+1} \;\theta \biggl(\sum^{n+1}_{k=1} y_k \biggr)\\ =   \int \biggl[y_{n+1}^ 2 + \biggl(\sum^{n}_{k=1} y_k \biggr)^2
    + 2 y_{n+1}\biggl(\sum^{n}_{k=1} y_k \biggr)  \biggl] dY_{n+1}\\
    = Q_{n+1} +\int \biggl(\sum^{n}_{k=1} y_k \biggr)^2 dY_{n}  \;  \biggl[ \theta \biggl(\sum^{n}_{k=1} y_k \biggr)  \biggl( 1+ \sum^{n}_{k=1} y_k \biggr) + \theta \biggl(- \sum^{n}_{k=1} y_k \biggr) e^{\sum^{n}_{k=1} y_k} \biggl] + \\
    + \int 2\biggl(\sum^{n}_{k=1} y_k \biggr) dY_{n}\biggl[  \theta \biggl(\sum^{n}_{k=1} y_k \biggr)\biggl( 1 -  \frac{1}{2}\biggl(\sum^{n}_{k=1} y_k \biggr)^2  \biggr) + \\ 
    + \theta \biggl(- \sum^{n}_{k=1} y_k \biggr) e^{\sum^{n}_{k=1} y_k}  \biggl(- \sum^{n}_{k=1} y_k +1 \biggr) \biggl] = Q_{n+1} \;,
\end{multline}
 for all $n$ integers. This completes the proof.

\section{Additional information on the benchmark tasks}
\label{app:benchmark-tasks-add-info}

\subsection{Benchmark models}
\label{app:benchmark-tasks-models}

\subsubsection{Hyperboloid}

The hyperboloid model \citep{forbes2022summary} is a 2-component mixture of tri-variate Student's $t$-distributions of the form
\begin{align*}
\bt &\sim \mathcal{U}_2(-2, 2) \\
\vc s \mid \bt &\sim 
    \frac{1}{2} t_3(\nu, F(\bt; \vc a_1, \vc a_2) \mathbb{I}, \sigma^2 \vc I) +
    \frac{1}{2} t_3(\nu, F(\bt; \vc b_1, \vc b_2) \mathbb{I}, \sigma^2 \vc I)
    \,,
\end{align*}
which are parameterized by degrees of freedom $\nu$, mean $$F(\bt; \vc x_1, \vc x_2) =\text{abs}\left( ||\bt - \vc x_1 ||_2 - ||\bt - \vc x_2 ||_2 \right)\,,
$$
and scale matrix $\sigma^2 \vc I$. 
$\mathbb{I}$ is a three-dimensional vector of ones. We follow \cite{forbes2022summary} and set\\ $\vc a_1 = [-0.5, 0.0]^T$, $\vc a_2 = [0.5, 0.0]^T$, $\vc b_1 = [0.0, -0.5]^T$, $\vc b_2 = [0.0, 0.5]^T$, $\nu = 3$ and $\sigma^2 = 0.01$ for our experiments. Hence, while the parameter vector $\bt$ is two-dimensional, $\vc s$ has three dimensions.

\subsubsection{Gaussian mixture model}
The Gaussian mixture model (GMM) uses the following generative process:
\begin{align*}
\bt & \sim \mathcal{U}_2(-10, 10) \\
\vc s \mid \bt & \sim \frac{1}{2} \mathcal{N}_2(\bt, \vc I) + \frac{1}{2} \mathcal{N}_2( \bt, \sigma^2 \vc I) 
\end{align*}
where $\sigma^2 = 0.01$, $\vc I$ is a unit matrix, and both $\bt \in \mathbb{R}^2$ and $\vc s \in \mathbb{R}^2$ are two-dimensional random variables. The GMM follows the representation in \cite{lueckmann2021benchmarking}.

\subsubsection{Mixture model with distractors}

We introduced a new benchmark model that, analogous to the SLCP task \citep{lueckmann2021benchmarking}, augments the data with dimensions that carry no information about the parameters. Specifically:
\begin{equation}
\begin{split}
    \theta & \sim \mathcal{U}(-10, 10) \\
    s_1, s_2 &\sim \alpha \mathcal{N}(\theta, 1)
    + (1 - \alpha) \mathcal{N}(-\theta, \sigma^2)\\
    s_3,\cdots, s_{11} &\sim \mathcal{N}(0, 1)\,,
\end{split}
\end{equation}
where we set $\alpha = \sigma = 0.3$. For each observation, the mixture model with distractors draws two independent samples from a Gaussian mixture.
It further generates 9 samples from a standard Gaussian, which carry no information about the parameters.
If $s_{1,obs}$ and $s_{2,obs}$ originate from the same mode (in our experiments, $s_{1,obs} = s_{2,obs} = 5$), the posterior over $\theta$ becomes approximately bimodal with highly imbalanced mass between the two modes.
This makes it a particularly challenging distribution to sample.

\subsubsection{Two moons}
The two-moons problem is a widely used benchmark in the SBI literature.
Its generative process is defined as:
\begin{align*}
\bt & \sim \mathcal{U}_2(-10, 10) \\
\alpha & \sim \mathcal{U}(-\pi / 2, \pi / 2) \\
r & \sim \mathcal{N}(0.1, 0.01^2) \\
\vc s \mid \bt & = 
\begin{pmatrix}
    r \cos{\alpha} + 0.25 \\ r \sin{\alpha}
\end{pmatrix}
 +
 \begin{pmatrix}
    -|\theta_1 + \theta_2| / \sqrt{2} \\ -(\theta_1 + \theta_2) / \sqrt{2}
\end{pmatrix}\,,
\end{align*}
where we are interested in inferring the two-dimensional posterior $p(\bt \mid \vc s)$ and treat $\alpha$ and $r$ as nuisance parameters.

\subsection{Experimental details}
\label{app:benchmark-tasks-experiments}

For the two moons and mixture model benchmark tasks, we used the reference posterior distributions of the \texttt{sbi} Python package \citep{tejero2020sbi}. For the hyperboloid model and mixture model with distractors, we draw a posterior sample of size $100\ 000$ using a slice sampler using the \texttt{sbijax} Python package \citep{dirmeier2024simulation}. We sample $10$ independent chains of length $20\ 000$ of which the first $10\ 000$ samples are discarded as warm-up. We then pool the samples of all $10$ chains. We used conventional MCMC sampler diagnostics (i.e., potential scale reduction factor and effective sample size) to monitor convergence.

We compare the inferred posterior distributions to the reference posterior distributions using C2ST
\citep{lopezpaz2017revisiting}, MMD \citep{sutherland2017generative} and H-Min \citep{zhao2022comparing} metrics. All metrics are computing by subsampling $10\ 000$ posterior samples without replacement from the inferred posterior and reference posterior distributions.

Conversely to the previous literature which evaluated their methods with only few simulations (e.g., up to $10\ 000$), we simulate data such that each method is trained to optimality. Since all our benchmarks are very low-dimensional, we trained each neural SBI method with $N=250\ 000$ simulated model outputs (we evaluated training each method using $500\ 000$ model outputs using two different random seeds, but found no inferential difference between $250\ 000$ and $500\ 000$ samples). For sampling-based methods, i.e., SMC-ABC and SABC, we evaluated $50$ million population updates
of an initial population of $10\ 000$ particles. While we did not conduct exhaustive experiments, we observed that much smaller number of population updates (i.e, $10$ - $20$ million) yielded comparable performance for both SMCABC and all SABC variants. Details for each method are shown below. If not noted otherwise, all neural methods use an Adam optimizer using a learning rate of $l=0.0001$ and are trained to convergence with a maximum of $2\ 000$ iterations. 

Each experiment was run on an AMD EPYCTM 7742 processor with 64 cores and 256 GB RAM. Runtimes have not been thoroughly monitored but lay between 2-8h. In total we computed $140$ experiments (5 seeds times 6 experimental models times 4 benchmark tasks). We uses the workflow tool Snakemake \citep{koster2012snakemake} to run all experiments automatically on a Slurm cluster.

\paragraph{SABC} 
SABC involves only a small number of tuning parameters and is quite robust to their selection.
Throughout this paper, we adopt the following standard choices:
The annealing speed is set to $v=1$. An importance sampling step decreasing the temperature by a factor $1+\delta=1.1$ (at the cost of effective sample size) is performed every $2*$(number of particles) successful updates, as well as at the beginning of the algorithm (See \cite{Albert_2013_ABC} for details). 
For the jumps in parameter space we replaced the original (adaptive) normal jump distribution by an interacting particle mechanism \citep{terbraak_2006_DEMC, goodman2010ensemble}. 
For all experiments, we used an Euclidean distance function as a metric.
We replace the steps in the approximation (\ref{eq:uapprox}) by linear ramps. This is important for particles not to get stuck - especially at small values. 
We parallelized the algorithm, updating the whole population in parallel once before updating the temperature(s). 
The code has been released as a Julia package: 
\url{https://github.com/Eawag-SIAM/SimulatedAnnealingABC.jl}
\paragraph{SMC-ABC} For SMC-ABC, we use an $\epsilon$-decay factor of $0.9$, which gave the best performance for $50$ million population updates. We resample the population when the relative effective sample size drops below $0.2$ which we empirically found to work well. As before, we use a Euclidean distance function.
\paragraph{APT} APT uses a neural spline flow \citep{durkan2019neural} consisting of $5$ normalizing flow layers. The flow uses a residual network with $10$ bins, $2$ blocks, and $64$ hidden nodes per block. 
\paragraph{BNRE} BRNE uses a residual network as a classifier using $5$ blocks of $64$ hidden nodes each.
\paragraph{FMPE} FMPE uses a residual network as a score network which consists of $5$ blocks of $64$ hidden nodes each. We used a time-embedding dimensionality of $32$.
\paragraph{NPSE} Since FMPE and NPSE are algorithmically identical and only differ in the definition of the forward process (see, e.g., \cite{tong2024improving}), we chose to use a MLP using $5$ layers with $64$ hidden nodes each for NPSE expecting otherwise very similar results to FMPE (we note that this is the same architecture as in the original publication \citep{sharrock2024sequential}). We used the variance-preserving SDE and a time-embedding dimensionality of $32$.

\clearpage
\subsection{Additional results}
\label{app:benchmark-tasks-add-results}
\subsubsection{Full table of results}

\begin{figure}[!h]
    \centering
    \includegraphics[width=\textwidth]{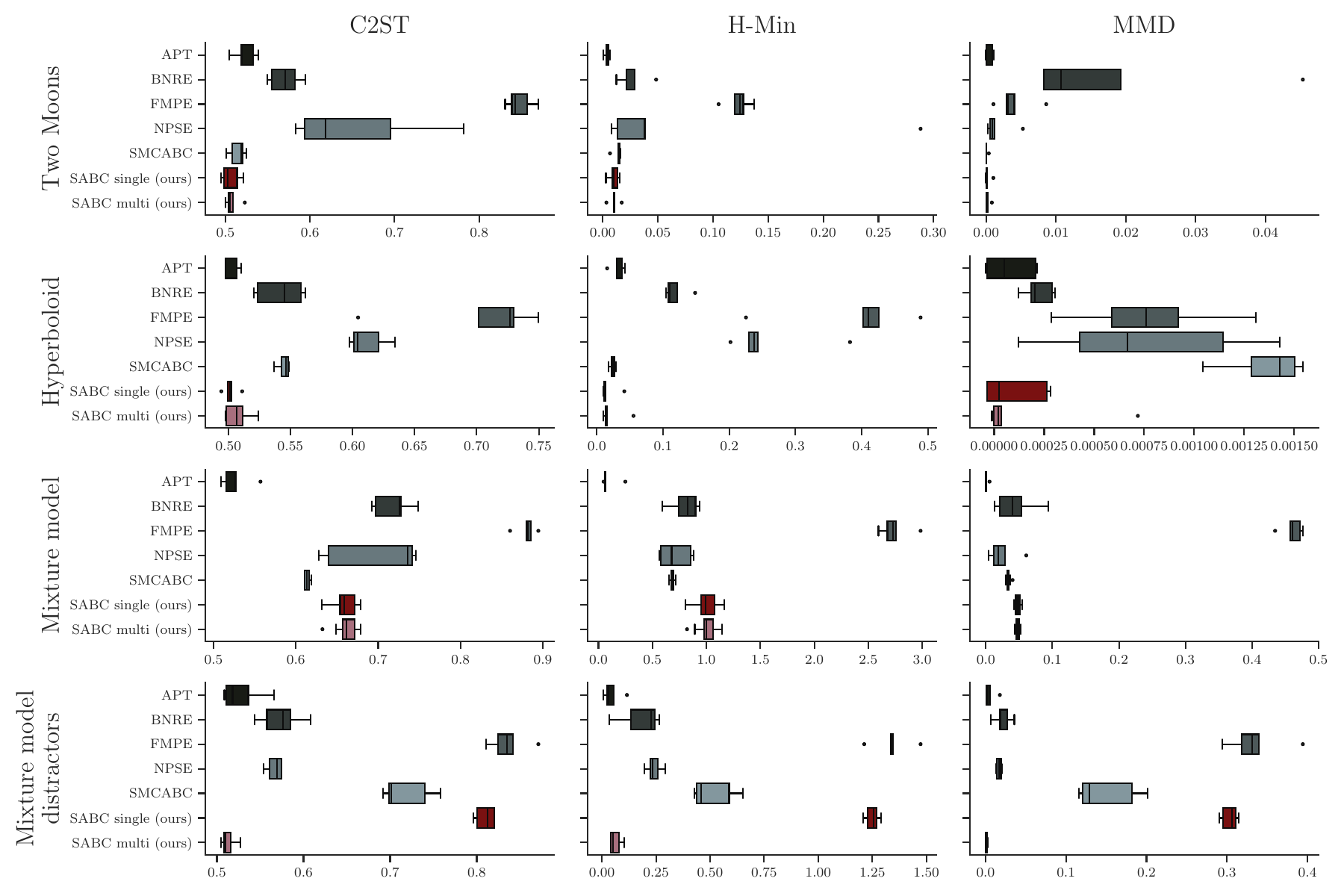}
    \caption{Evaluation of SABC and baseline methods on benchmark tasks using C2ST, H-min, and MMD metrics.}
    \label{app:sabc-bechmark-tasks}
\end{figure}

\clearpage
\subsubsection{Posterior distributions}
\label{app:sabc-benchmark-results-all-posteriors}


\begin{figure}[!ht]
\centering
\begin{subfigure}[b]{0.3\textwidth}
\includegraphics[width=1\textwidth]{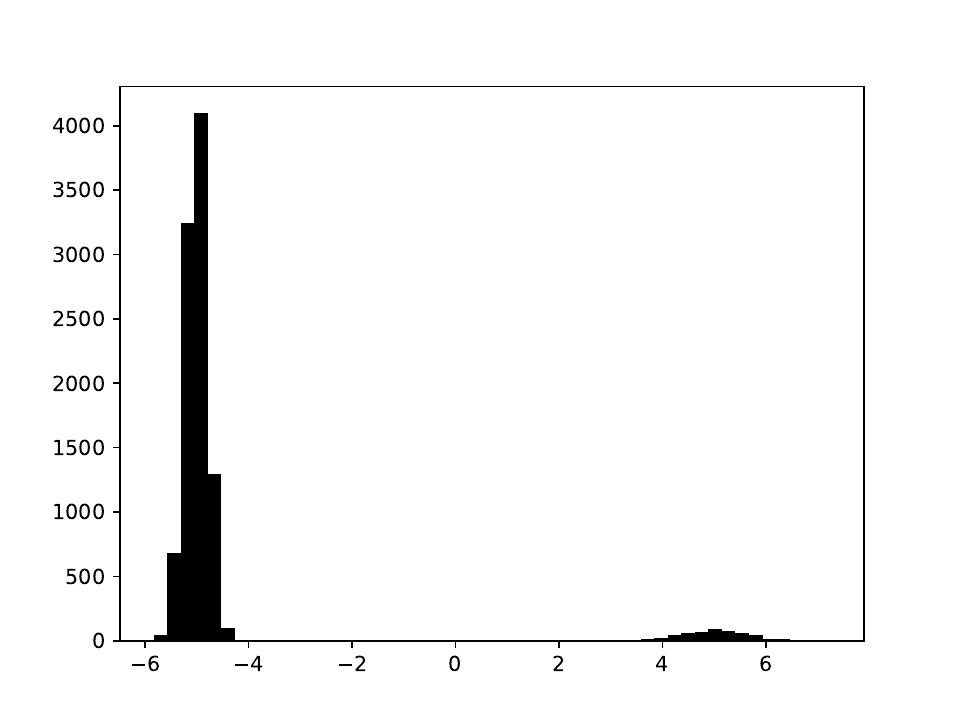}
\caption{Reference.}
\end{subfigure}
\begin{subfigure}[b]{0.3\textwidth}
\includegraphics[width=1\textwidth]{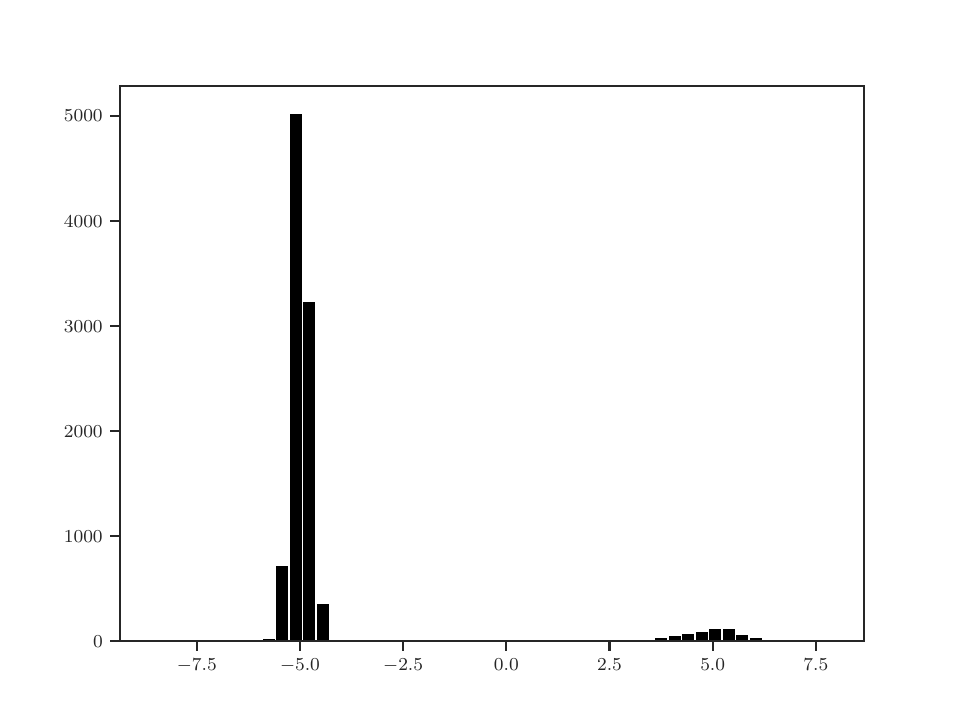}
\caption{APT.}
\end{subfigure}
\begin{subfigure}[b]{0.3\textwidth}
\includegraphics[width=1\textwidth]{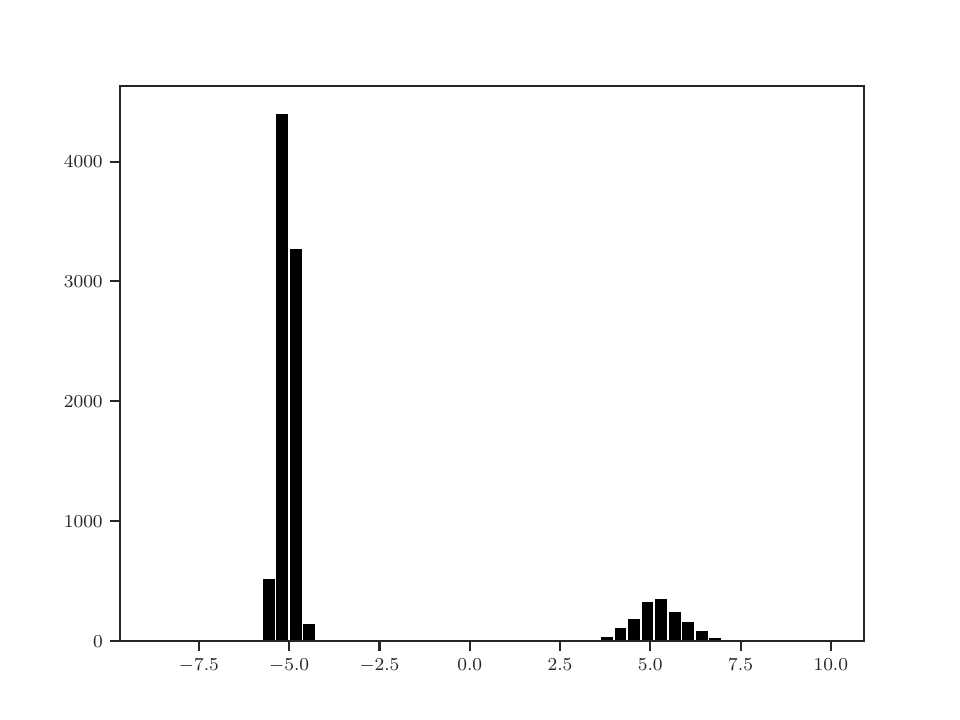}
\caption{BNRE.}
\end{subfigure}
~

\begin{subfigure}[b]{0.3\textwidth}
\includegraphics[width=1\textwidth]{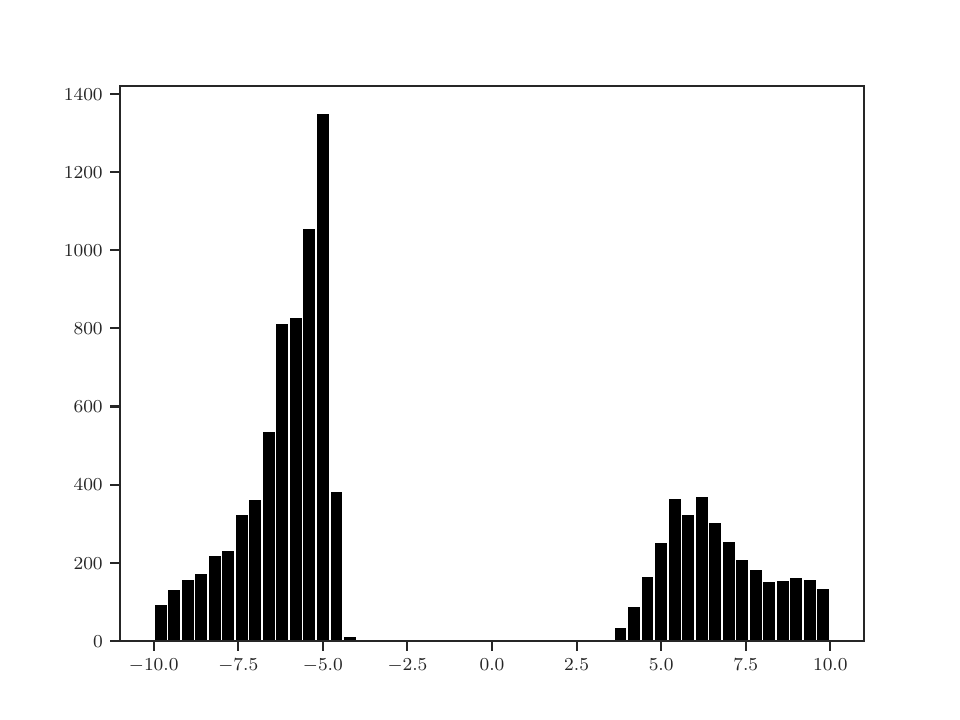}
\caption{FMPE.}
\end{subfigure}
\begin{subfigure}[b]{0.3\textwidth}
\includegraphics[width=1\textwidth]{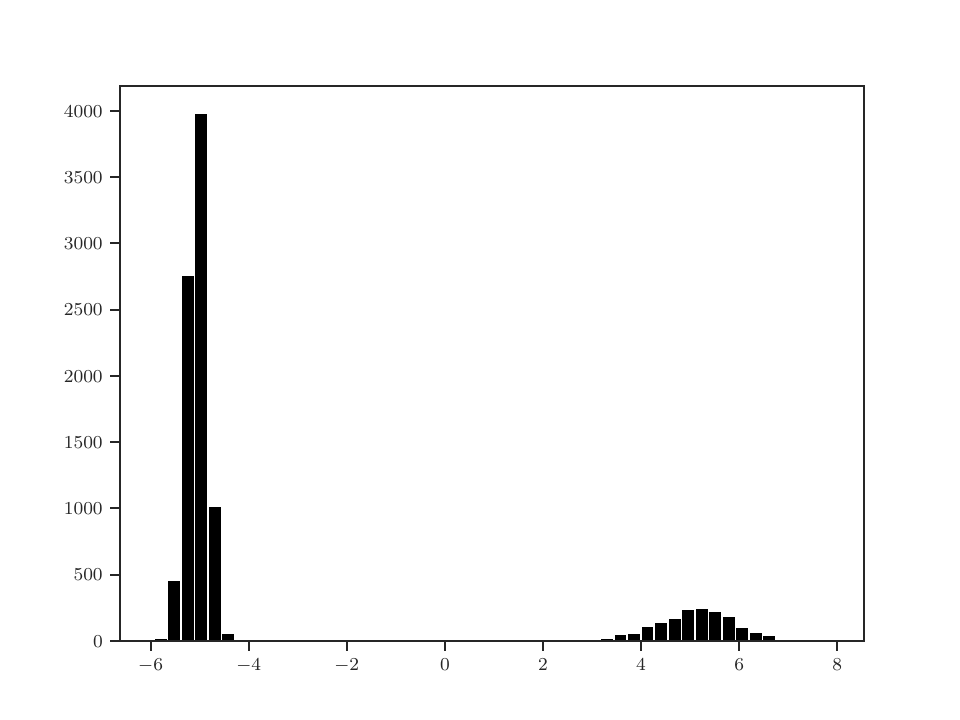}
\caption{NPSE.}
\end{subfigure}
\begin{subfigure}[b]{0.3\textwidth}
\includegraphics[width=1\textwidth]{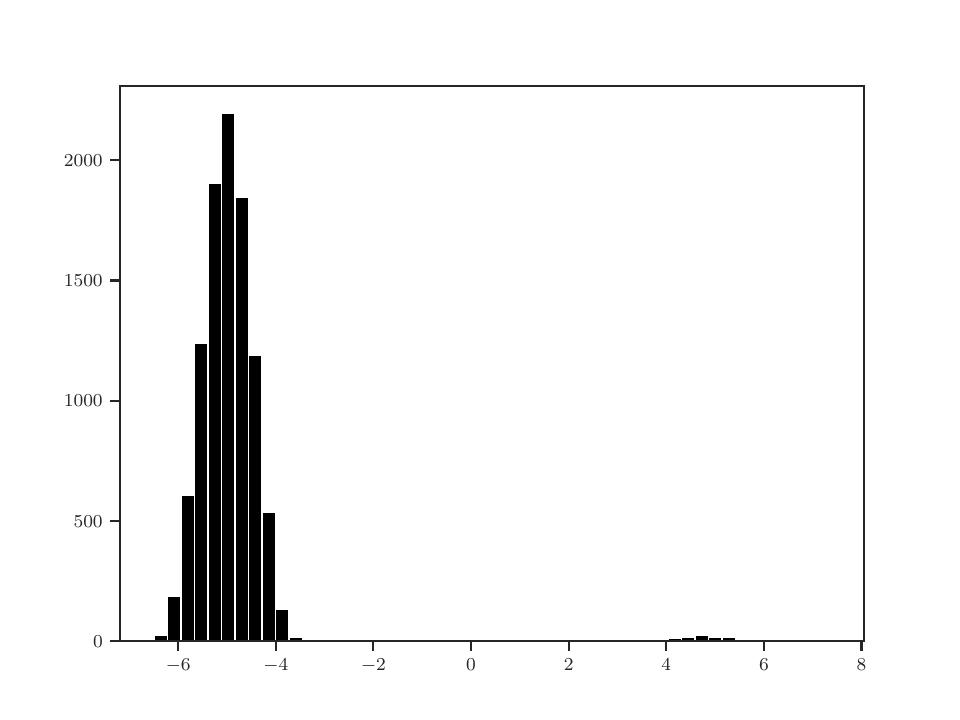}
\caption{SMCABC.}
\end{subfigure}
~
\begin{subfigure}[b]{0.3\textwidth}
\includegraphics[width=1\textwidth]{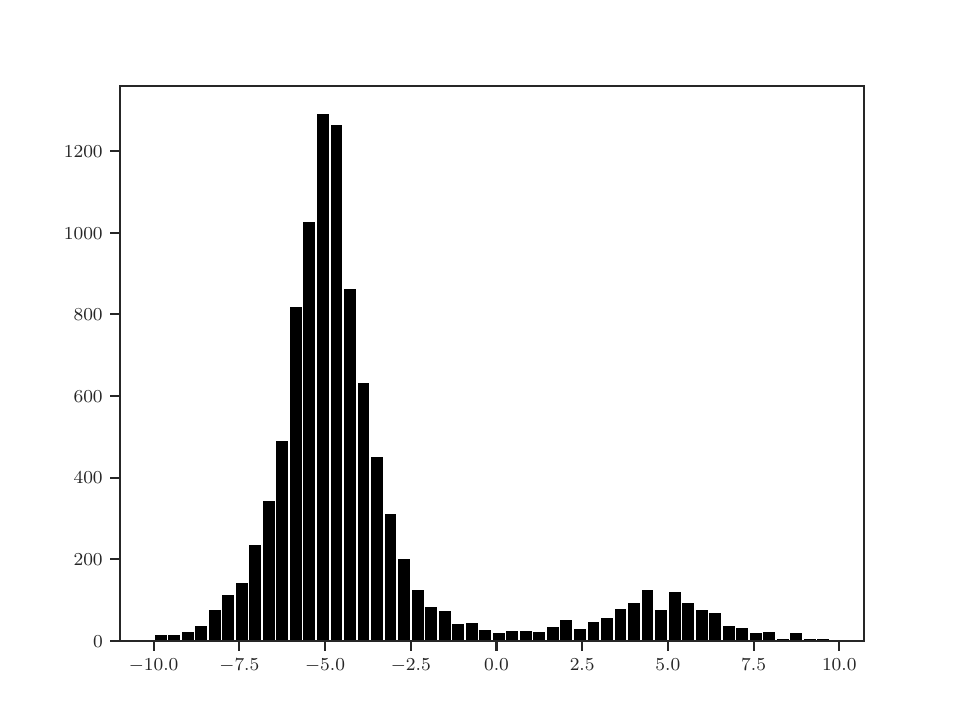}
\caption{SABC single (ours).}
\end{subfigure}
\begin{subfigure}[b]{0.3\textwidth}
\includegraphics[width=1\textwidth]{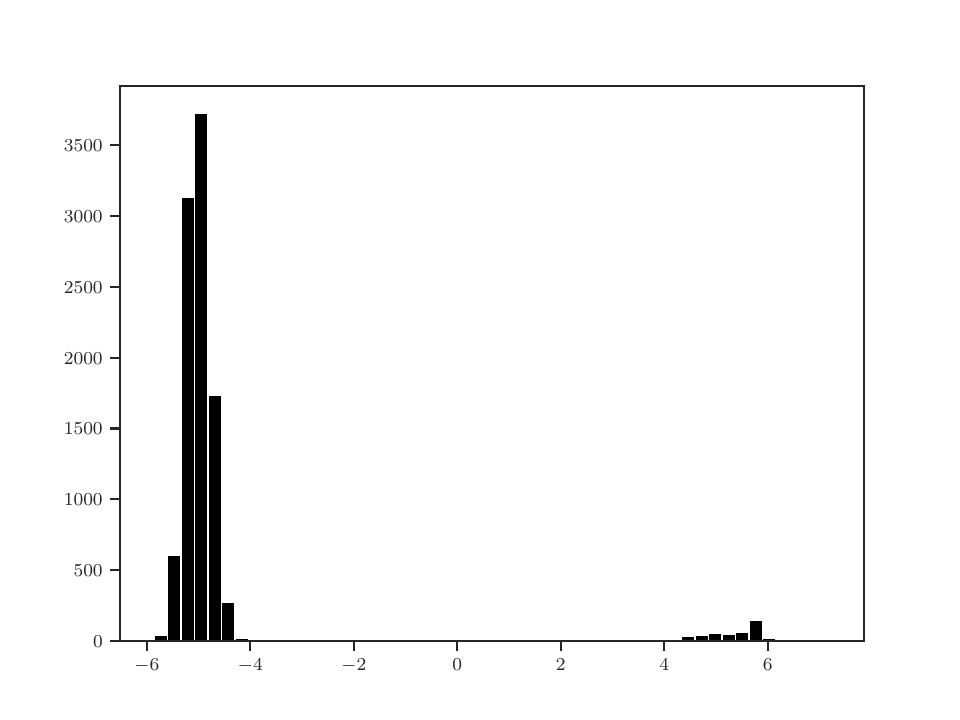}
\caption{SABC multi (ours).}
\end{subfigure}
\caption{Posterior distributions for the mixture model with distractors example using a specific seed.}
\end{figure}


\begin{figure}[!ht]
\centering
\begin{subfigure}[b]{0.3\textwidth}
\includegraphics[width=1\textwidth]{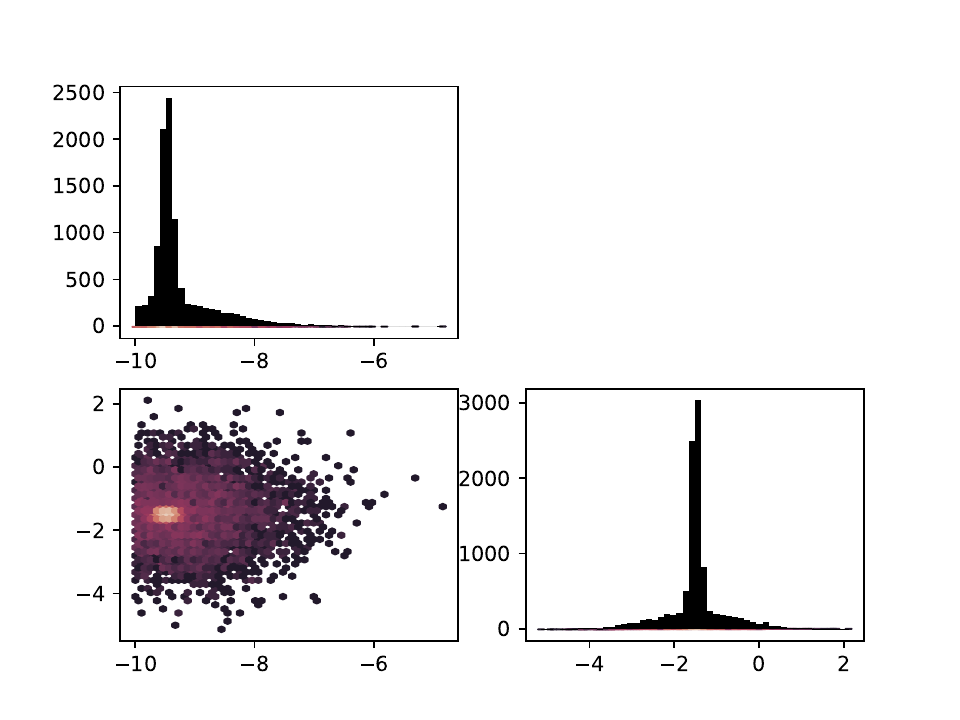}
\caption{Reference.}
\end{subfigure}
\begin{subfigure}[b]{0.3\textwidth}
\includegraphics[width=1\textwidth]{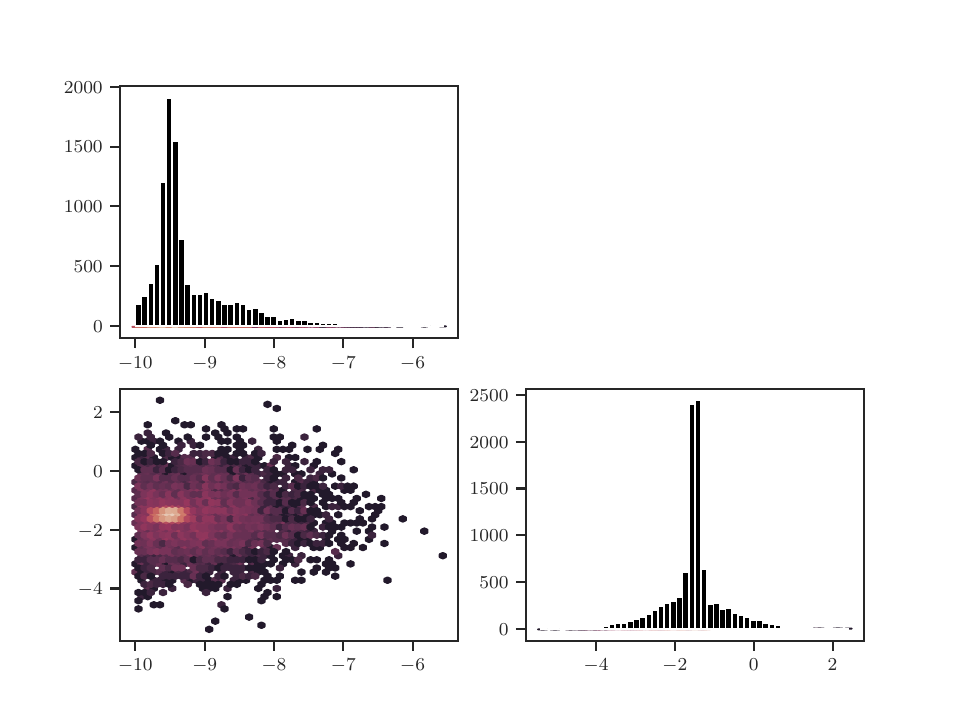}
\caption{APT.}
\end{subfigure}
\begin{subfigure}[b]{0.3\textwidth}
\includegraphics[width=1\textwidth]{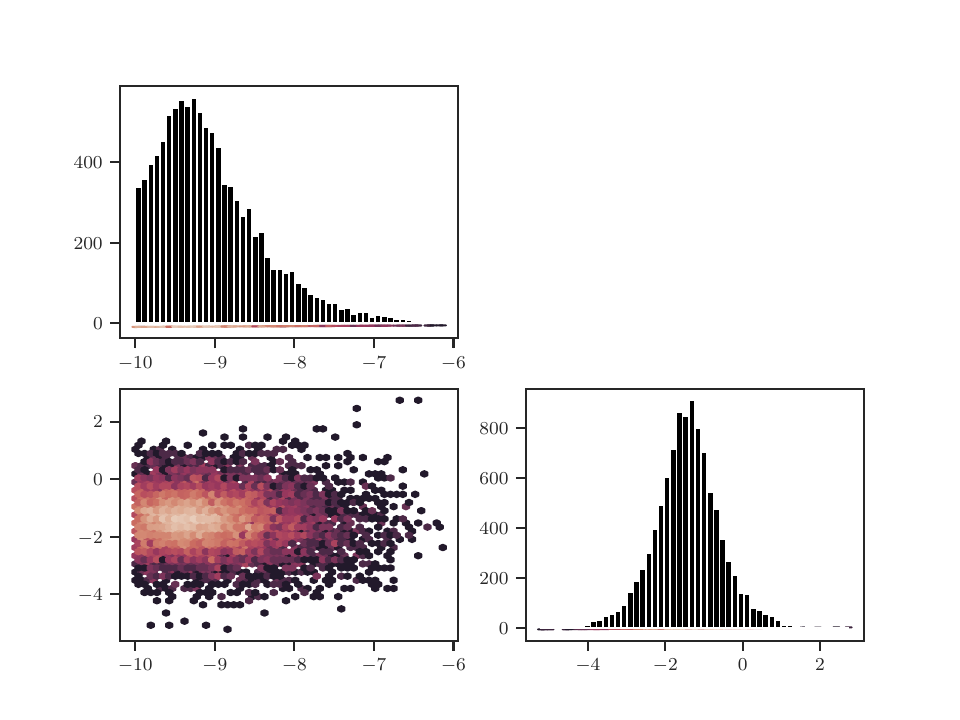}
\caption{BNRE.}
\end{subfigure}
~
\begin{subfigure}[b]{0.3\textwidth}
\includegraphics[width=1\textwidth]{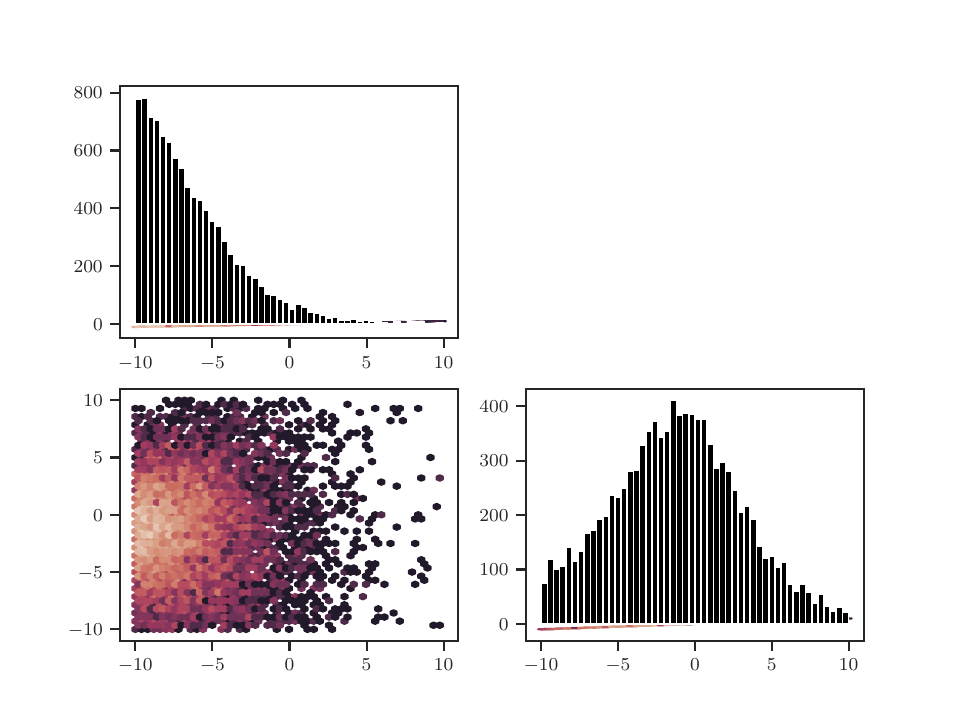}
\caption{FMPE.}
\end{subfigure}
\begin{subfigure}[b]{0.3\textwidth}
\includegraphics[width=1\textwidth]{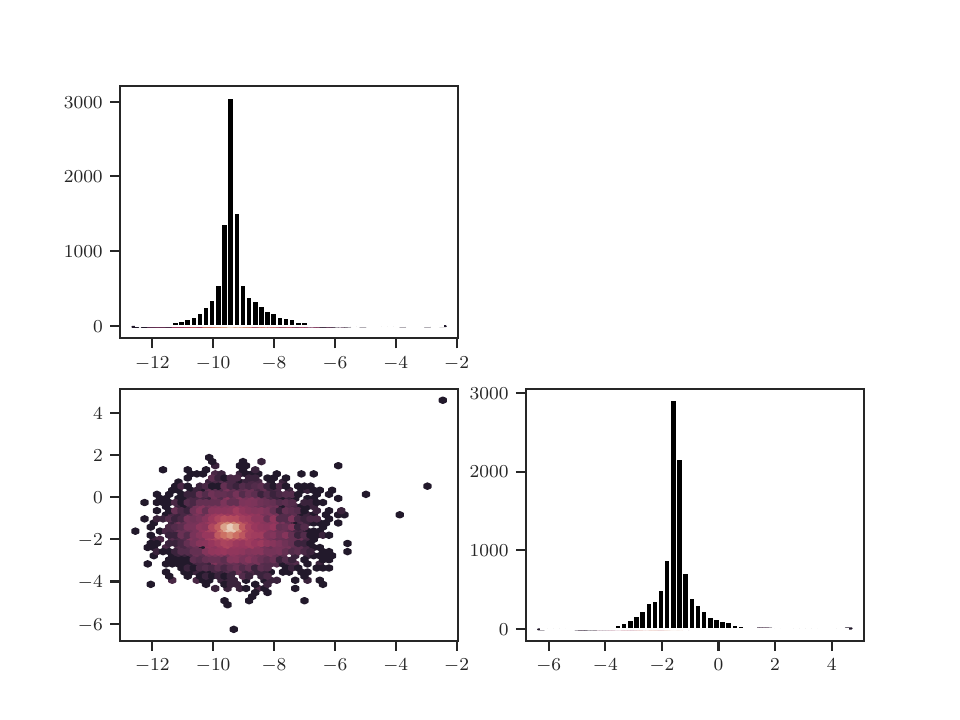}
\caption{NPSE.}
\end{subfigure}
\begin{subfigure}[b]{0.3\textwidth}
\includegraphics[width=1\textwidth]{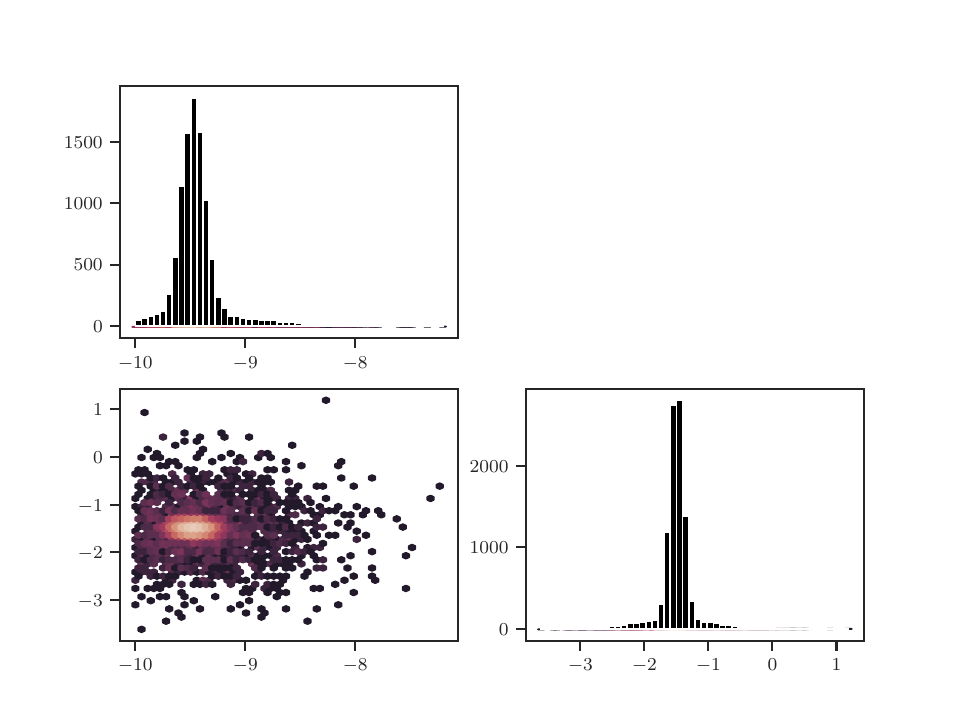}
\caption{SMCABC.}
\end{subfigure}
~
\begin{subfigure}[b]{0.3\textwidth}
\includegraphics[width=1\textwidth]{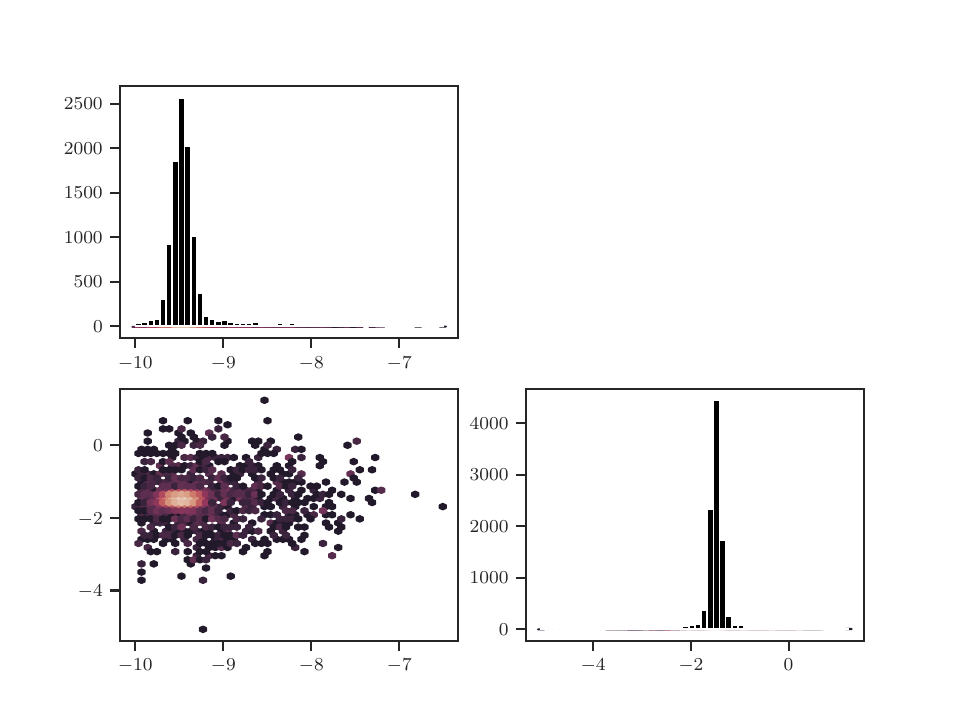}
\caption{SABC single (ours).}
\end{subfigure}
\begin{subfigure}[b]{0.3\textwidth}
\includegraphics[width=1\textwidth]{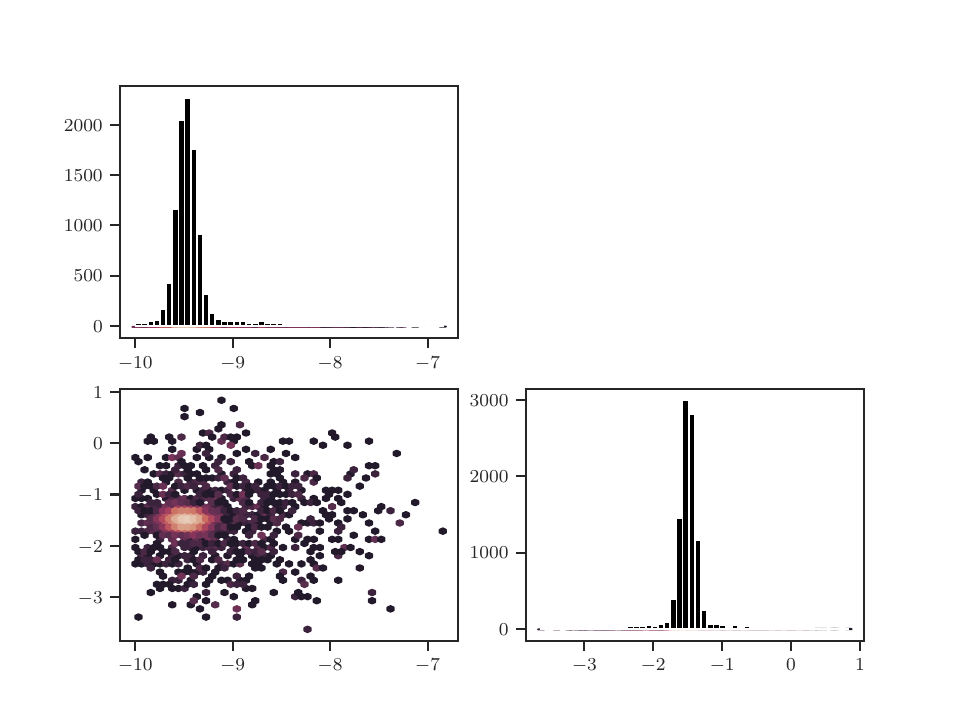}
\caption{SABC multi (ours).}
\end{subfigure}
\caption{Posterior distributions for the mixture model example using a specific seed.}
\end{figure}

\begin{figure}[!ht]
\centering
\begin{subfigure}[b]{0.3\textwidth}
\includegraphics[width=1\textwidth]{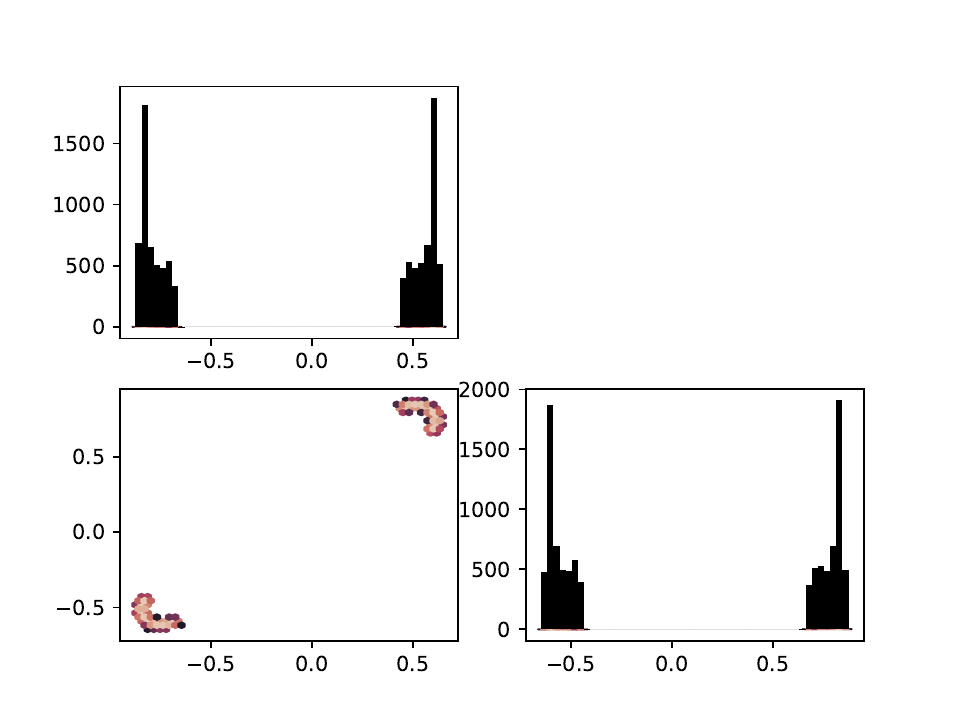}
\caption{Reference.}
\end{subfigure}
\begin{subfigure}[b]{0.3\textwidth}
\includegraphics[width=1\textwidth]{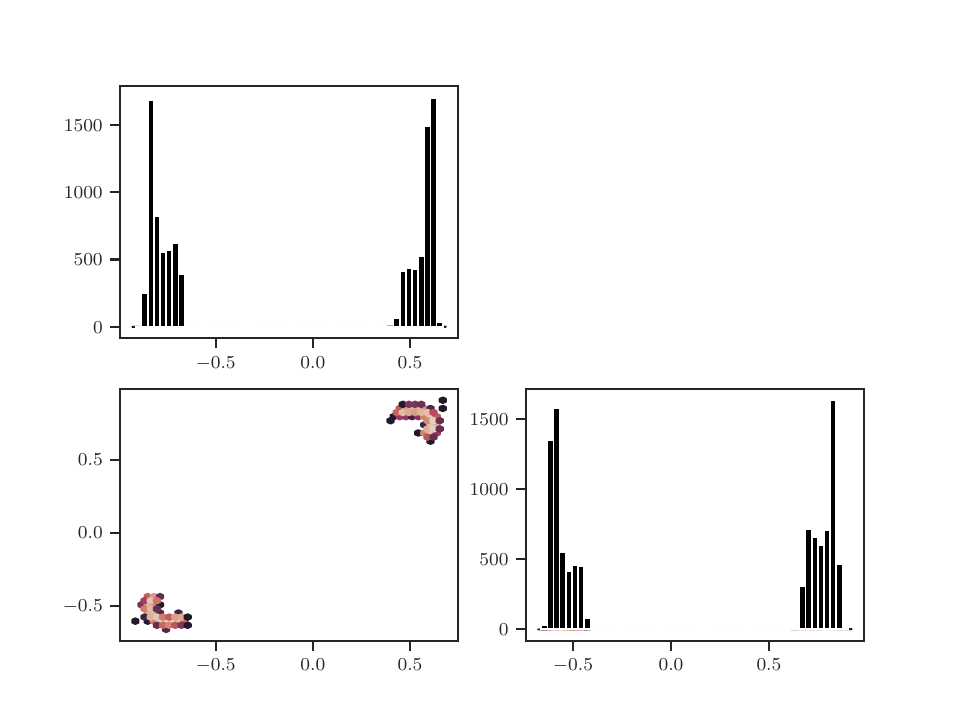}
\caption{APT.}
\end{subfigure}
\begin{subfigure}[b]{0.3\textwidth}
\includegraphics[width=1\textwidth]{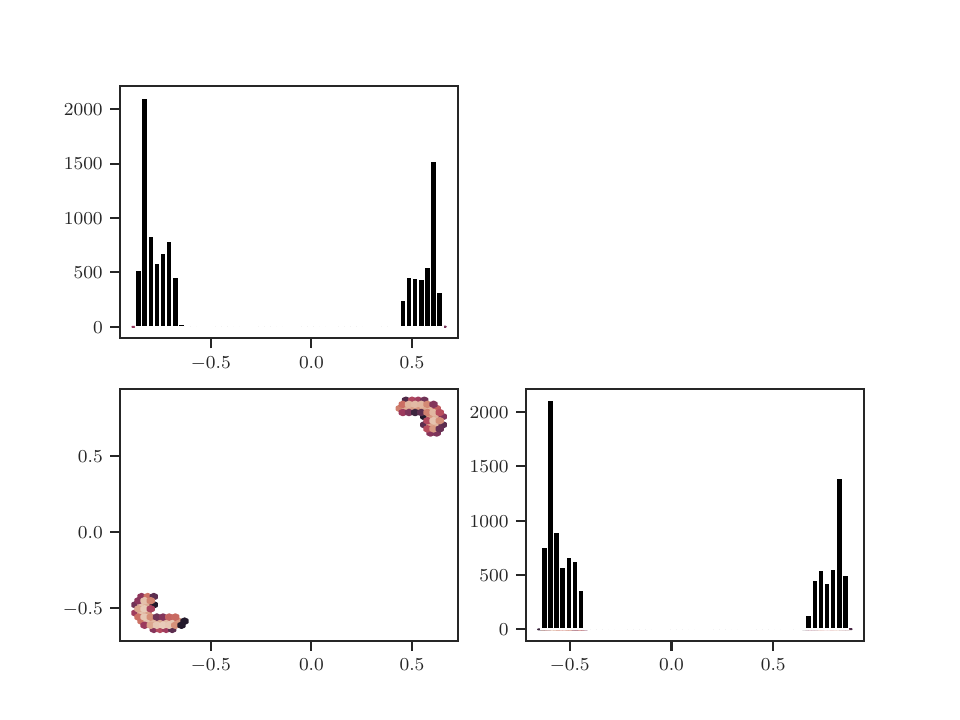}
\caption{BNRE.}
\end{subfigure}
~
\begin{subfigure}[b]{0.3\textwidth}
\includegraphics[width=1\textwidth]{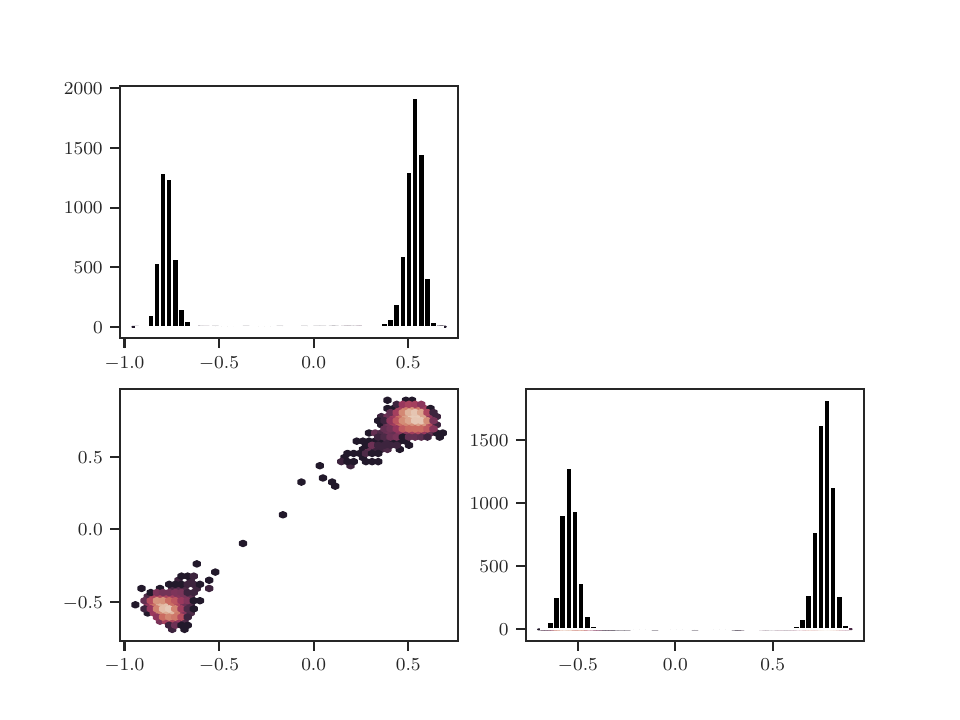}
\caption{FMPE.}
\end{subfigure}
\begin{subfigure}[b]{0.3\textwidth}
\includegraphics[width=1\textwidth]{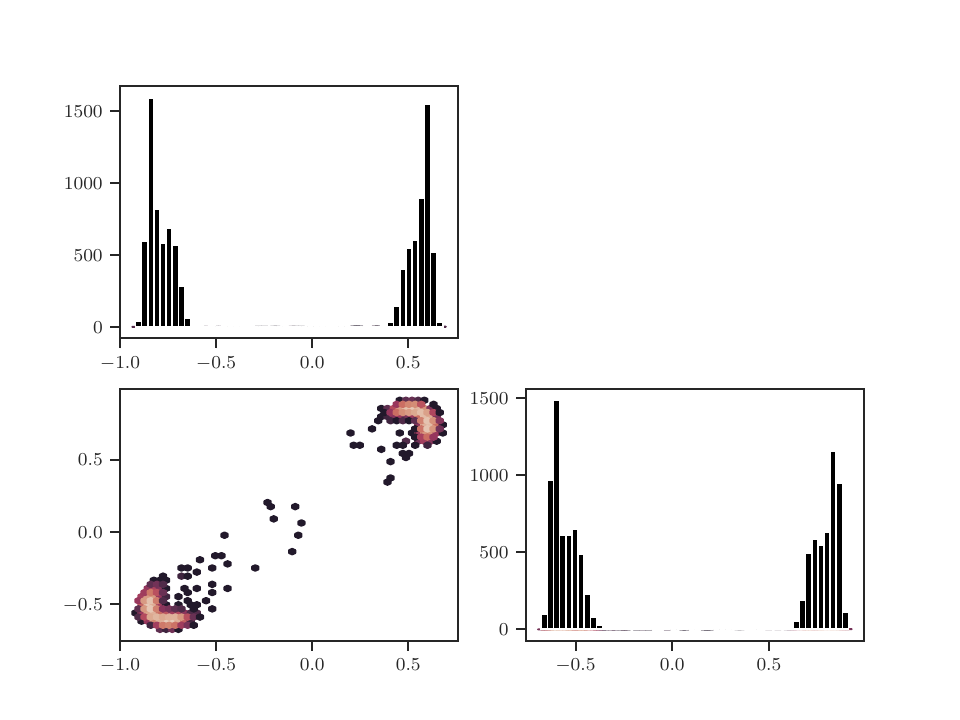}
\caption{NPSE.}
\end{subfigure}
\begin{subfigure}[b]{0.3\textwidth}
\includegraphics[width=1\textwidth]{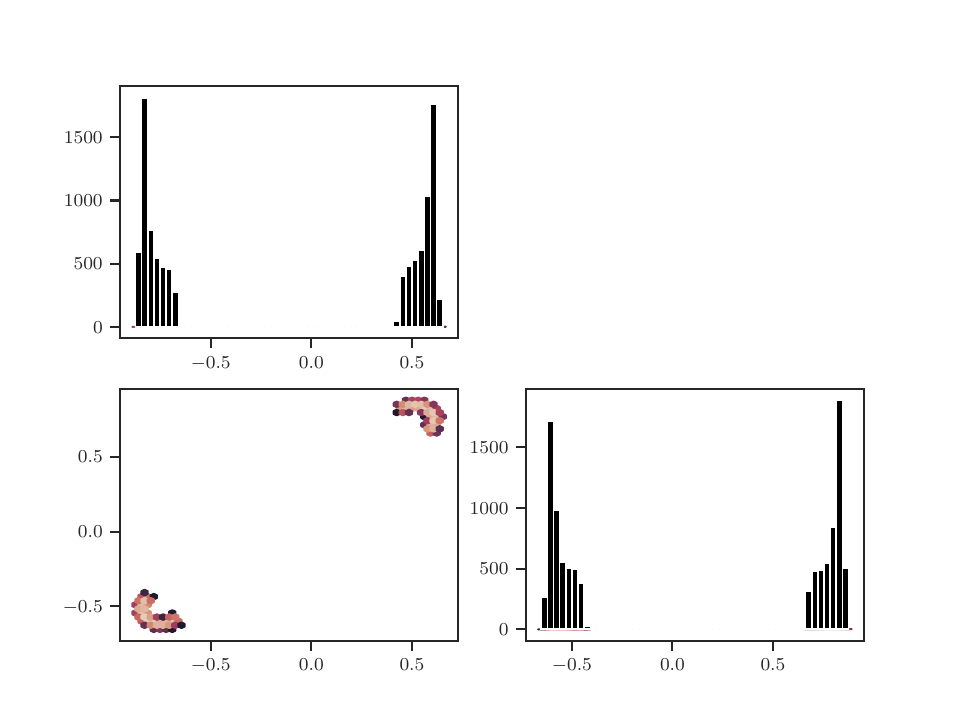}
\caption{SMCABC.}
\end{subfigure}
~
\begin{subfigure}[b]{0.3\textwidth}
\includegraphics[width=1\textwidth]{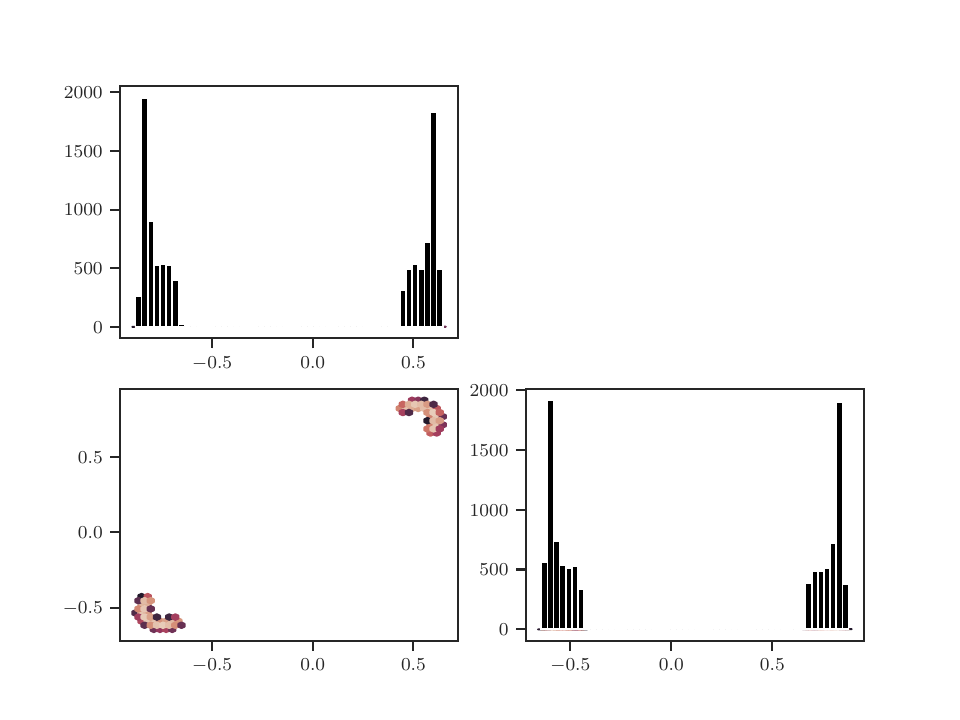}
\caption{SABC single (ours).}
\end{subfigure}
\begin{subfigure}[b]{0.3\textwidth}
\includegraphics[width=1\textwidth]{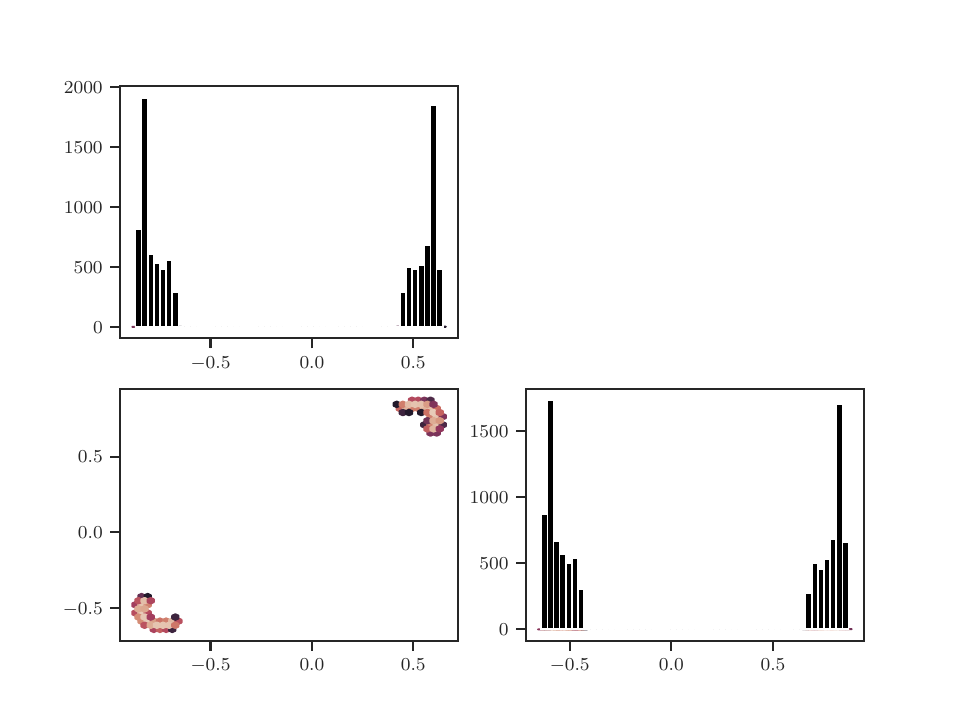}
\caption{SABC multi (ours).}
\end{subfigure}
\caption{Posterior distributions for the two moons example using a specific seed.}
\end{figure}

\clearpage
\subsubsection{Rho trajectories}

\begin{figure}[!ht]
\centering
\begin{subfigure}[b]{0.35\textwidth}
\includegraphics[width=1\textwidth]{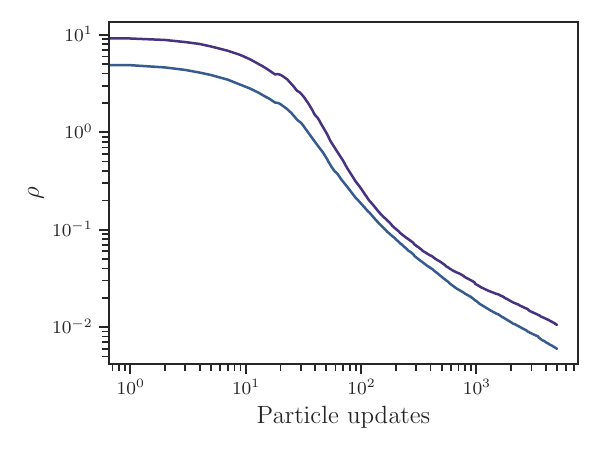}
\end{subfigure}
\begin{subfigure}[b]{0.35\textwidth}
\includegraphics[width=1\textwidth]{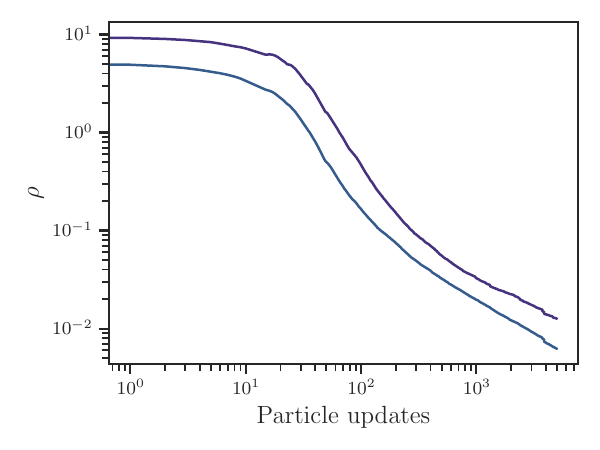}
\end{subfigure}
\begin{subfigure}[b]{0.35\textwidth}
\includegraphics[width=1\textwidth]{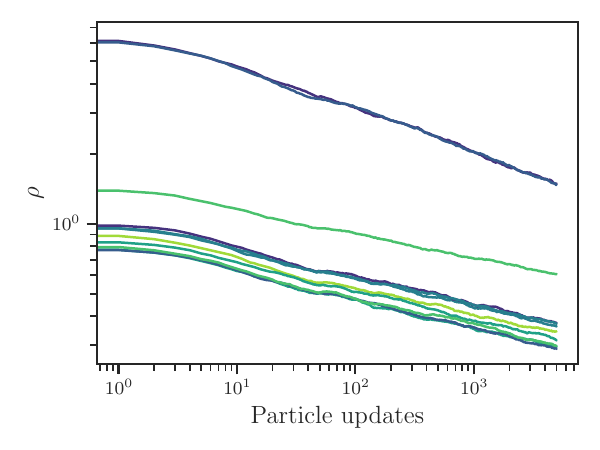}
\end{subfigure}
\begin{subfigure}[b]{0.35\textwidth}
\includegraphics[width=1\textwidth]{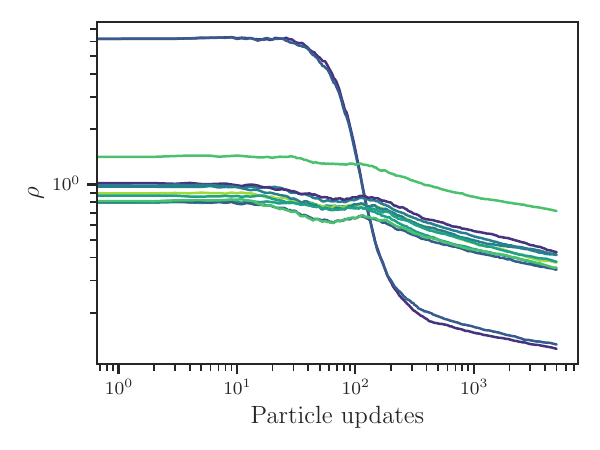}
\end{subfigure}
\begin{subfigure}[b]{0.35\textwidth}
\includegraphics[width=1\textwidth]{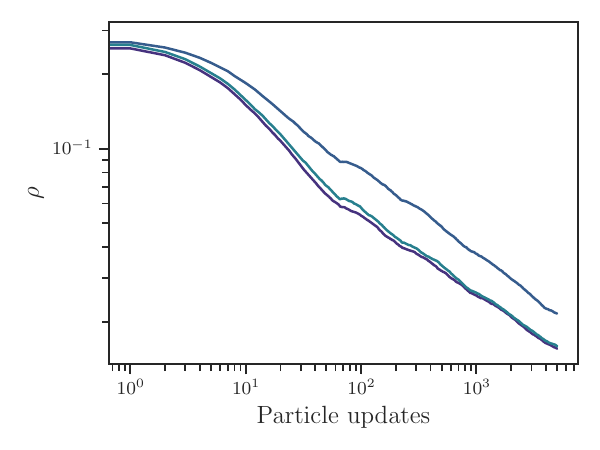}
\end{subfigure}
\begin{subfigure}[b]{0.35\textwidth}
\includegraphics[width=1\textwidth]{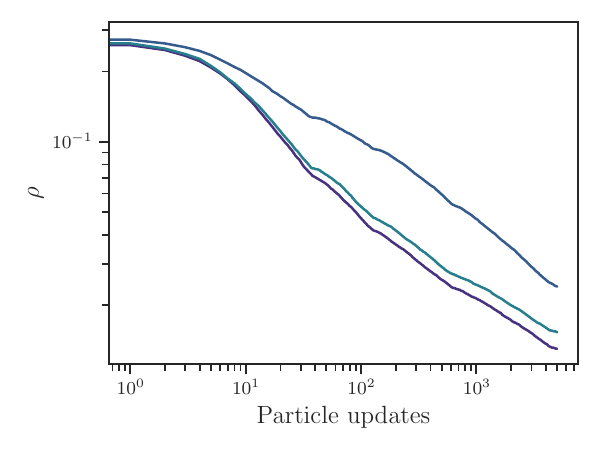}
\end{subfigure}
\begin{subfigure}[b]{0.35\textwidth}
\includegraphics[width=1\textwidth]{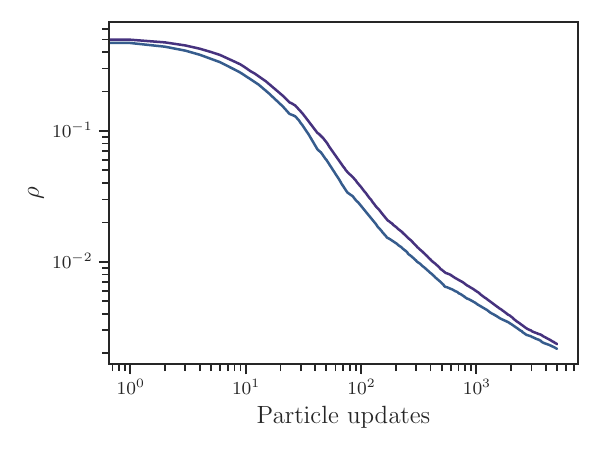}
\end{subfigure}
\begin{subfigure}[b]{0.35\textwidth}
\includegraphics[width=1\textwidth]{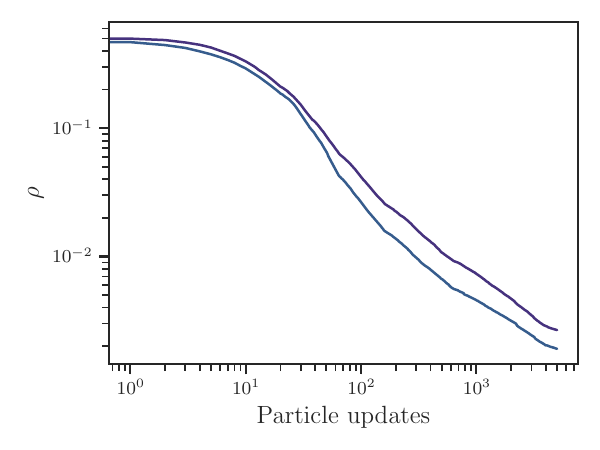}
\end{subfigure}
\caption{Individual distance trajectories $\rho_i$ during SABC sampling for the mixture model, mixture with distractors, hyperboloid, and two moons (top to bottom).
Left: single temperature; Right: multiple temperatures.
Single-temperature SABC keeps distances aligned, while multiple temperatures allow informative statistics to converge faster (notably in the distractor case).}
\label{fig:rho-trajectories}
\end{figure}

\clearpage

\section{Additional information on the SIR model}

\subsection{Model description}\label{AppSIR}

The SIR model is a common SBI benchmark model \citep{lueckmann2021benchmarking}. The ODE is defined as follows:
\begin{equation}
\begin{split}
\frac{\mathrm{d}S}{\mathrm{d}t}  &= - \theta_1 \frac{SI}{N} \,, \\
\frac{\mathrm{d}I}{\mathrm{d}t}  &= \theta_1 \frac{SI}{N}  -\theta_2 I \,,\\
\frac{\mathrm{d}R}{\mathrm{d}t}  &=  \theta_2 I \,, \\
\end{split}
\end{equation}
where $N$ is a constant. We aim to compute the posterior, for the two parameters $\theta_1$ and $\theta_2$, given a time series of observations 
\begin{equation}
 Y_t \sim \text{Binomial}(1000, \tfrac{I_t}{N})\,,  
 \quad 
 t=1,\dots,T\,.
\end{equation}
Following \cite{lueckmann2021benchmarking}, we simulate the SIR model 
using a random prior draw from
\begin{equation}
\begin{split}
\theta_1 \sim \text{LogNormal}(\log(0.4), 0.5)\\
\theta_2 \sim \text{LogNormal}(\log(1/8), 0.25)\,.
\end{split}
\end{equation}
As initial conditions we set $S=999999$, $I=1$ and $R=0$. We set $T=160$ and evaluate the ODE at 100 equidistant time points.

\subsection{Experimental details}

We obtain reference posterior distributions using a Hit-and-Run slice sampler.
We run four chains for $10{,}000$ iterations each, discarding the first $5{,}000$ as warmup.
Standard MCMC diagnostics are used to assess convergence.

For the baseline methods, we adopt the same hyperparameter configurations as in Section~\ref{app:benchmark-tasks-experiments}. 
For SABC, we first sample a data set $\{(\vc y_n, \bt_n) \}_{n=1}^N$ from the prior of size $N=10000$. We then train a summary network as described in \cite{chen2023learning}. Their method makes use of three different kinds of networks:
\begin{itemize}
    \item A (primary) summary network $g$ which we parameterize using an MLP with two hidden layers of 128 nodes each which reduces the dimensionality of the data to $K$ summaries where $K$ is a hyperparameter which we set to $4$.
    \item A secondary summary network $h$ which we parameterize using another MLP with the same architecture as $g$.
    \item A MLP network acting as a "critic" with a single hidden layer with 128 nodes.
\end{itemize}
After training the network to convergence, $g$ is used within SABC to compute summary statistics.
We run both SABC variants for $50 \ 000 \ 000$ iterations yielding $10\ 000$ posterior samples.

\section{Additional information on the Jansen-Rit model}

\subsection{Model description}

The Jansen-Rit neural mass model is a 6-dimensional SDE of the form:

\begin{equation*}
\begin{split}
\mathrm{d}Y_0(t) \ &= \ Y_3(t)\mathrm{d}t \\
\mathrm{d}Y_1(t) \ &= \ Y_4(t)\mathrm{d}t \\
\mathrm{d}Y_2(t) \ &= \ Y_5(t)\mathrm{d}t \\
\mathrm{d}Y_3(t) \ &= \ \bigg[ Aa \Big[  \text{sigm}\Big(Y_1(t) - Y_2(t) \Big)  \Big] - 2aY_3(t) - a^2Y_0(t) \bigg]  \mathrm{d}t +\sigma_3 \mathrm{d}W_3(t)\\
\mathrm{d}Y_4(t) \ &= \ \bigg[ Aa \Big[ \mu + C_2\text{sigm}\Big(C_1 Y_0(t) \Big) \Big] - 2aY_4(t) - a^2Y_1(t) \bigg]  \mathrm{d}t +\sigma_4 \mathrm{d}W_4(t)\\
\mathrm{d}Y_5(t) \ &= \ \bigg[ Bb \Big[ C_4\text{sigm}\Big(C_3 Y_0(t) \Big) \Big] - 2bY_5(t) - b^2Y_2(t)  \bigg]  \mathrm{d}t +\sigma_5 \mathrm{d}W_5(t)\,,\\
\end{split}
\end{equation*}
where
\begin{equation*}
    \text{sigm}(y) = \frac{v_{\text{max}}}{1 + \exp(r(v_0 - y))}\,,
\end{equation*}
and where the observed data is
\begin{equation}
y(t) = 10^{g/10}\left( y_1(t) - y_2(t) \right)\,.
\end{equation}
We choose all constants following \cite{ableidinger2017stochastic} setting $A=3.25$, $B=22$, $a=100$, $b=50$, $v_0=6$, $v_{\text{max}}=5$, $r=0.56$, $\sigma_3=0.01$, $\sigma_5=1.0$ and $\sigma_4 = \sigma$ (i.e., one of the free parameters of the model). The parameters $C_i$ are related via $C_1 = C, C_2 = 0.8 C, C_3 = C_4 = 0.25 C$. For inferring the posterior distribution, we compute the PSD of the signal using Welch's method evaluated at frequencies from $0Hz$ to $64Hz$ and use it as a set of summary statistics.

\subsection{Experimental details}

For SABC and all baseline methods, we employed the same hyperparameter settings as in the benchmark tasks.(Appendix~\ref{app:benchmark-tasks-experiments}).

\section{Additional information on the solar dynamo model}

\subsection{Model description}
\label{app:solardynamo-models}

We consider a second-order stochastic delay differential equation model describing the time evolution of the solar magnetic field strength $B(t)$ \citep{wilmot_2006_delayODEsolarDynamo, albert2021SR,ulzega2025shedding}:
\begin{equation}\label{eq:dynamo_model}
\left( \tau \frac{d}{dt} + 1\right)^2 B(t) = - \mathcal{N}_d F\left( B(t-T) \right) + \sqrt{\tau} B_{\mathrm{max}} \sigma \eta(t) \,,
\end{equation}
where $\tau$ is the magnetic diffusion timescale, $\mathcal{N}_d$ is the dimensionless dynamo number, quantifying the efficiency of the field-generating dynamo process, which amplifies and sustains magnetic fields against field-destroying diffusive decay in the solar interior, $T$ is a time-delay parameter accounting for the finite timescales associated with the transport of magnetic fields by meridional circulation and the buoyant rise of magnetic flux tubes through the solar convection zone, and 
\begin{equation}\label{eq:quenching_function}
    F(B)=\frac{B}{4} \left( 1+ \mathrm{erf}\left(B^2-B^2_{\mathrm{min}}\right) \right)\left( 1- \mathrm{erf}\left(B^2-B^2_{\mathrm{max}}\right) \right) \,,
\end{equation}
is a nonlinear quenching function that restricts the action of the field-generating mechanisms to field strengths in the range $B_{\mathrm{min}} \lesssim B \lesssim B_{\mathrm{max}}$. The last term on the right-hand side of Equation~(\ref{eq:dynamo_model}) is an additive white-noise contribution, with $\langle \eta(t) \eta(t') \rangle = \delta(t-t') $ and standard deviation $\sqrt{\tau} B_{\mathrm{max}} \sigma$. Here $\sigma$ is the dimensionless noise amplitude. For further details on the model and its parameters, see \cite{albert2021SR} and references therein.
The vector of parameters to be inferred is $\boldsymbol{\theta} = \left\{B_{\mathrm{max}} , \mathcal{N}_d, T, \sigma, \tau  \right\}$ for which we use the following prior distributions
\begin{equation*}
\begin{split}   
B_{\mathrm{max}}  &\sim \mathcal{U}(1.0, 15.0), \\
\mathcal{N}_d & \sim \mathcal{U}(1.0, 15.0), \\
T &\sim \mathcal{U}(0.1, 120), \\
\sigma &\sim \mathcal{U}(0.01, 0.3), \\
\tau &\sim \mathcal{U}(0.1, 120). 
\end{split}
\end{equation*}
The lower threshold $B_{\mathrm{min}}$ is not inferred and is set to $B_{\mathrm{min}} = 1$.

We use two independent datasets, namely the official monthly mean SN record from \cite{SILSO_Sunspot_Number}, which consists of 3251 observations covering the time span 1749-2019, and the high-resolution $^{14}$C-based SN reconstruction of \cite{usoskin2021solarActivityResonstructed}. The latter time series comprises 929 data points and represents the first physics-based quantitative reconstruction of solar magnetic activity over a millennial timescale (971-1899), with an exceptional annual resolution that allows individual solar cycles to be resolved. We assume that both the observed and reconstructed SN values are proportional to the squared magnetic field $B^2(t)$. 
Observations $\vc x_{obs}$ are sampled at $N+1$ time points $n \Delta t$, with $n = 0,1,...,N$, and with the sampling interval $\Delta t = 1$ month.
In the SABC inference we compare the summary statistics $\vc s_{obs}$ to the corresponding features $\vc s=\vc s(\vc x)$ where $\vc x = \mathbf{B}^2$ and $\mathbf{B}$ is the model output sampled at the same $N+1$ time points as the observed data. The comparison is made using a standard Euclidean distance between the individual summary statistics. 

Given the oscillatory nature of the system, it is convenient to focus on specific spectral features of the data. Therefore, we consider selected Fast Fourier Transform (FFT) components as summary statistics.
However, applying the FFT requires windowing the data to reduce spectral leakage. We use the discrete Hann window, a well-established method in signal processing,
\begin{equation}
    w_n = \sin^2 \left( \frac{\pi n}{N} \right) \,, \quad 0\leq n \leq N  \,.
\end{equation}
It can be shown that the Hann window introduces correlations between Fourier modes whose index difference is less than 3. Here we adopt the absolute values of 20 FFT components with indexes varying in steps of 6, that is, $ \left\{0,6,12,18,24,...,114 \right\}$, as summary statistics. This choice allows us to avoid the correlations between Fourier modes introduced by the Hann window, while retaining enough information to characterize the spectral properties of the data. Moreover, ignoring higher-order modes allows us to discard high-frequency noise that may not be adequately captured by the model.
For more information, see \cite{ulzega2025shedding}.

\subsection{Experimental details}
\label{app:solardynamo-experimental-details}

For the solar dynamo model, we evaluated the SABC (single) version against the sequential neural SBI methods SNLE \citep{papamakarios_2019_sequentialNeuralLikelihood} and APT \citep{greenberg2019automatic}. We also conducted extensive experiments using SNLE, APT, FMPE, NPSE, and the dimensionality-reducing SSNLE \citep{dirmeier2023simulation} on the raw solar dynamo data, but found that a) using manually crafted summary statistics works better than evaluating the aforementioned methods on the raw data, and b) that SNLE and APT produced significantly better posterior inferences than the other previously used baselines (data not shown).


For both APT and SNLE, we use masked autoregressive flows with $10$ flow layers \citep{papamakarios2017masked}. Each flow layer uses an autoregressive network with $2$ layers and $64$ hidden nodes. Both APT and SNLE are trained sequentially for $10$ rounds, where for each round we generate $20\ 000$ novel outputs using the proposal posteriors obtained from the previous round. We use slice samplers for both APT and SNLE to draw samples from the proposal posterior distributions. We use an Adam optimizer with a learning rate of $0.0001$ and train until convergence or until $2\ 000$ iterations are reached, respectively. After the last round of training, we draw a posterior sample of size $20\ 000$ which we then use to draw $20\ 000$ realizations from the PPD (eq. \ref{eq:PPD}) (i.e., one draw for each posterior sample).

For SABC, we have opted for the yearly mean SN dataset~\citep{SILSO_Sunspot_Number} over the higher-resolution monthly mean SN, as it retains sufficient resolution to capture the spectral features of interest while substantially reducing the high computational cost typical of ABC-type algorithms. The yearly SN dataset comprises $271$ data points. We have initially run the inference using $10\ 000$ particles. Then, to refine the posterior sample, we have performed a final importance sampling step by discarding slowly converging, less informative particles with the largest distances between observed and simulated data. To this end, we impose an arbitrary cut-off distance such that the least well-converged $30\%$ of the particles are removed from the posterior sample. The latter thus consists of about $7000$ particles. The hyperparameters are the same as the benchmark tasks.

For visualization (i.e., Figure~\ref{fig:sabc-solardynamo-ppd}), we sort the PPD samples by their Euclidean distance to the observed summary statistics $\vc s(\vc x_{obs})$ and take $50$ samples corresponding to equidistantly-spaced quantiles of the distances.

The SNLE and APT experiments were conducted on an AMD EPYCTM 7742 processor with 64 cores and 256 GB RAM. Runtimes were roughly 24 hours in both cases. The SABC algorithm was run on a single HPE ProLiant XL230k Gen10 node using 32 Xeon-Gold 6142 2.6GHz processors with 196 GB of RAM. The full inference, with $5\cdot 10^8$ particle updates, required about 120 hours.

It is worth noting that the SABC posterior distribution was generated using $10\ 000$ particles to provide a comparison consistent with the other machine-learning–based methods, SNLE and APT. Nonetheless, we emphasize that using only $1\ 000$ particles yields a posterior distribution that is practically indistinguishable from the one obtained with $10\ 000$ particles. Accordingly, the computing time can be reduced proportionally to just 12 hours.

\clearpage
\subsection{Additional results}
\label{app:solardynamo-additional-results}

\begin{figure}[h]
\centering
\includegraphics[width=0.8\textwidth]{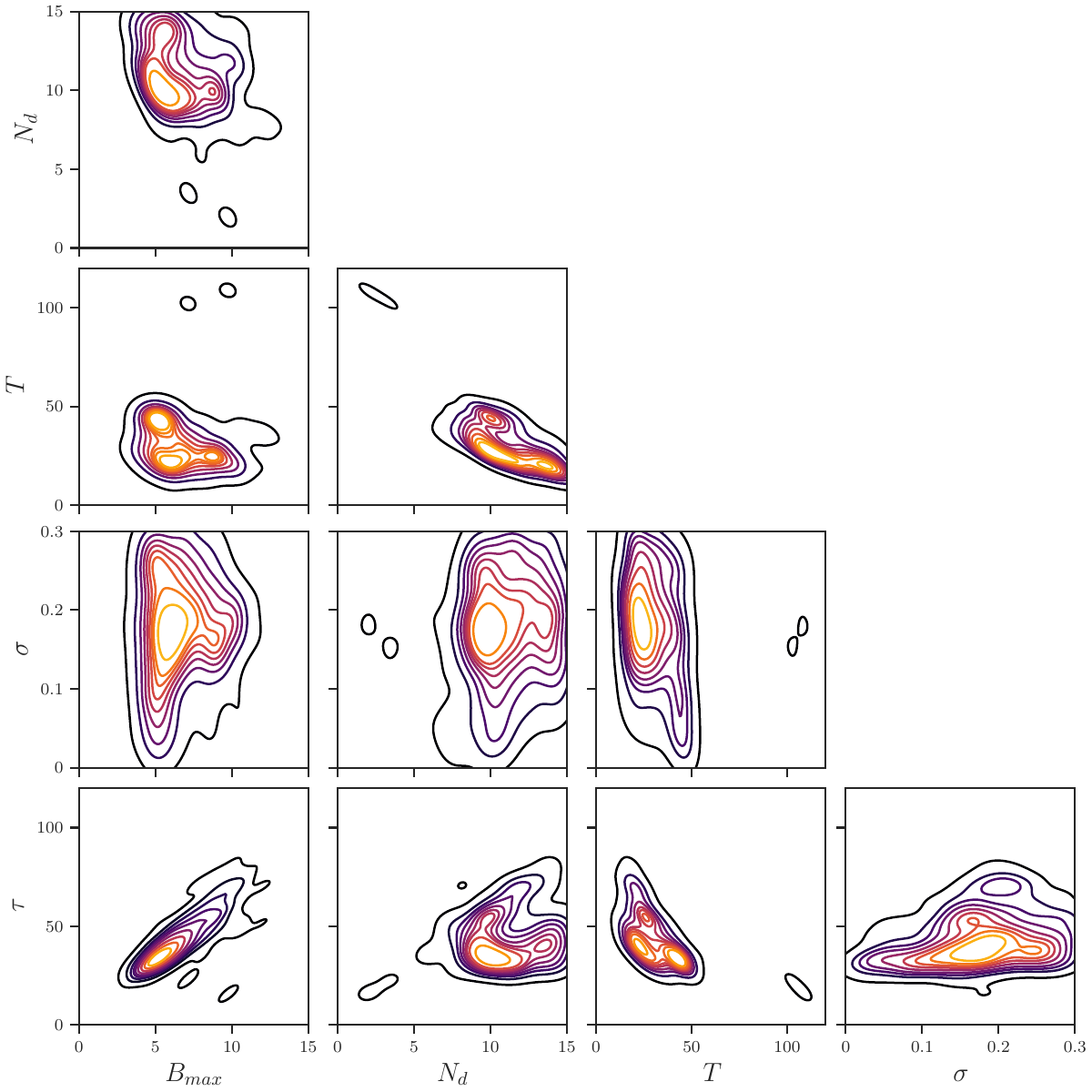}
\caption{SABC bivariate densities for the SN record.}
\label{fig:sabc-solardynamo-sabc-bivariate}
\end{figure}

\begin{figure}
\centering
\includegraphics[width=0.8\textwidth]{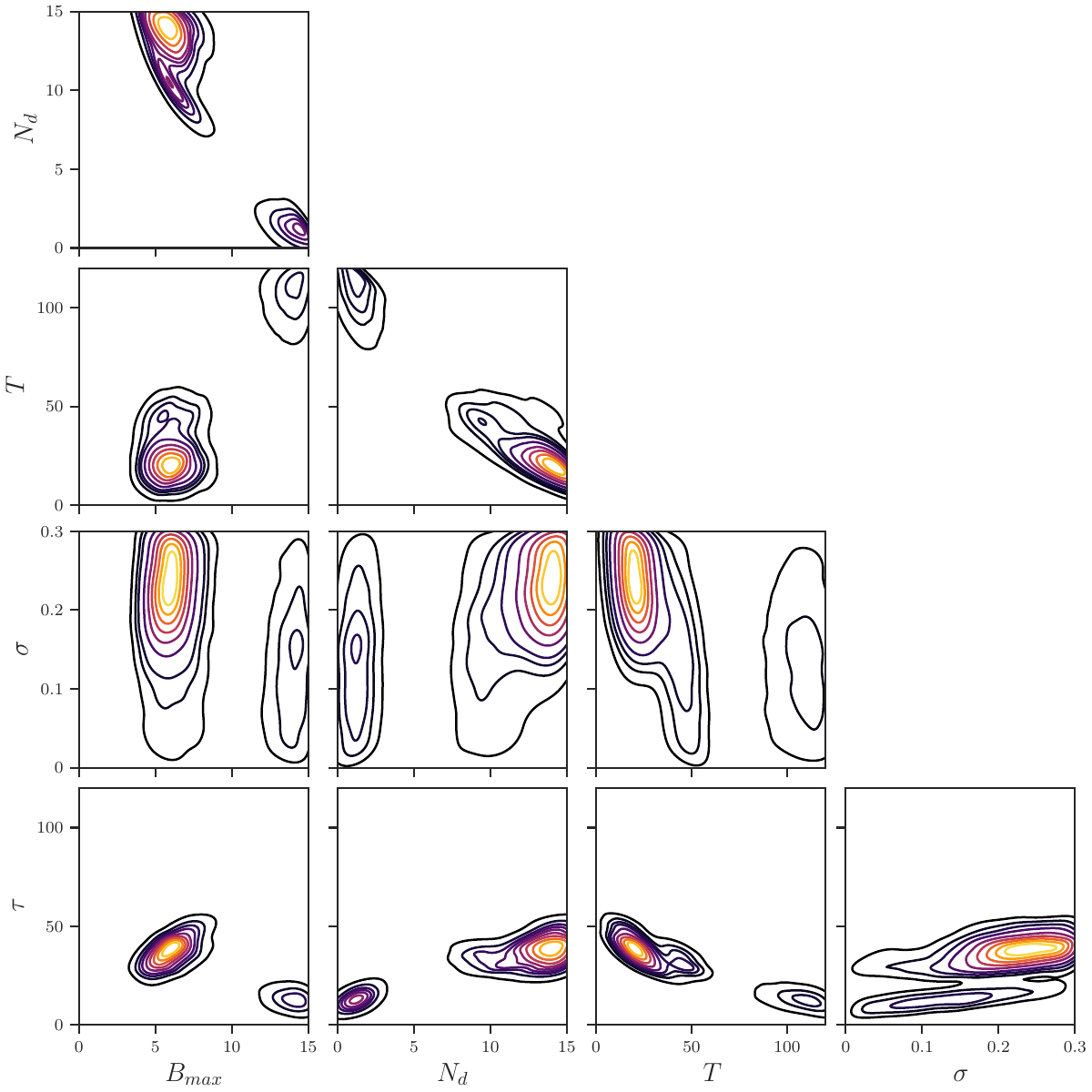}
\caption{Sequential APT bivariate densities for the SN record after $10$ rounds.}
\label{fig:sabc-solardynamo-apt-bivariate}
\end{figure}

\begin{figure}
\includegraphics[width=0.8\textwidth]{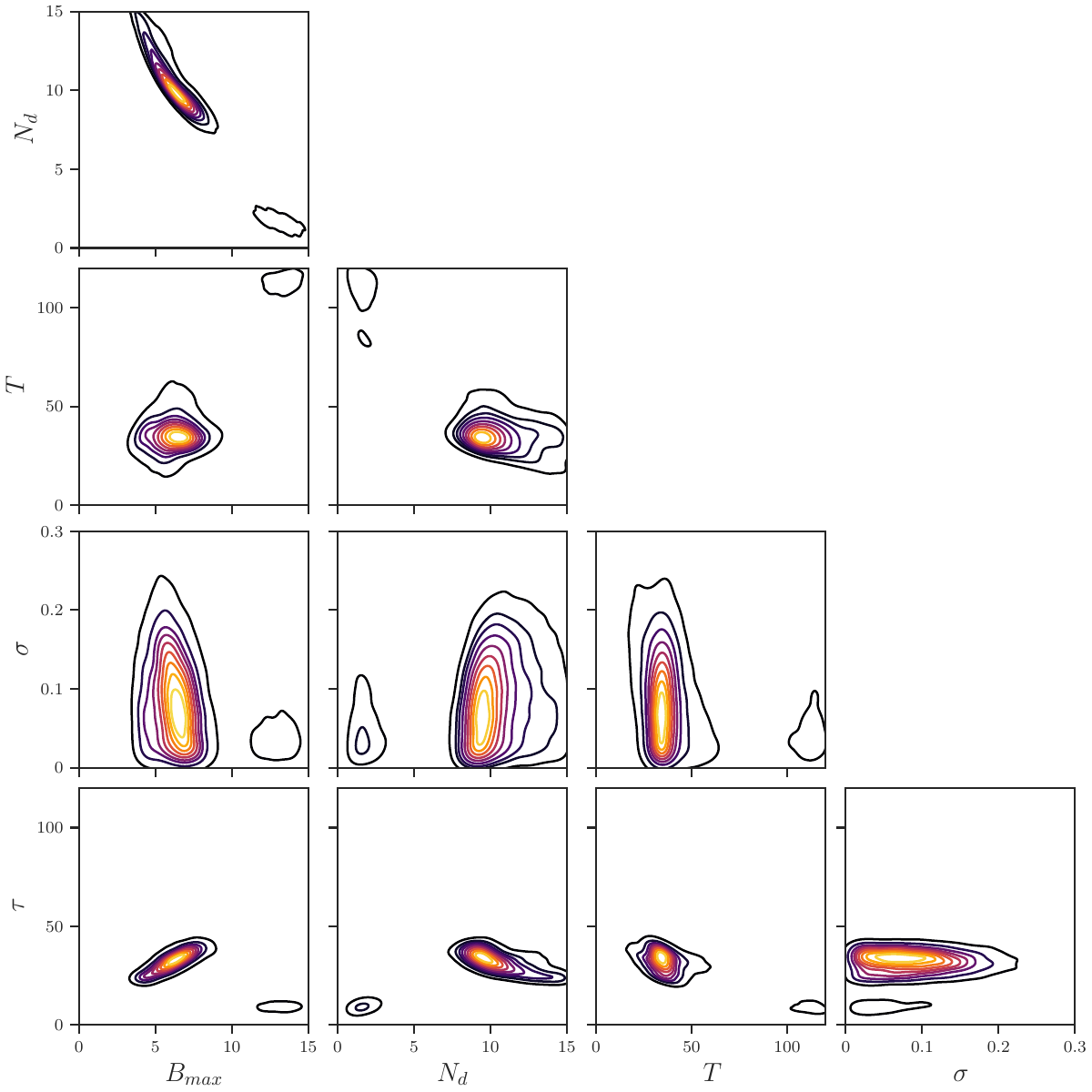}
\caption{Sequential NLE bivariate densities for the SN record after $10$ rounds.}
\label{fig:sabc-solardynamo-snle-bivariate} 
\end{figure}

\begin{figure}
\includegraphics[width=0.8\textwidth]{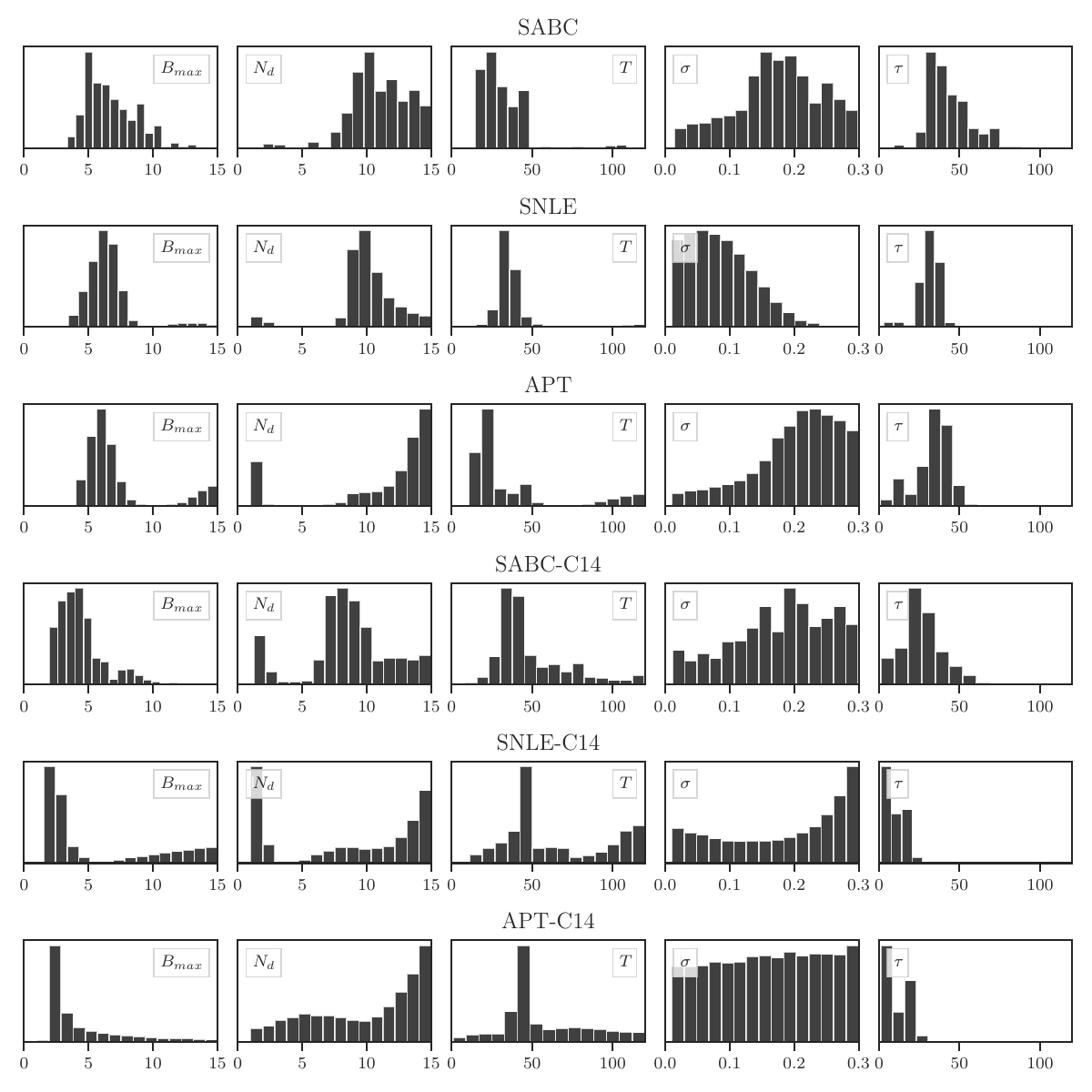}
\caption{Marginal distributions of the posterior of SABC, SNLE and APT for the sun spots (top three rows) and $^{14}$C (bottom three rows), respectively.}
\label{fig-solar-AllMarginals} 
\end{figure}

\begin{figure}
\includegraphics[width=0.8\textwidth]{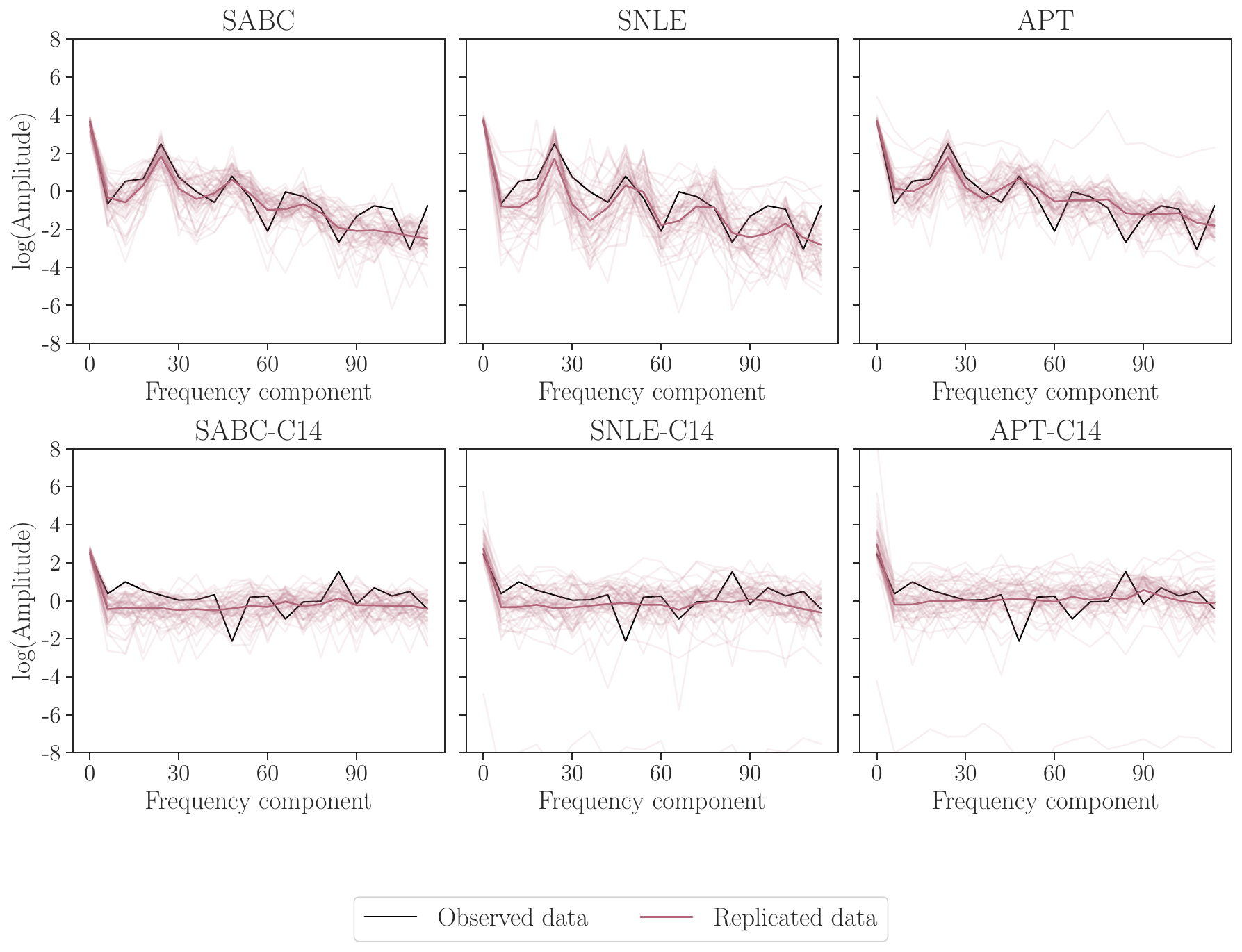}
\caption{Posterior predictive distributions of SABC, SNLE and APT, for the sun spots (top row) and $^{14}$C (bottom row), respectively.}
\label{fig-solar-AllPPD} 
\end{figure}

\begin{figure}
\includegraphics[width=1\textwidth]{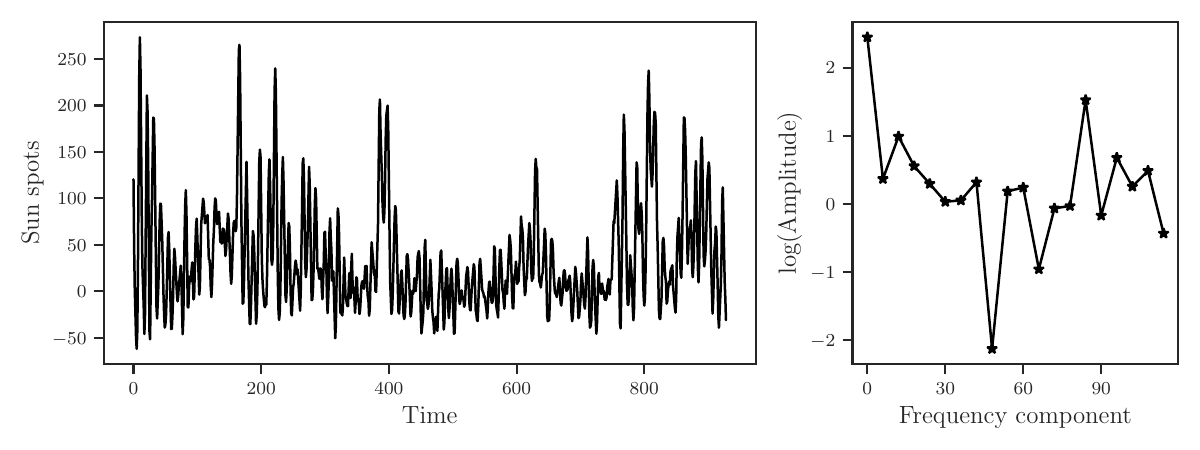}
\caption{$^{14}$C data and the 20 FFT components used for posterior inference. The dataset comprises 929 yearly observations over the period 971-1899.}
\label{fig:C14data}
\end{figure}

\begin{figure}
\begin{subfigure}[b]{0.45\textwidth}
\includegraphics[width=1\textwidth]{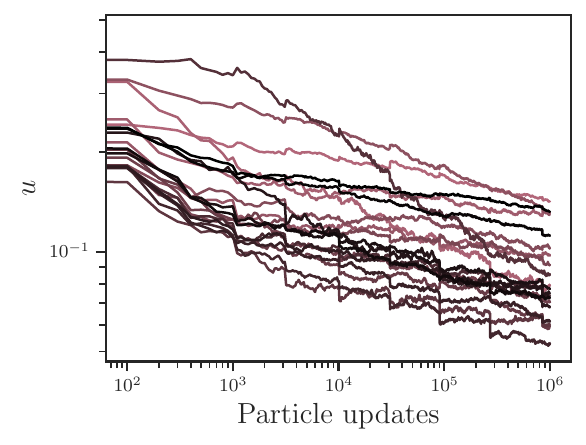}
\end{subfigure}
~
\begin{subfigure}[b]{0.45\textwidth}
\includegraphics[width=1\textwidth]{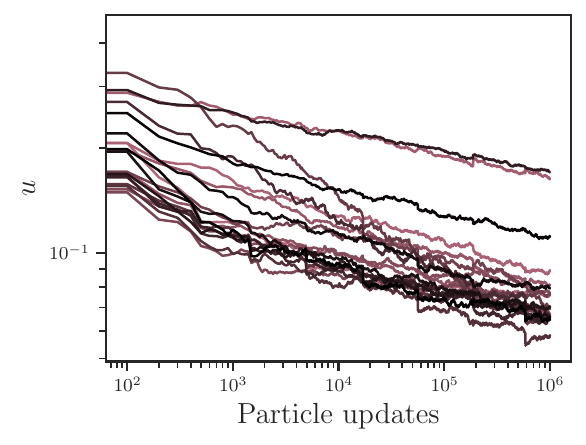}
\end{subfigure}
\caption{Convergence of energies ($u$´s) of SABC for sunspots (left) and $^{14}$C (right). The colors represent the different summary statistics (i.e., the 20 FFT components). The two trailing energies in the right panel correspond to Fourier components that are most out-of-sample.}
\label{fig-solar-uCurves} 
\end{figure}